\documentclass[useAMS,usenatbib]{mn2e}
\usepackage{epsfig}
\usepackage{graphicx}
\usepackage{times}
\usepackage{color}

\def\cm3{cm$^{-3}$}
\def\kms{km~s$^{-1}$}

\def\rsun{R$_{\odot}$}
\def\rphot{$R_{\rm phot}$}
\def\tphot{$T_{\rm phot}$}
\def\vphot{$V_{\rm phot}$}

\def\msun{M$_{\odot}$}

\def\one{{\,\sc i}}
\def\two{{\,\sc ii}}
\def\three{{\,\sc iii}}

\def\ergs{\,erg\,s$^{-1}$}
\def\foe{10$^{51}$\,erg}

\def\beq{\begin{equation}}
\def\eeq{\end{equation}}

\def\lesssim{\mathrel{\hbox{\rlap{\hbox{\lower4pt\hbox{$\sim$}}}\hbox{$<$}}}}
\def\gtrsim{\mathrel{\hbox{\rlap{\hbox{\lower4pt\hbox{$\sim$}}}\hbox{$>$}}}}

\def\aj{AJ}

\def\apj{ApJ}

\def\apjl{ApJL}
\def\aap{A\&A}

\def\mnras{MNRAS}

\def\pasa{Publications of the Astron. Soc. of Australia}

\def\isoni{$^{56}{\rm Ni}$}

\def\cmfgen{{\sc cmfgen}}

\def\kepler{{\sc kepler}}
\def\v1d{{\sc v1d}}

\newcommand{\iso}[2]{\ensuremath{^{#1}\rm{#2}}}

\title[Radiative transfer of SNe IIb/Ib/Ic]{Radiative-transfer models for supernovae IIb/Ib/Ic from binary-star progenitors}
\author
[Luc Dessart et al.]
{\vspace{0.3cm} Luc Dessart,$^1$\thanks{email: Luc.Dessart@oca.eu}
D. John Hillier,$^{2}$
Stan Woosley,$^3$
Eli Livne,$^4$
Roni Waldman,$^4$
\newauthor
Sung-Chul Yoon,$^5$
and Norbert Langer.$^6$
\newauthor
\\
$^{1}$:
Laboratoire Lagrange, Universit\'e C\^{o}te d'Azur, Observatoire de la C\^{o}te
d'Azur, CNRS, Boulevard de l'Observatoire, CS 34229, 06304 Nice cedex 4, France. \\
$^2$: Department of Physics and Astronomy \& Pittsburgh Particle Physics,
Astrophysics, and Cosmology Center (PITT PACC),  University of Pittsburgh, \\
3941 O'Hara Street, Pittsburgh, PA 15260, USA. \\
$^3$: Department of Astronomy and Astrophysics, University of California,
Santa Cruz, CA 95064, USA. \\
$^4$: Racah Institute of Physics, The Hebrew University, Jerusalem 91904, Israel. \\
$^5$: Department of Physics and Astronomy, Seoul National University, Gwanak-ro 1, Gwanak-gu, Seoul, 151-742, Republic of Korea \\
$^6$: Argelander-Institut f\"{u}r Astronomie, Universit{\"a}t Bonn, Auf dem H\"{u}gel 71, 53121, Bonn, Germany \\
}
\date{Accepted 2015 July 28.  Received 2015 July 27; in original form 2015 June 11.}

\pagerange{\pageref{firstpage}--\pageref{lastpage}} \pubyear{2015}

\begin{document}

\maketitle

\label{firstpage}

\begin{abstract}
We present 1-D non-Local-Thermodynamic-Equilibrium time-dependent
radiative-transfer simulations for supernovae (SNe) of type  IIb, Ib, and Ic that result from
the terminal explosion of the mass donor in a close-binary system.
Here, we select three ejecta with a total kinetic energy of
$\approx$\,1.2$\times$10$^{51}$\,erg,
but characterised by different ejecta masses (2-5\,\msun), composition, and chemical mixing.
The type IIb/Ib models correspond to the progenitors that have retained their He-rich shell
at the time of explosion.
The type Ic model arises from a progenitor that has lost its helium shell, but retains
0.32\,\msun\ of helium in a CO-rich core of 5.11\,\msun.
We discuss their photometric and spectroscopic properties during the first
2-3 months after explosion, and connect these to their progenitor and ejecta properties
including chemical stratification.
For these three models, Arnett's rule overestimates the \isoni\ mass by $\approx$\,50\% while the
procedure of Katz et al., based on an energy argument, yields a
more reliable estimate. The presence of strong C\one\ lines around 9000\AA\  prior to maximum is an
indicator that the pre-SN star was under-abundant in helium. As noted by others, the
1.08$\mu$m feature is a complex blend of C\one, Mg\two, and He\one\ lines, which
makes the identification of He uncertain in SNe Ibc unless other He\one\ lines can be identified.
Our models show little scatter in ($V-R$) colour 10\,d after $R$-band maximum.
We also address a number of radiative transfer properties of SNe Ibc,
including the notion of a photosphere, the inference of a representative ejecta expansion rate,
spectrum formation, blackbody fits and ``correction factors''.
\end{abstract}


\section{Introduction}

  SNe IIb/Ib/Ic are understood to arise from the collapse of the iron core of massive stars having lost
  most (type IIb) or all (types Ib and Ic) of their envelope exterior to the He core. Wolf-Rayet stars are therefore
  the natural progenitors of these transient phenomena.
  However in the 80's, after the first well sampled light curves of type I core-collapse SNe were obtained, it became evident that
  their ejecta were of low mass \citep{ensman_woosley_88, WLW95}, in accord with the binary evolution scenario and moderate
  progenitor masses \citep{wellstein_langer_99}.
  As wind mass loss rates for Wolf-Rayet stars were revised downwards (see, e.g., \citealt{hillier_91}, \citealt{nugis_lamers_00})
  and inferences for low SN Ib/c ejecta strengthened
  \citep{woosley_94_93j,dessart_11_wr,drout_11_ibc,bersten_etal_12_11dh,hachinger_13_he,benvenuto_etal_13_11dh,
  ergon_14_11dh,jerkstrand_15_iib},
  the binary star evolution channel became the main scenario for the production of type I core collapse SNe
  (see, e.g., \citealt{yoon_ibc_15}).

  An extensive study of binary evolution, and its implications for SN types, was made by  \citet{podsiadlowski_92}.
  As part of their study, they estimated probabilities for various evolutionary scenarios for massive stars with masses between
  8 and 20\,\msun. They concluded that a complete understanding of massive star evolution will only be obtained when we
  understand binary star evolution. This also requires a census of binaries and their parameters. A more recent study
  is that of \cite{claeys+11} who studied the roles of binaries in producing type IIb SNe.

 In this first paper of a series we perform 1-D non-Local-Thermodynamic-Equilibrium (non-LTE) time-dependent
radiative-transfer simulations for supernovae of type  IIb, Ib, and Ic that result from
the terminal explosion of the primary star in a close-binary system.  The progenitor models are taken from
the study of \citet{yoon_etal_10}, which treats the evolution from the main sequence until neon burning.
The later stages of evolution, and the subsequent explosion is done with \kepler\
(Woosley et al., in preparation; \citealt{kepler_78}; Section~\ref{sect_setup}).

This endeavour extends beyond the previous studies of \citet{dessart_11_wr, dessart_etal_12}
by using better physics, a broader range of progenitors, and a broader range of explosion energies and \iso{56}Ni mass.
In \citet{dessart_11_wr}, non-thermal effects and chemical mixing were ignored. In \citet{dessart_etal_12}, the focus
was on mixing alone, based on only two explosion models with the same energy.
Both progenitor models had a large helium mass.
Here, the full set comprises about 30 model sequences covering a much broader parameter space.
We compute  full time sequences in non-LTE, with allowance for time dependence, non-thermal effects
and non-local energy deposition (Section~\ref{sect_setup}).
Throughout this series, we limit the discussion to the photospheric phase. The nebular phase properties,
because they are so strongly connected to the nature of mixing, will be treated in a separate study.
Our goal is to characterise the light curve and spectral properties of SNe IIb/Ib/Ic and correlate them
to the explosion model and to the progenitor star. This includes a discussion of the bolometric
light curves, multi-band light curves, colour evolution (Section~\ref{sect_phot}),
spectral evolution (Section~\ref{sect_spec}), the properties of near-IR He\one\ lines
(Section~\ref{sect_hei}), the influence of mixing (Section~\ref{sect_mixing}),
the properties of spectrum formation and the notion of a photosphere in type I SNe (Section~\ref{sect_spec_form}),
the inference of the ejecta expansion rate and the constraint on chemical stratification (Section~\ref{sect_vm}),
and the correction factors to be used when approximating the synthetic spectra of SNe IIb/Ib/Ic near maximum light
with a blackbody (Section~\ref{sect_corfac}).

In Section~\ref{sect_conc}, we summarise the results of the present paper.
In  Paper\two, we will do a statistical study based on the entire grid of about 30 model sequences
to examine correlations/distinction between the models.
In Paper\three, we will use our grid of models to compare to observations of SNe IIb/Ib/Ic.

\begin{figure*}
\epsfig{file=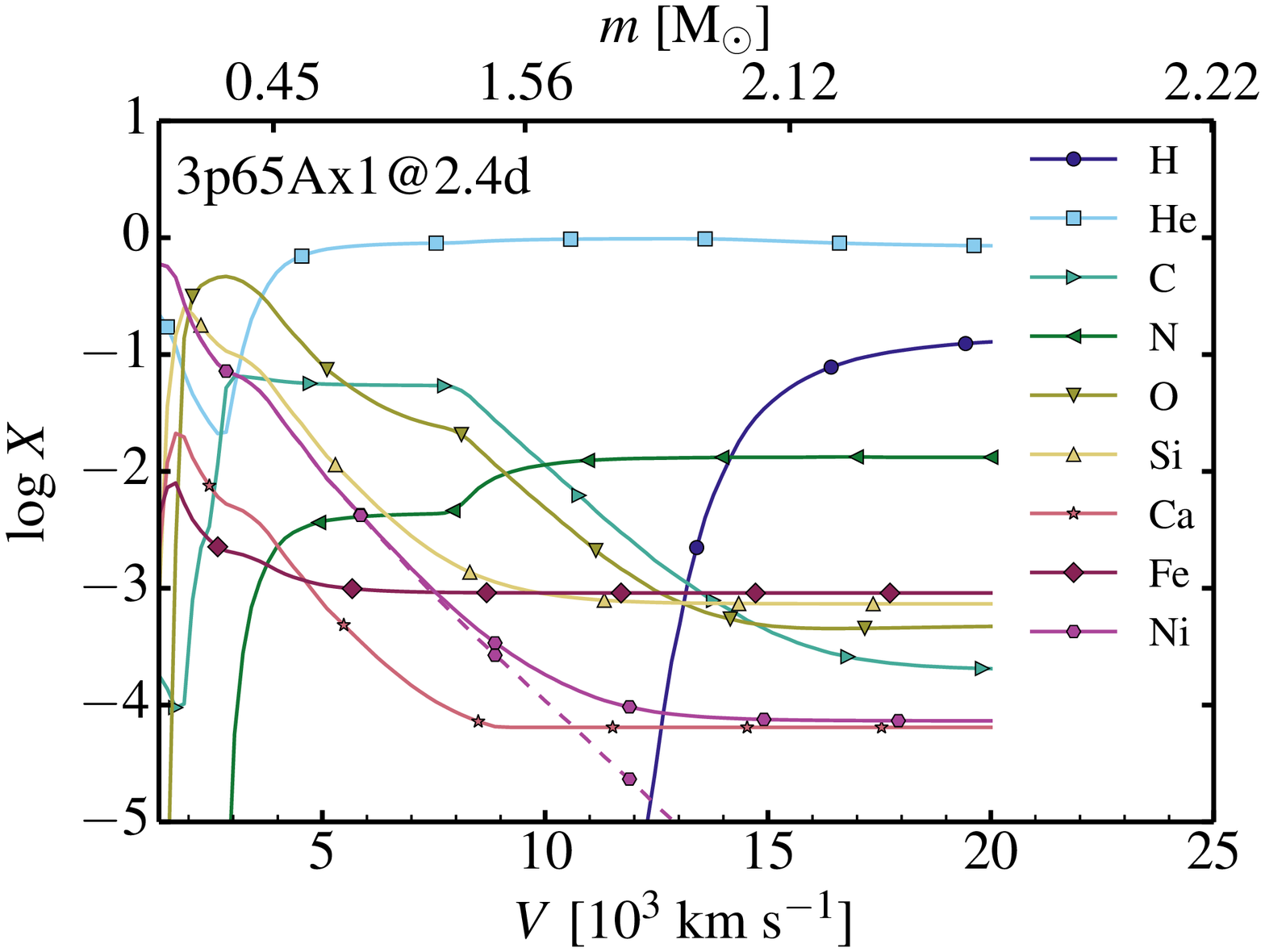,width=8.75cm}
\epsfig{file=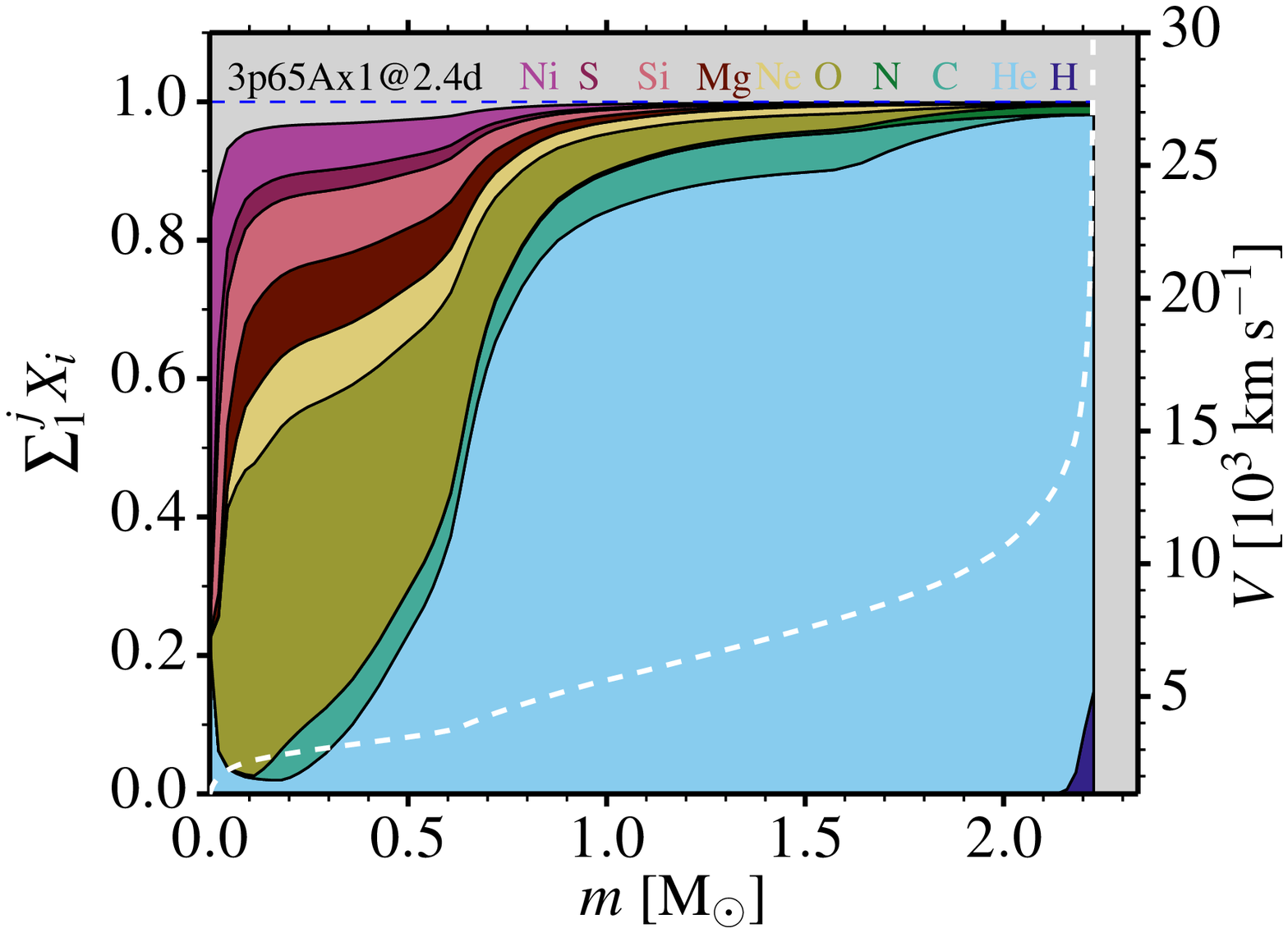,width=8.75cm}
\epsfig{file=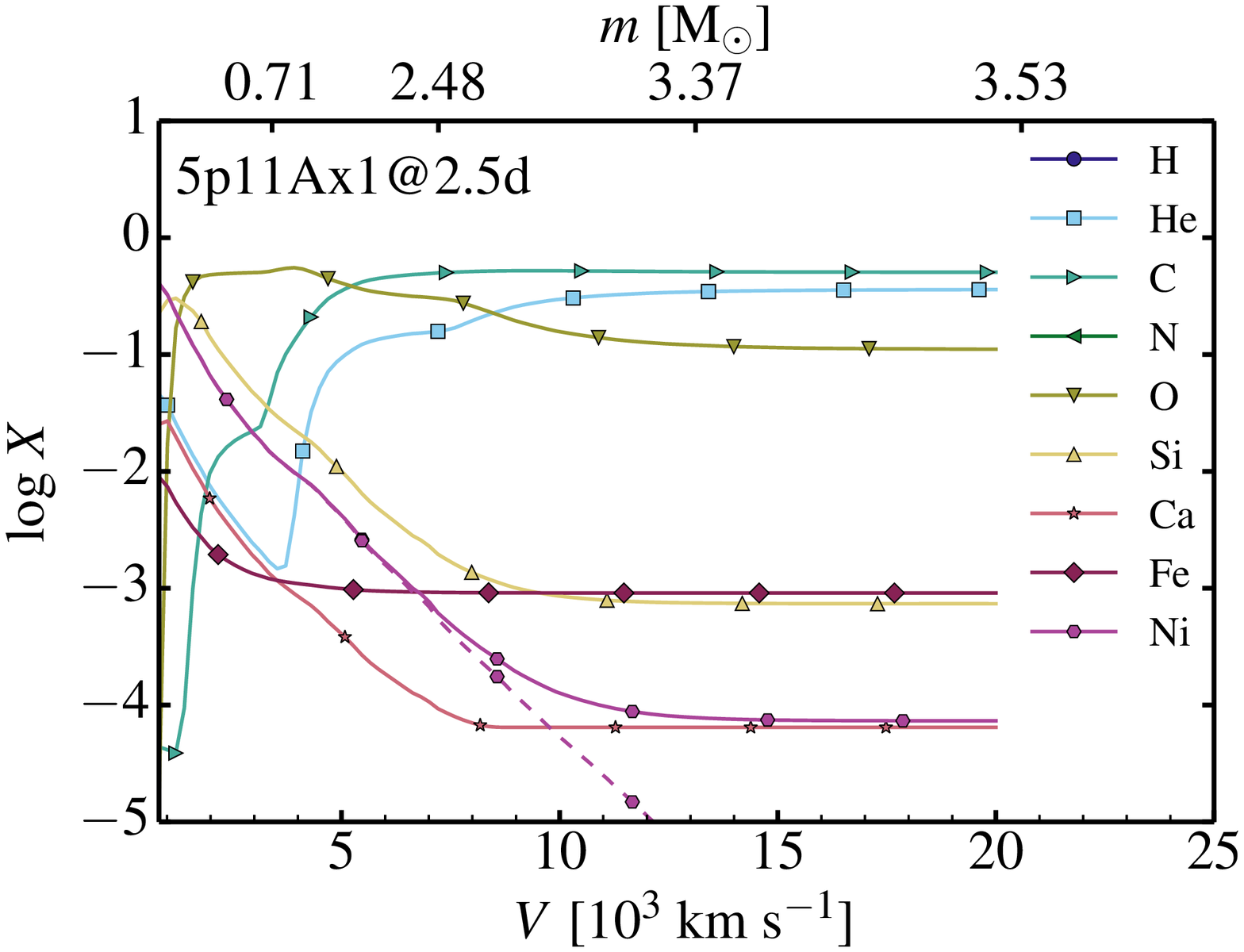,width=8.75cm}
\epsfig{file=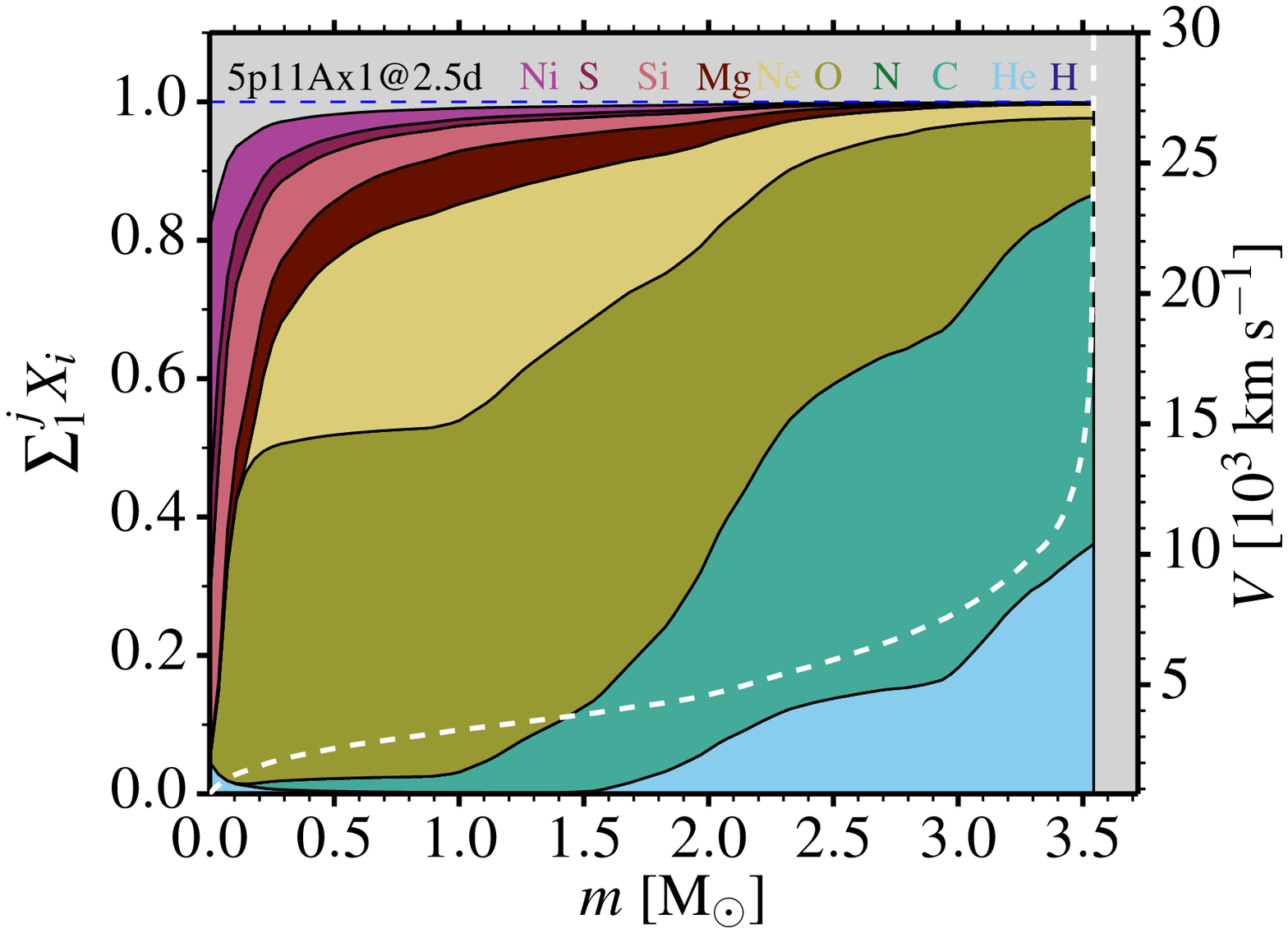,width=8.75cm}
\epsfig{file=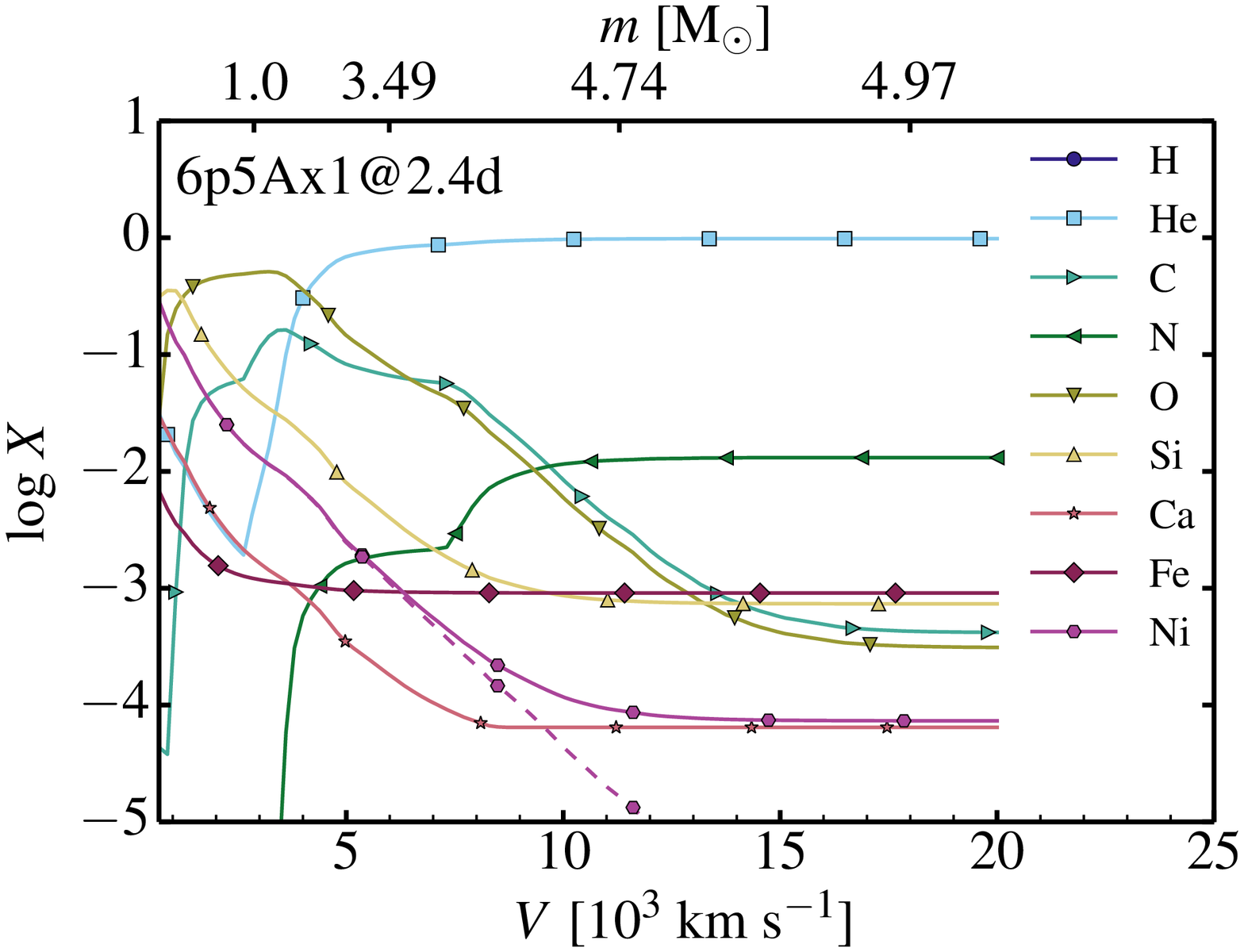,width=8.75cm}
\epsfig{file=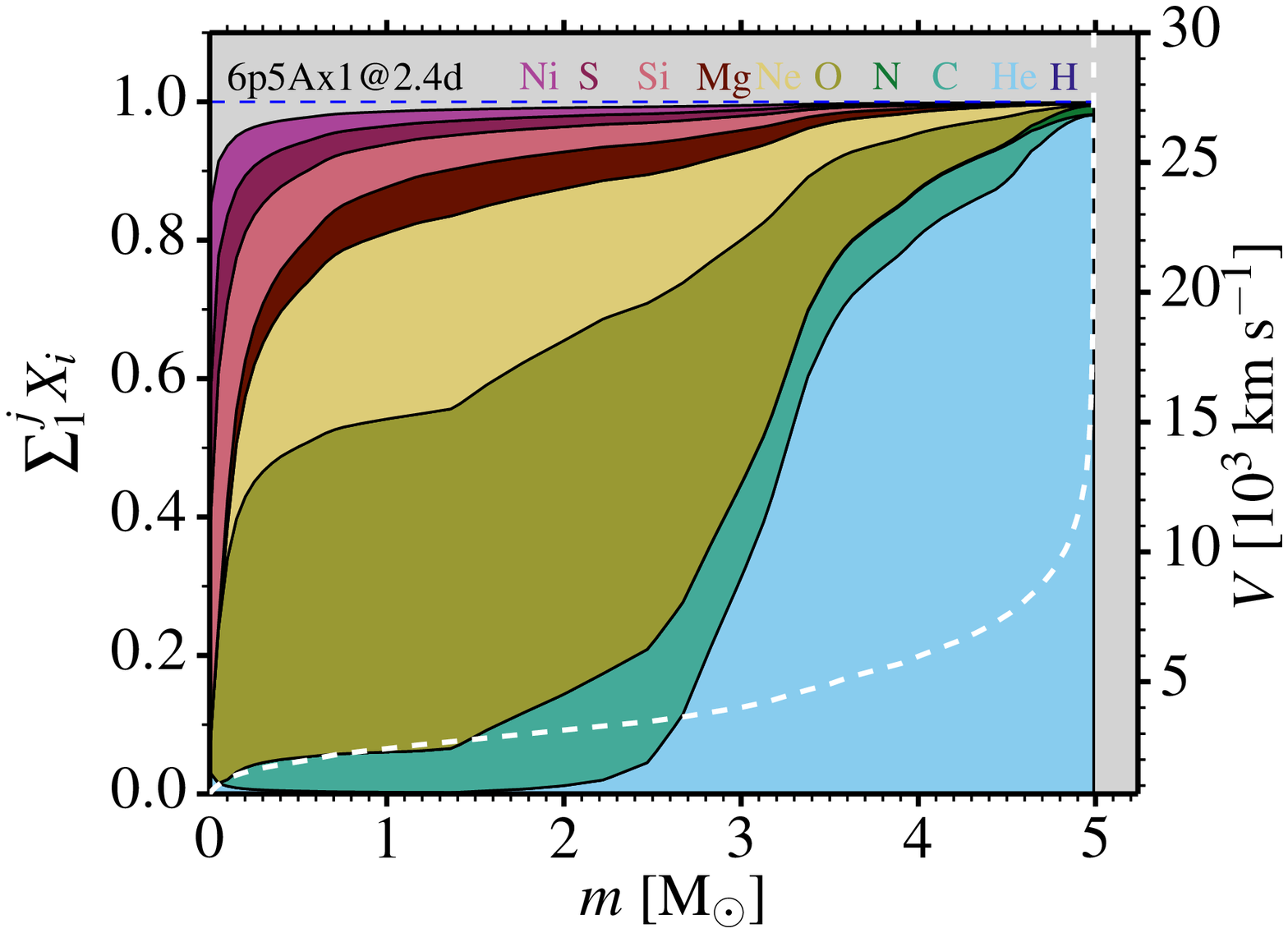,width=8.75cm}
\caption{
Ejecta composition at $\approx$\,2.5\,d after explosion for models 3p65Ax1 (top row), 5p11Ax1 (middle row),
and 6p5Ax1 (bottom row). In the left column, we show the mass fraction versus velocity
for a selection of species. The dashed line corresponds to \iso{56}Ni. The top axis
gives the lagrangian mass coordinate. For clarity, the curve
for \iso{56}Co is not shown; at 2.5\,d it amounts to  $\approx$\,1/4 of the \iso{56}Ni mass.
In the right column, we show for each mass shell $m$ the cumulative mass fraction
$\sum_1^j X_i(m)$, where $i$ covers species H, He, C, N, O, Ne, Mg, Si, S, and Ni
(for Ni, we show the current total mass fraction, which therefore accounts for decays).
We also show the velocity (dashed line, right axis).
This illustration reveals more clearly the relative fraction occupied by abundant species.
The variable size of the mass buffer between the He-rich layers and the \iso{56}Ni rich layers
is key to understand the production of He\one\ lines.
\label{fig_comp_set}}
\end{figure*}

\begin{table}
\begin{center}
\caption{Summary of progenitor properties at the time of core collapse. The last column gives the iron-core mass.
\label{tab_presn}}
\begin{tabular}{
l@{\hspace{2mm}}c@{\hspace{2mm}}c@{\hspace{2mm}}c@{\hspace{2mm}}
c@{\hspace{2mm}}c@{\hspace{2mm}}c@{\hspace{2mm}}c@{\hspace{2mm}}}
\hline
Model    &    $M_{\rm i}$  &     $M_{\rm f}$ & L$_{\star}$ & T$_{\rm eff}$ & R$_{\star}$ &  M$_{\rm CO}$  & M$_{\rm Fe}$ \\
               &        [\msun]      &        [\msun]      &      [\ergs]   & [K]   & [cm]   &  [\msun]  &  [\msun]  \\
\hline
3p65  & 16.0  &   3.65  &  2.16(38)    &     21\,700  &    1.17(12)    &  2.03  &  1.43 \\
5p11  & 60.0  &   5.11  &  5.42(38)      &   130\,000  &    5.19(10)  &  3.64 &  1.52 \\
6p5    &  25.0  &   6.50  &   6.40(38)       & 68\,700    &  2.01(11)   &  4.21 &    1.48  \\
\hline
\end{tabular}
\end{center}
\end{table}

%
% Following tables are made with ~/python/cmfgen/prop_model_list_initial.py
%
\begin{table}
\begin{center}
\caption{Summary of ejecta properties. The last three columns give the ejecta velocity that bounds 99\%
of the corresponding species total mass. The integration is done inwards in velocity space for H and He, and
outwards for \iso{56}Ni.
\label{tab_ejecta_glob}}
\begin{tabular}{
l@{\hspace{3mm}}c@{\hspace{3mm}}c@{\hspace{3mm}}c@{\hspace{3mm}}
c@{\hspace{3mm}}c@{\hspace{3mm}}c@{\hspace{3mm}}c@{\hspace{3mm}}
c@{\hspace{3mm}}
}
\hline
Model    &    $M_{\rm r}$ & $M_{\rm e}$ &        $E_{\rm kin}$  &  $V_{\rm 99, H}$&  $V_{\rm 99, He}$&  $V_{\rm 99, Ni}$  \\
               &               [\msun] &        [\msun] &      [B] &      [\kms]  &      [\kms]  &      [\kms]      \\
%              Mremnant          Mejecta                Z         Ekin [B]          V99 HYD           V99 HE         V99 NICK
\hline
3p65Ax1  &    1.43  &   2.22  &        1.24  &   1.34(4)  &   3.06(3)  &   7.00(3)  \\
3p65Ax2  &    1.43  &   2.22  &        1.22  &   1.24(4)  &   2.62(3)  &   1.01(4)  \\
5p11Ax1  &    1.57  &   3.54  &      1.25  &    \dots  &   1.68(3)  &   6.20(3)  \\
5p11Ax2  &    1.52  &   3.59  &      1.29  &    \dots  &   2.71(3)  &   9.18(3)  \\
6p5Ax1  &     1.53 &   4.97  &       1.26  &    \dots  &   3.28(3)  &   5.98(3)  \\
6p5Ax2  &     1.55 &   4.95  &       1.25  &    \dots  &   2.55(3)  &   8.60(3)  \\
\hline
\end{tabular}
\end{center}
\end{table}

\begin{table*}
\begin{center}
    \caption{
Cumulative yields for our grid of models at $\approx$\,2\,d after explosion. For \iso{56}Ni, we give the original mass,
i.e., prior to decay. Models with different mixing show slightly different levels of fallback, and consequently
show slightly different yields.
\label{tab_ejecta_mass}}
\begin{tabular}{
l@{\hspace{5mm}}c@{\hspace{5mm}}c@{\hspace{5mm}}c@{\hspace{5mm}}
c@{\hspace{5mm}}c@{\hspace{5mm}}c@{\hspace{5mm}}c@{\hspace{5mm}}
c@{\hspace{5mm}}c@{\hspace{5mm}}c@{\hspace{5mm}}c@{\hspace{5mm}}
c@{\hspace{5mm}}c@{\hspace{5mm}}
}
\hline
Model    &   H   &   He  &   C  &   N  &   O  &   Si  &    S &   Ca  &     ($^{56}$Ni)$_0$ \\
         &        [\msun] &        [\msun] &        [\msun] &        [\msun] &        [\msun] &        [\msun] &        [\msun] &        [\msun] &               [\msun]       \\
\hline
%              M H             M He              M C              M N              M O             M Si              M S             M Ca             M Fe         M  56 Ni
3p65Ax1  &  4.99(-3)  & 1.49(0)   & 9.40(-2)  & 1.09(-2)  & 3.02(-1)  & 7.38(-2)  & 2.36(-2)  & 3.94(-3)  &  7.42(-2) \\
3p65Ax2  &  4.72(-3)  & 1.48(0)   & 9.37(-2)  & 1.08(-2)  & 3.01(-1)  & 7.30(-2)  & 2.33(-2)  & 3.91(-3) & 7.66(-2) \\
5p11Ax1  &     0      & 3.15(-1)  & 8.92(-1)  &    0      & 1.42(0)   & 1.28(-1)  & 4.00(-2)  & 5.98(-3)  &  8.94(-2) \\
5p11Ax2  &     0      & 3.26(-1)  & 9.23(-1)  &    0      & 1.44(0)   & 1.28(-1)  & 4.04(-2)  & 6.13(-3)  &  9.46(-2) \\
6p5Ax1  &      0      & 1.67(0)   & 4.13(-1)  & 7.59(-3)  & 1.57(0)   & 2.12(-1)  & 9.31(-2)  & 9.30(-3)  &  9.90(-2) \\
6p5Ax2  &      0      & 1.66(0)   & 4.11(-1)  & 7.52(-3)  & 1.57(0)   & 2.14(-1)  & 9.40(-2)  & 9.51(-3)  &  1.02(-1) \\
\hline
\end{tabular}
\end{center}
\end{table*}

\begin{table*}
\begin{center}
    \caption{
Mass fractions for some important species in the outermost mass shell for our grid of models.
This shell corresponds to the progenitor surface. The same outermost-shell composition
is obtained for the x2-model counterparts (hence not shown).
\label{tab_ejecta_surf}}
\begin{tabular}{
l@{\hspace{5mm}}c@{\hspace{5mm}}c@{\hspace{5mm}}c@{\hspace{5mm}}
c@{\hspace{5mm}}c@{\hspace{5mm}}c@{\hspace{5mm}}c@{\hspace{5mm}}
c@{\hspace{5mm}}c@{\hspace{5mm}}}
\hline
%             Xs H            Xs He             Xs C             Xs N             Xs O            Xs Si             Xs S            Xs Ca            Xs Fe
Model    &   $X_{\rm H,s}$ &  $X_{\rm He,s}$  &   $X_{\rm C,s}$ &   $X_{\rm N,s}$ &   $X_{\rm O,s}$ &   $X_{\rm Si,s}$ &   $X_{\rm S,s}$ &   $X_{\rm Ca,s}$ &  $X_{\rm Fe,s}$  \\
\hline
3p65Ax1  &  1.480(-1)  & 8.332(-1)  & 1.980(-4)  & 1.320(-2)  & 4.831(-4)  & 7.352(-4)  & 3.651(-4)  & 6.442(-5)  & 9.086(-4)  \\
5p11Ax1  &     0     & 3.611(-1)  & 5.051(-1)  &    0     & 1.100(-1)  & 7.362(-4)  & 3.651(-4)  & 6.442(-5)  & 9.097(-4)  \\
6p5Ax1  &      0     & 9.813(-1)  & 4.147(-4)  & 1.309(-2)  & 3.098(-4)  & 7.345(-4)  & 3.647(-4)  & 6.435(-5)  & 9.086(-4)  \\
\hline
\end{tabular}
\end{center}
\end{table*}

\section{Numerical setup and initial conditions}
\label{sect_setup}

\subsection{Pre-supernova evolution, explosion, and supernova radiative transfer}

  \citet{yoon_etal_10} computed the evolution from the main sequence until the end of neon burning,
  including differential rotation, tides and mass and angular momentum transfer, for a
  limited set of close-binary systems (orbital period of $\approx$\,4\,d), covering a range of primary-star masses
  (12 to 60\,\msun) and mass ratios ($\gtrsim$1-1.5).  These simulations were then evolved to core collapse
  (Woosley et al., in preparation), and subsequently exploded with \kepler\ using a piston.

  We have performed a grid of explosion models, corresponding to the binary models in the Yoon et al. study
  that are numbered Yoon\_13 (the final mass of the primary star, $M_{\rm 1,f}$, is 3.0\,\msun),
  Yoon\_9 ($M_{\rm 1,f}=$\,3.65\,\msun),
  Yoon\_35 ($M_{\rm 1,f}=$\,4.64\,\msun; $Z=$\,0.004), Yoon\_31\footnote{This model has a final mass of 5.11\,\msun,
  and not 4.96\,\msun\ as given in Table~1 of \citet{yoon_etal_10}. This is because the remapping of this model
  into \kepler\ was done at an earlier stage and the final evolution to core collapse neglected wind mass loss.}
  ($M_{\rm 1,f}=$\,5.11\,\msun),
  and Yoon\_30 ($M_{\rm 1,f}=$\,6.5\,\msun). All models use the solar composition initially, except model Yoon\_35.
  This choice includes models with residual surface hydrogen (Yoon\_13, 9, 35), no hydrogen but high helium
  and nitrogen abundances  (Yoon\_30), and little helium but high abundance of carbon and oxygen (Yoon\_31).

   In this paper, we limit the discussion to just three progenitor models, 3p65, 5p11, and 6p5 (Table~\ref{tab_presn}).
   Their diversity
   in envelope composition leads them to produce, after explosion, the three common type I core-collapse SN
   IIb, Ib, and Ic. Models 3p0 and 4p64 will be discussed in Papers\two\ and \three.

  Upon reaching core collapse, each model was exploded with \kepler\ to produce ejecta with an asymptotic kinetic energy of
  0.6 (suffix C), 1.2 (suffix A), 2.4 (suffix B), and 5$\times$\foe\ (suffix D).
  In practice, the lower mass models were limited to lower energy explosions (0.6--2.4$\times$\foe).
  Explosive nucleosynthesis is treated in \kepler\ and therefore these simulations naturally produce some \iso{56}Ni.
  Higher energy explosions exhibit higher post-shock temperatures, which favour \iso{56}Ni production.
  The full set of explosion models is characterised by an original \iso{56}Ni mass of 0.05--0.2\,\msun.
  Here, we only discuss ``A'' models, characterised by a kinetic energy at infinity of $\approx$\,1.2$\times$\,\foe.

  These explosion models were then remapped into \v1d\ \citep{livne_93,DLW10b,DLW10a} at 10$^4$\,s and evolved to 10$^5$\,s.
  This additional step is to explore a different mixing approach from that employed in \kepler.
  In our 1-D approach, we take the multi-dimensional process of mixing
  \citep{fryxell_etal_91,kifonidis_etal_06,hammer_15_3d,wongwathanarat_15_3d} into account by applying a simple
  algorithm. Starting from the ejecta base, we make homogeneous all mass shells within 1000 (x1) or 2000\,\kms\ (x2),
  and sweep this mixing-box through all mass shells all the way to the outermost ejecta
  layers.\footnote{In \kepler, the width of the mixing box
  is specified in mass space rather than velocity space. This is problematic for species like H and He that may be present
  only in the outermost layers of the progenitor star, potentially introducing some mixing out to the largest velocities
  in low mass ejecta, which is probably unphysical.} This mixing box is thus applied hundreds of times (i.e., the number
  of times is equal to the number of grid points in the model).
  Because of the overwhelming evidence that SN Ib/c ejecta are mixed
  \citep{lucy_91,dessart_etal_12}, we do not include unmixed models. But because the magnitude of
  mixing is very uncertain, we use two values corresponding to moderate and strong mixing.
  Our approach enforces a macroscopic mixing (large scale mixing of species over a wide range of velocities)
  as well as a microscopic mixing (each mass shell is homogeneous with no segregation of species at a given velocity).
  Multi-dimentional simulations of core-collapse SNe predict strong macroscopic mixing, but very limited microscopic mixing
  \citep[e.g.,][and references there in]{wongwathanarat_15_3d}. Observationally, this distinction becomes evident
  at nebular times \citep{jerkstrand_15_iib}.
  Models with different mixing produce ejecta that suffer a slightly different fallback. This leads to small variations in yields
  (Table~\ref{tab_ejecta_mass}).

   Our nomenclature is to name each model according to its final mass. For example, model 3p65Ax1 corresponds to model Yoon\_9
   which died as a 3.65\,\msun\ star. It was exploded to yield an asymptotic kinetic energy of 1.2$\times$\foe, and was mixed with
   a velocity width of 1000\,\kms.
   %Our full set includes models 3p0, 3p65, 4p64, 5p11, and 6p5.

  At 10$^5$\,s, all models are remapped onto the grid of the non-LTE radiative transfer code \cmfgen\ and evolved
  until nebular times \citep{HD12}. The entire ejecta is treated, and each model is evolved with the same physics.
  We treat non-thermal processes \citep{dessart_etal_12,li_etal_12} and time dependence \citep{DH08}.
  The $\gamma$-ray energy deposition is computed using a Monte Carlo approach \citep{HD12}, and for this study
  only the decay of \iso{56}Ni is taken into account.
  We enforce a minimum mass fraction for metals not included in the small network of 19 isotopes used
  in the pre-SN simulations, and assign them their solar metallicity value.
    The model atoms are the same as those used in \citet{dessart_etal_13b}, with the updates for forbidden
  line transitions discussed in \citet{d14_tech}.

  Convolving the emergent flux computed with {\sc cmf\_flux} \citep{cmf_flux}
  over filter bandpasses and integrating over the full frequency range, we directly extract the
  photometric and bolometric properties of all model sequences.
  Because of the different physics treated in \v1d\ and \cmfgen, a few time steps are needed to relax the
  ejecta initially, in particular in the regions with a Rosseland mean optical depth less than a few.
  Hence, we only show the results for expansion ages greater than $\approx$\,2\,d.

  The main deficiencies of the present simulations are the limitation to close-binary progenitors
  and the use of a mixing procedure that is simplistic. We are in the process of curing these deficiencies.

\begin{figure*}
\epsfig{file=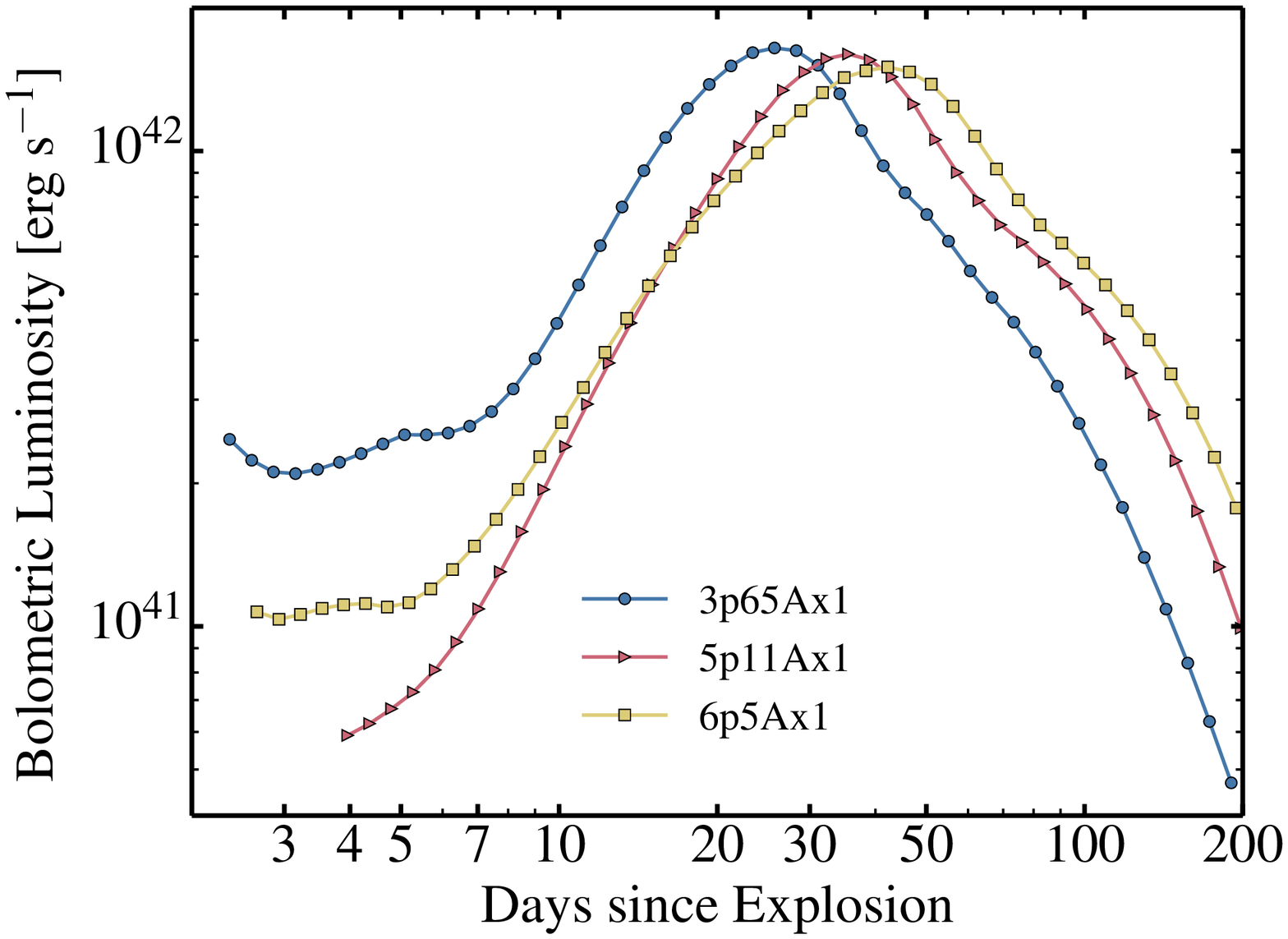,width=8.5cm}
\epsfig{file=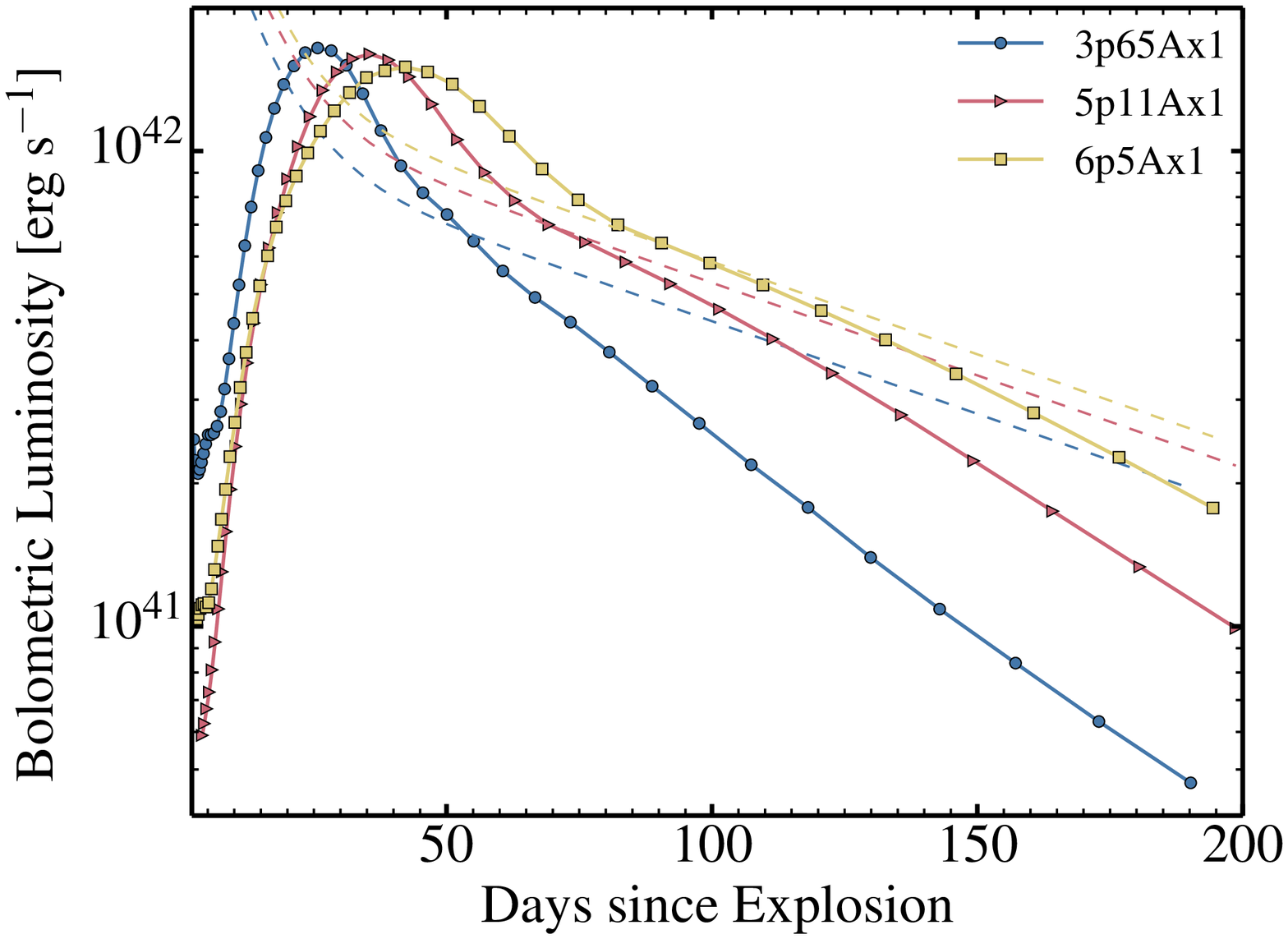,width=8.5cm}
\caption{Bolometric luminosity light curve for models 3p65Ax1, 5p11Ax1,
and 6p5Ax1. We use a logarithmic scale for the $x$-axis in the left panel to show the
contrast in early-time luminosity and the possible existence of a short-lived post-breakout plateau
(even in the presence of mixing). The right panel shows the full evolution, and in particular
the  increase in the peak delay, light curve width, and nebular brightness for higher mass ejecta
(associated in this context with the higher $\gamma$-ray trapping efficiency).
For each model, we overlay the decay power associated with \iso{56}Ni and its daughter
isotope \iso{56}Co (dashed line).
\label{fig_lbol}
}
\end{figure*}

\begin{table*}
\begin{center}
    \caption{
    Luminosity and photometric properties for our model set.
 For each entry, we give the rise time to maximum, the value at peak, and the corresponding
 magnitude change between the time of peak and 15\,d later. The R-band maximum occurs
 within 1 day of the bolometric maximum.
\label{tab_lc_mod1}}
\begin{tabular}{
l@{\hspace{3mm}}
c@{\hspace{3mm}}c@{\hspace{3mm}}c@{\hspace{3mm}}
c@{\hspace{3mm}}c@{\hspace{3mm}}c@{\hspace{3mm}}
c@{\hspace{3mm}}c@{\hspace{3mm}}c@{\hspace{3mm}}
c@{\hspace{3mm}}c@{\hspace{3mm}}c@{\hspace{3mm}}
c@{\hspace{3mm}}c@{\hspace{3mm}}c@{\hspace{3mm}}
}
\hline[mag]
Model   & \multicolumn{3}{c}{$L_{\rm bol}$}   & \multicolumn{3}{c}{$L_{\rm UVOIR}$}   & \multicolumn{3}{c}{$U$}   & \multicolumn{3}{c}{$B$}   \\
               & $t_{\rm rise}$ & Max. & $\Delta M_{15}$ & $t_{\rm rise}$ & Max. & $\Delta M_{15}$ & $t_{\rm rise}$ & Max. & $\Delta M_{15}$ & $t_{\rm rise}$ & Max. & $\Delta M_{15}$ \\
               & [d]  & [erg\,s$^{-1}$] & [mag]  & [d]  & [erg\,s$^{-1}$] & [mag]    & [d]  &[mag] & [mag]  & [d]  & [mag] & [mag]  \\
\hline
3p65Ax1     &   2.631(1)   &   1.648(42)  &   6.157(-1)  &   2.579(1)   &   1.290(42)  &   6.956(-1)  &   2.387(1)   &  -1.576(1)   &   1.157(0)   &   2.410(1)   &  -1.662(1)   &   1.028(0) \\
5p11Ax1     &   3.584(1)   &   1.590(42)  &   3.939(-1)  &   3.506(1)   &   1.182(42)  &   4.234(-1)  &   3.208(1)   &  -1.524(1)   &   6.838(-1)  &   3.256(1)   &  -1.633(1)   &   5.941(-1) \\
6p5Ax1      &   4.212(1)   &   1.495(42)  &   2.262(-1)  &   4.057(1)   &   1.079(42)  &   2.449(-1)  &   3.634(1)   &  -1.497(1)   &   2.554(-1)  &   3.668(1)   &  -1.606(1)   &   3.623(-1) \\
\hline
\end{tabular}
\end{center}
\end{table*}

\begin{table*}
\begin{center}
    \caption{Same as Table~\ref{tab_lc_mod1}, but now for the $V$, $R$, and $I$ bands.
\label{tab_lc_mod2}}
\begin{tabular}{
l@{\hspace{3mm}}
c@{\hspace{3mm}}c@{\hspace{3mm}}c@{\hspace{3mm}}
c@{\hspace{3mm}}c@{\hspace{3mm}}c@{\hspace{3mm}}
c@{\hspace{3mm}}c@{\hspace{3mm}}c@{\hspace{3mm}}
c@{\hspace{3mm}}c@{\hspace{3mm}}c@{\hspace{3mm}}
c@{\hspace{3mm}}c@{\hspace{3mm}}c@{\hspace{3mm}}
}
\hline
Model    & \multicolumn{3}{c}{$V$}   & \multicolumn{3}{c}{$R$}   & \multicolumn{3}{c}{$I$}   \\
      & $t_{\rm rise}$ & Max. & $\Delta M_{15}$ & $t_{\rm rise}$ & Max. & $\Delta M_{15}$ & $t_{\rm rise}$ & Max. & $\Delta M_{15}$ \\
                & [d]  & [mag] & [mag]  & [d]  &[mag] & [mag]   & [d]  & [mag] & [mag]  \\
\hline
3p65Ax1     &   2.551(1)   &  -1.723(1)   &   7.828(-1)  &   2.668(1)   &  -1.734(1)   &   6.099(-1)  &   2.972(1)   &  -1.740(1)   &   3.855(-1)  \\
5p11Ax1     &   3.442(1)   &  -1.715(1)   &   4.745(-1)  &   3.616(1)   &  -1.734(1)   &   4.108(-1)  &   3.920(1)   &  -1.747(1)   &   2.772(-1)  \\
6p5Ax1      &   3.941(1)   &  -1.702(1)   &   2.804(-1)  &   4.274(1)   &  -1.731(1)   &   2.656(-1)  &   4.564(1)   &  -1.754(1)   &   2.064(-1)  \\
\hline
\end{tabular}
\end{center}
\end{table*}

\begin{table*}
\begin{center}
    \caption{
   Same as Table~\ref{tab_lc_mod1}, but now for the $J$, $H$, and $K$ bands.
\label{tab_lc_mod3}}
\begin{tabular}{
l@{\hspace{3mm}}
c@{\hspace{3mm}}c@{\hspace{3mm}}c@{\hspace{3mm}}
c@{\hspace{3mm}}c@{\hspace{3mm}}c@{\hspace{3mm}}
c@{\hspace{3mm}}c@{\hspace{3mm}}c@{\hspace{3mm}}
c@{\hspace{3mm}}c@{\hspace{3mm}}c@{\hspace{3mm}}
c@{\hspace{3mm}}c@{\hspace{3mm}}c@{\hspace{3mm}}
}
\hline
\hline
Model    & \multicolumn{3}{c}{$J$}   & \multicolumn{3}{c}{$H$}   & \multicolumn{3}{c}{$K$}   \\
      & $t_{\rm rise}$ & Max. & $\Delta M_{15}$ & $t_{\rm rise}$ & Max. & $\Delta M_{15}$ & $t_{\rm rise}$ & Max. & $\Delta M_{15}$ \\
                & [d]  &[mag] & [mag]  & [d]  &[mag] & [mag]   & [d]  & [mag] & [mag]  \\
\hline
3p65Ax1       &   2.751(1)   &  -1.731(1)   &   4.328(-1)  &   3.041(1)   &  -1.751(1)   &   2.992(-1)  &   2.862(1)   &  -1.764(1)   &   4.406(-1)  \\
5p11Ax1       &   3.786(1)   &  -1.746(1)   &   3.279(-1)  &   3.977(1)   &  -1.772(1)   &   2.557(-1)  &   3.668(1)   &  -1.768(1)   &   3.500(-1)  \\
6p5Ax1        &   4.655(1)   &  -1.752(1)   &   2.208(-1)  &   4.769(1)   &  -1.777(1)   &   1.514(-1)  &   4.610(1)   &  -1.776(1)   &   2.127(-1)  \\
\hline
\end{tabular}
\end{center}
\end{table*}

  \subsection{Summary of ejecta properties}

Although model ejecta masses are lower than $\approx$\,5\,\msun,
there is great disparity in the ejecta composition.
Model 3p65 has some residual hydrogen in its outermost layers,
with a total mass of 0.005\,\msun\ and a maximum mass fraction of 0.14.
It constitutes a prototype for a type IIb event \citep{dessart_11_wr}.

Models 3p65 and 6p5 have a massive He-rich shell of about
1\,\msun\ in which the helium mass fraction is $\gtrsim$\,90\%. This shell represents
a large fraction of the total ejecta mass, from $\approx$\,70\% in model 3p65 down to 35\% in model 6p5.
As we discuss in \citet{dessart_etal_12}, the larger that fraction the greater the likelihood of
exciting He\,\one\ lines and producing a SN Ib (or IIb if hydrogen is present).
Model 5p11 is hydrogen deficient and exhibits a balanced mixture of He, C, and O at its surface.

If all these progenitors were luminous enough to drive an optically-thick wind and produce an
emission-line spectrum, models 3p65 and 6p5 would correspond to a WN or WNh type, but model 5p11
would clearly be of WC type. Model 5p11 has lost its He-rich shell (that shell made of $\gtrsim$\,90\% helium). Its
surface is within the C-rich part of the CO core. This model appears more suitable to produce a type Ic SN.

 In this paper we focus on
  three models, endowed with the same ejecta kinetic energy at infinity of $\approx$\,1.2$\times$\foe,
  but different masses and composition.
  These models, in order of increasing pre-SN mass,  are 3p65Ax1, 5p11Ax1, and 6p5Ax1.
  The models have initial \iso{56}Ni masses of  0.074 to 0.099\,\msun. This is somewhat smaller than
  the mean of 0.2\,\msun\ inferred by \cite{drout_11_ibc} for SNe Ibc, but there is considerable scatter
  within the class.
  Progenitor and ejecta properties are summarised in Tables~~\ref{tab_ejecta_glob}--\ref{tab_ejecta_surf}, while
 the composition for each SN model is shown in velocity and Lagrangian-mass spaces in Fig.~\ref{fig_comp_set}.

  The goal with this restricted set is to discuss the salient features of type IIb, Ib, and Ic rather than
  correlate the properties of their light curves and spectra with those of the progenitor and explosion
  --- this will be done in Paper\two.
  We also address the influence of mixing by comparing these models to the ``x2"
  counterpart, i.e.,  3p65Ax2, 5p11Ax2, 6p5Ax2 (Section~\ref{sect_mixing}).
  The description of the full model set is deferred to Paper\two.

\begin{figure*}
\epsfig{file=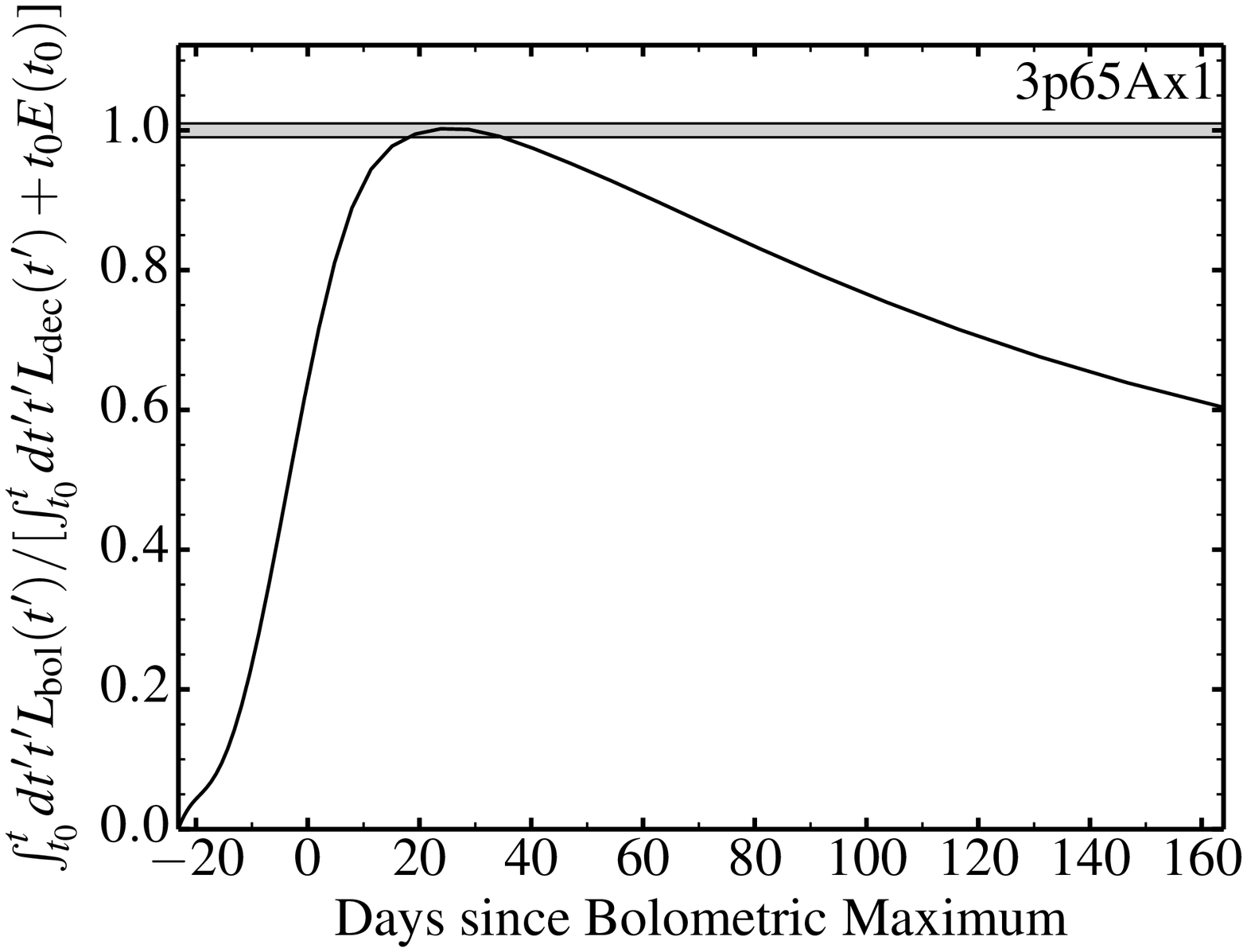,width=5.8cm}
\epsfig{file=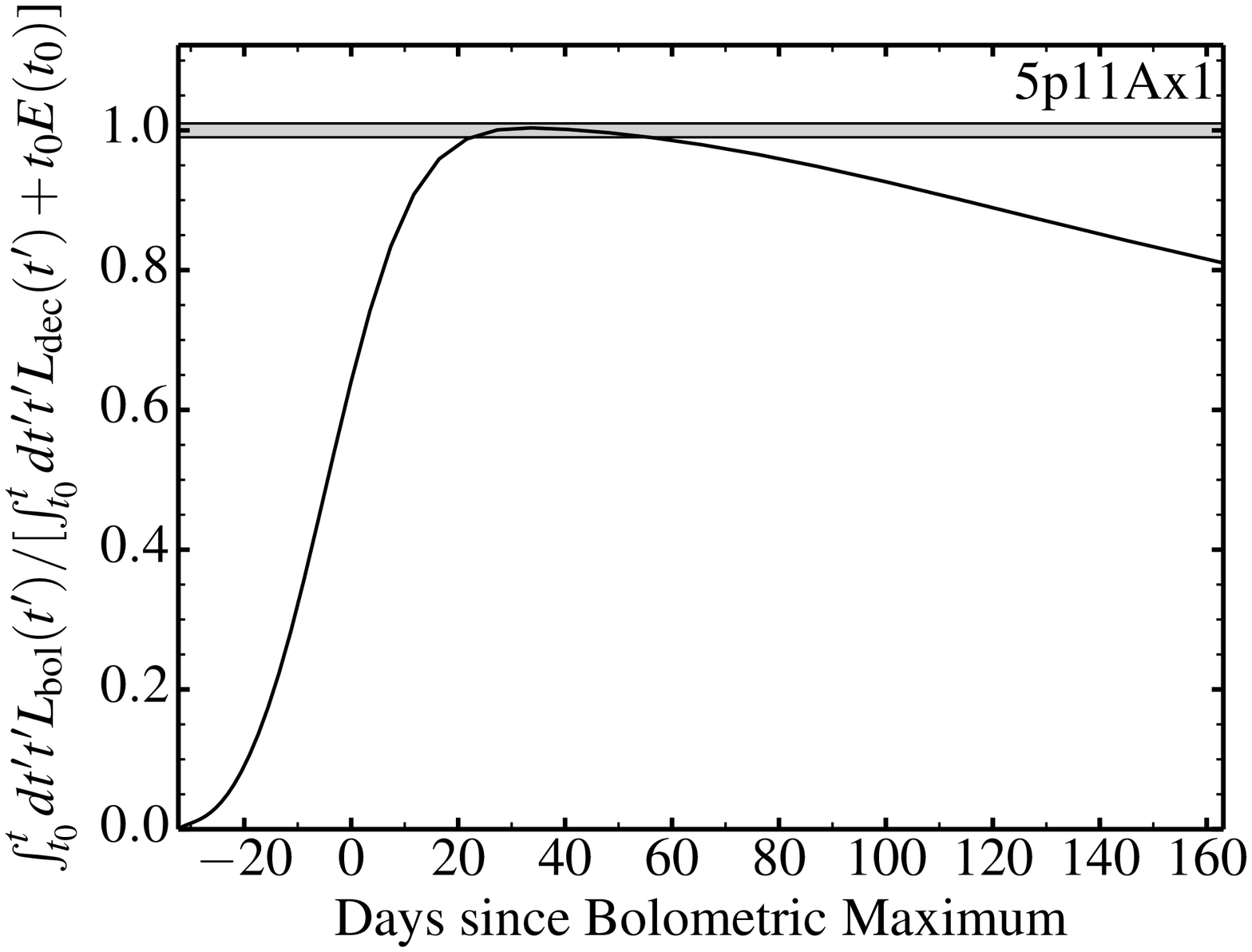,width=5.8cm}
\epsfig{file=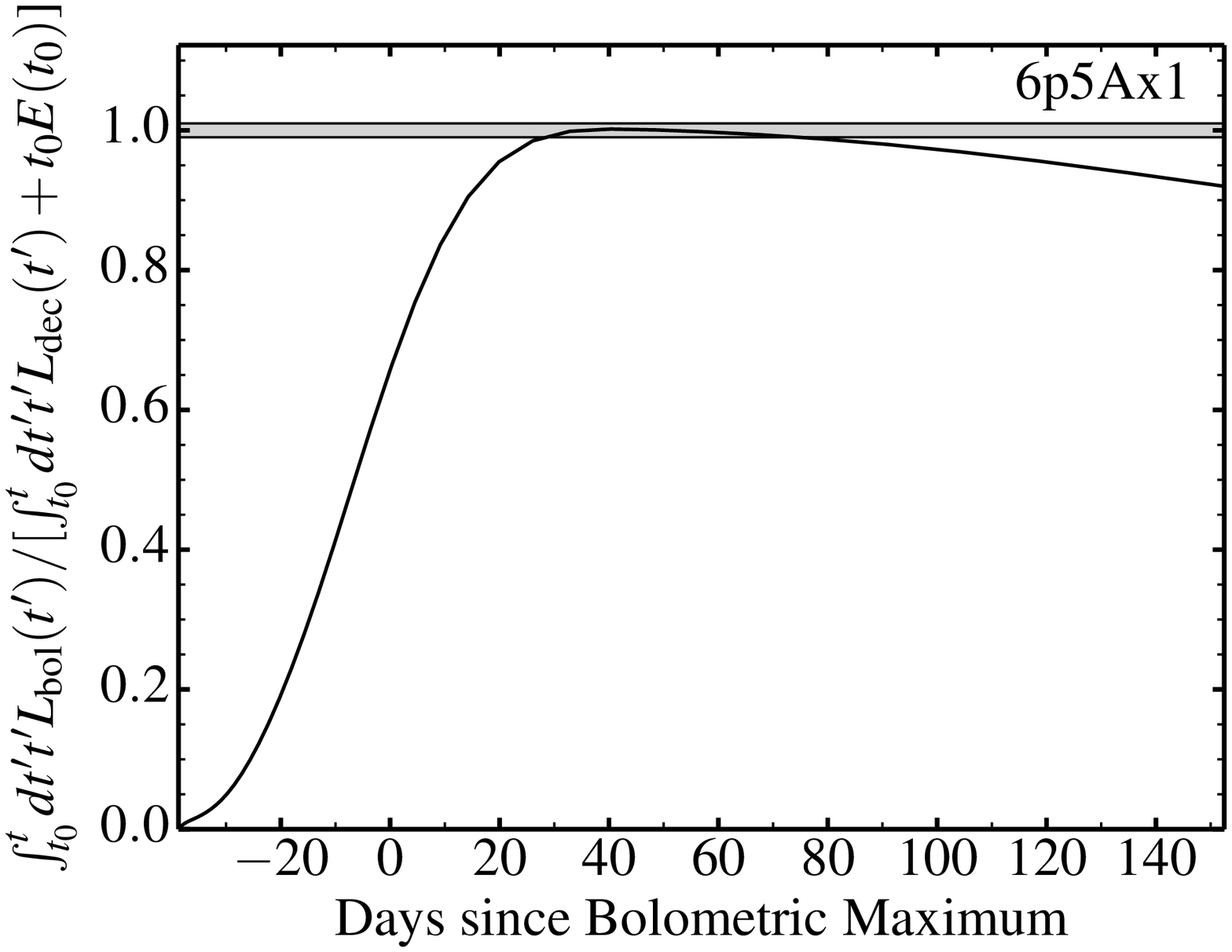,width=5.8cm}
\caption{Illustration of the variation of the ratio
$\int_{t_0}^t dt' t' L_{\rm bol}(t')$ /[$\int_{t_0}^t dt' t' L_{\rm dec}(t') + t_0E(t_0)]$
versus time since bolometric maximum (we use $t_0=3$\,d). As pointed out by \citet{katz_13_56ni},
this ratio eventually approaches unity at late times if one uses the trapped decay energy.
Since the trapped fraction is not directly inferred, we find more convenient to check this expression
using the total decay energy. In that case, the ratio is unity to within a few per cent about 10-20\,d
after maximum. At later times, $\gamma$-ray escape makes this ratio drop below unity,
the more so as time passes and the ejecta thins out. Higher mass ejecta exhibit a longer
period in which the ratio sustains its maximum value.
\label{fig_boaz}
}
\end{figure*}

\section{Photometric evolution}
\label{sect_phot}
  \subsection{Bolometric light curve}

   After shock emergence, the light curve of a SN is controlled by the
   simultaneous effects of cooling, heating, and energy transport.\footnote{By definition,
   the bolometric luminosity includes contributions at all wavelengths. However, here we adopt the usual
   convention in SN modeling,
   and exclude the contribution of gamma-rays which generally cannot be measured.}
   The cooling is initially primarily caused by expansion until the ejecta reaches a large
   radius. When the ejecta turn optically thin, cooling is primarily by radiation.
   The dominant heating source is radioactive decay, with a small contribution
   from recombination at early times.

   The shocked envelope holds a large internal energy at shock breakout (about 50\% of the total
   ejecta energy at that time). Despite the degradation by expansion of this shock-deposited energy,
   the SN is still significantly luminous a few days after the explosion. A post-breakout
   plateau of the order of 10$^{41}$\ergs\ is seen in the larger mass model 6p5Ax1,
   while mixing in the lower mass models leads to a gradual brightening by 2-3 days (Fig.~\ref{fig_lbol}).

   This post-breakout luminosity is function of the progenitor radius, ejecta mass, and the amount and spatial distribution
   of \iso{56}Ni \citep{woosley_94_93j,dessart_11_wr,bersten_etal_12_11dh,piro_nakar_13,nakar_piro_14}.
   In this work, all progenitors have surface radii of $\lesssim$\,20\,\rsun, which causes the faintness
   of the post-breakout phase relative to the subsequent peak.

   The SNe IIb/Ib/Ic models systematically brighten a few days after explosion; this stems
   from energy deposited into the ejecta by the decay of \iso{56}Ni exclusively (other unstable isotopes
   are ignored in this work).
   The three models have $\approx$\,0.09\,\msun\ of \iso{56}Ni and radiate about the same
   amount of energy. They do so over a longer time for higher mass ejecta, and consequently
   reach a smaller maximum luminosity.
   For our model set, the rise time to bolometric maximum increases from 26.3\,d  for model 3p65Ax1
   to 41.9\,d for model 6p5Ax1, while the peak luminosity drops from 1.65 to 1.49$\times$10$^{42}$\,\ergs,
   in the same model order.\footnote{If we scale the peak luminosity to exactly the same \iso{56}Ni
   mass of 0.0742\,\msun,  the peak luminosity drops from 1.65 to 1.12$\times$10$^{42}$\,\ergs.}

   The post-maximum bolometric light curve shows a more rapid fading for smaller mass ejecta
   (i.e., 0.61\,mag for model 3p65Ax1, and only 0.22\,mag for the higher mass model 6p5Ax1).
   For a higher ejecta mass
   there is more efficient trapping of stored
energy (primarily in the form of moderate-energy ``thermal" photons) within the ejecta,
as well as more efficient trapping of the $\gamma$-rays released by the decay of \iso{56}Ni and \iso{56}Co.
The latter becomes more evident as time progresses. At 140\,d past maximum, the bolometric light curve
drops at 0.0196\,mag per day for model 3p65Ax1, 0.0178\,mag for model 5p11Ax1, and
0.0153\,mag for model 6p5Ax1.

%Model delta_m at time past maximum:  3p65Ax1 0.0196179275778 140.0
%Model delta_m at time past maximum:  5p11Ax1 0.0178029901603 140.0
%Model delta_m at time past maximum:  6p5Ax1 0.0153218228953 140.0

\subsection{Peak bolometric luminosity and \iso{56}Ni mass}

   In our three models the peak luminosity is $\approx$\,50\% larger than the instantaneous decay rate at maximum, significantly
   offset from the prediction of Arnett's rule (which states that the bolometric luminosity is equal to the total decay
   rate at that time; \citealt{arnett_82}).
   This ``rule", identified in a simplified analytical model of SN Ia light curves, is not a fundamental law
   that should hold exactly because, for example, of energy conservation. There
   is thus no fundamental reason why it should apply accurately to type Ibc SNe.
   In the delayed-detonation models of \citet{blondin_etal_13}, which cover a range of \iso{56}Ni masses,
   the offset is up to $\approx$\,10\%, and Arnett's rule yields a reliable estimate of the \iso{56}Ni mass.  However, for
   the models presented here,  Arnett's rule overestimates the \iso{56}Ni mass by 44\%, 53\%, and 52\%  for models 3p65Ax1,
   5p11Ax1, and 6p5Ax1, respectively.

% Using routines: ~/python/cmfgen/comp_lbol_to_arnett.py
%Ratio of Lbol_max / Lbol_edecay 6p5Ax1_lc.dat 1.43989398722
%Ratio of Lbol_max / Lbol_edecay 5p11Ax1_lc.dat 1.53188822477
%Ratio of Lbol_max / Lbol_edecay 3p65Ax1_lc.dat 1.52055731845

 To estimate the \iso{56}Ni mass from the bolometric luminosity, one may instead use
the idea of  \citet{katz_13_56ni}, which is robustly based on an energy-conservation argument.
Considering heating from radioactive decay
and cooling through expansion and radiation, one can solve the internal energy equation out to late
times to find
$$  t E(t) -t_0E(t_0)  =  \int_{t_0}^t dt' t' L_{\rm dec}(t')- \int_{t_0}^t dt' t' L_{\rm bol}(t')  \;\;,\nonumber$$
where $L_{\rm dec}(t)$ is the decay luminosity and  
$E(t)$ is the total radiation energy trapped within the ejecta at $t$.
At late times, as the ejecta becomes optically thin, there is essentially no stored radiation energy
so we neglect the term $t E(t)$.
To test this expression, we use $t_0=$\,3\,d. At that time, $E$ is of the order of 10$^{48}$\,erg, and is
about a tenth of the time-integrated bolometric luminosity. Furthermore, as expected, the ratio
$$\int_{t_0}^t dt' t' L_{\rm bol}(t') \Big/ \left(\int_{t_0}^t dt' t' L_{\rm dec}(t') + t_0 E(t_0) \right)$$
is within a few percent of unity (a superior accuracy than achieved with Arnett's rule) at 10-20\,d after maximum
(Fig.~\ref{fig_boaz}). Neglecting the term $t_0 E(t_0)$ increases the offset by a few per cent.
At early times, the ratio is much below unity because the decay energy
is deposited at large optical depth, too large to influence the rate at which radiation escapes from the photosphere.
At late times,  the ratio drops below unity again because of $\gamma$-ray escape --- the value of unity would
be recovered if we used the decay energy effectively deposited within the ejecta, or if we included
the escaping $\gamma$-ray luminosity in the bolometric luminosity.
At  10-20\,d after maximum, $\gamma$ rays are fully trapped
in typical SNe IIb/Ib/Ic and so the relation holds satisfactorily without any detailed calculation at that time.
The downside of this approach, which also affects Arnett's rule (although only at maximum),
is the difficulty of constraining the bolometric luminosity from limited multi-band photometry.

\begin{figure*}
\epsfig{file=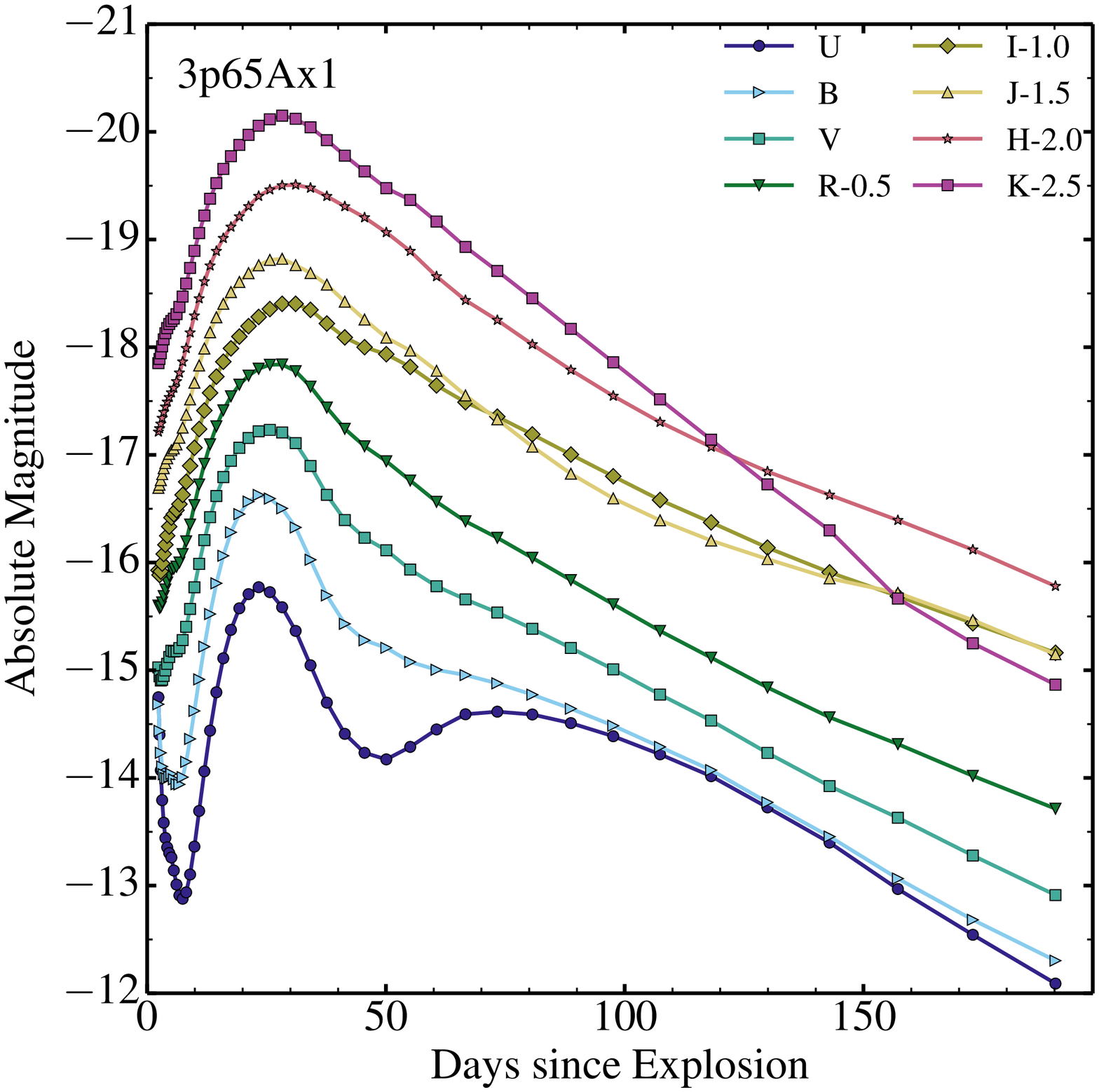,width=8.5cm}
\epsfig{file=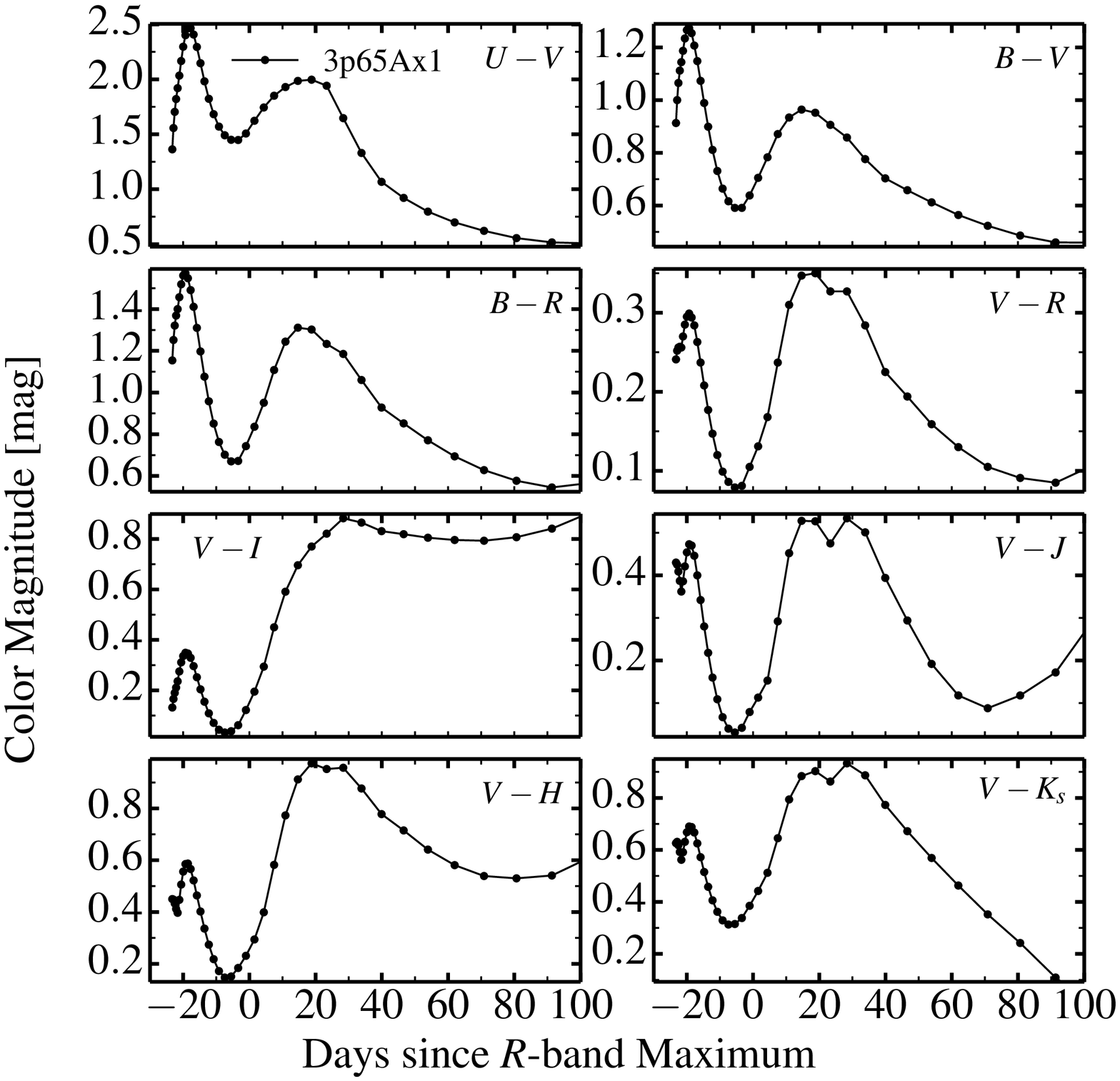,width=8.5cm}
\caption{Multi-band light curves (left) and colour evolution for a selection of optical
and near-IR bands (right) for model 3p65Ax1. Redder bands tend to peak at later times
while the colour evolution shows a complex behaviour. Notice how the minimum colour, for each
plotted filter combination, occurs prior to $R$-band (and hence bolometric) maximum.
\label{fig_photometry_3p65ax1}}
\end{figure*}

\begin{figure*}
\epsfig{file=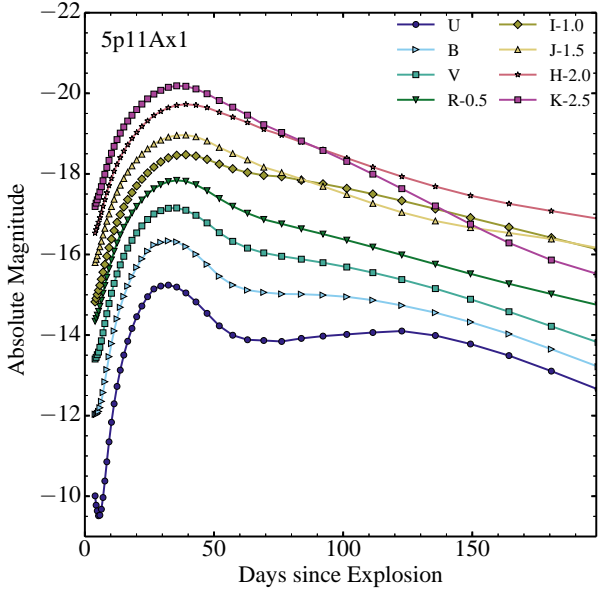,width=8.5cm}
\epsfig{file=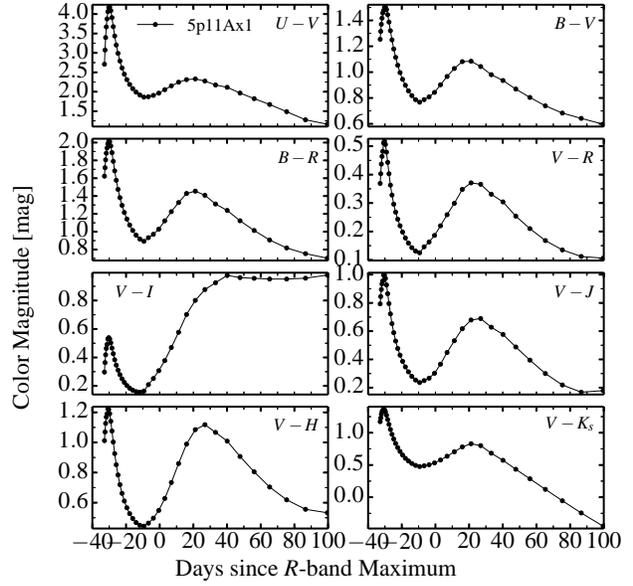,width=8.5cm}
\caption{Same as Fig.~\ref{fig_photometry_3p65ax1}, but now for model 5p11Ax1.
\label{fig_photometry_5p11ax1}}
\end{figure*}

\begin{figure*}
\epsfig{file=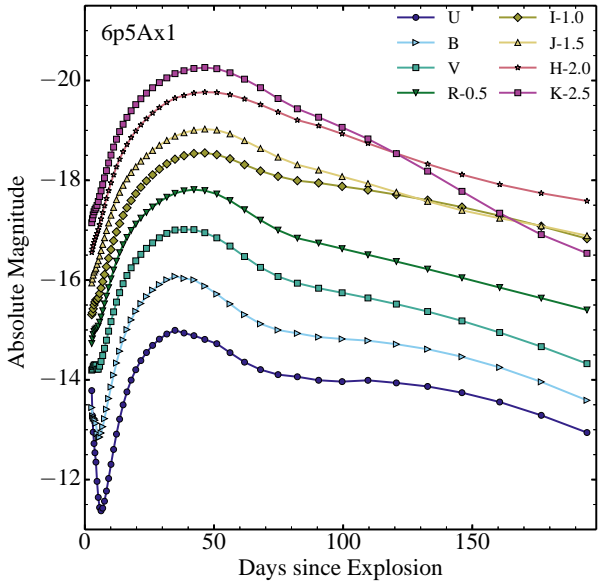,width=8.5cm}
\epsfig{file=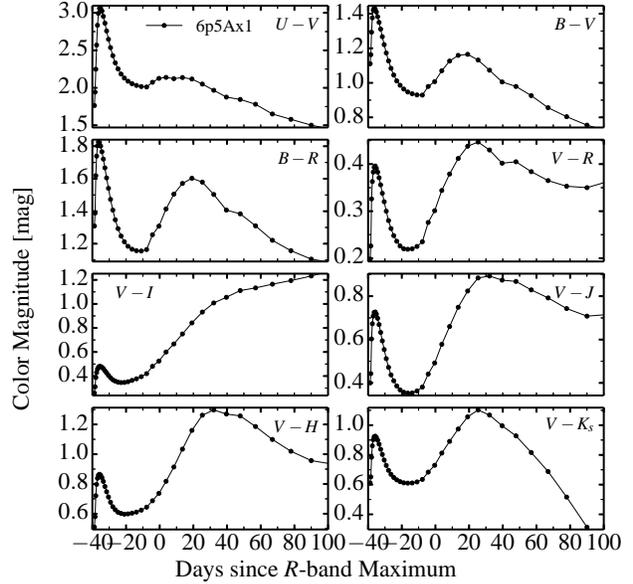,width=8.5cm}
\caption{Same as Fig.~\ref{fig_photometry_3p65ax1}, but now for model 6p5Ax1.
The curves are similar to those shown for model 3p65Ax1, but the timescales for
the variations are longer.
\label{fig_photometry_6p5ax1}
}
\end{figure*}

\subsection{Multi-band light curves}

In this section, we use multi-band light curves to discuss the distribution
of the emergent flux with wavelength and time. Results, including colours,
are shown in Figs.~\ref{fig_photometry_3p65ax1}--\ref{fig_photometry_6p5ax1},
while  rise times, peak magnitudes, and post-maximum decline rates (the change
in magnitude 15\,d after the time of peak) are provided
in Tables~\ref{tab_lc_mod1}--\ref{tab_lc_mod3}.

Generally we obtain shorter rise times in bluer optical bands, although an exception is that
 the $I$-band peaks later than the $J$ band (likely because of the strengthening Ca\two\ triplet
lines; see Section~\ref{sect_spec}). The $R$ band peaks within about a day
of the bolometric light curve maximum. As for the bolometric maximum, models with
a larger ejecta mass peak later.  Higher mass models also show a larger
spread in the times at which each magnitude peaks.
Because of the comparable \iso{56}Ni mass,
the peak magnitudes are, however, similar, band-by-band, for all three models.
For example, we obtain a peak $R$ band magnitude of $-$17.34, $-$17.34, and $-$17.31
for models 3p65Ax1, 5p11Ax1, and 6p5Ax1 (a closer correspondence than for the
bolometric luminosity maximum).

\begin{figure}
\epsfig{file=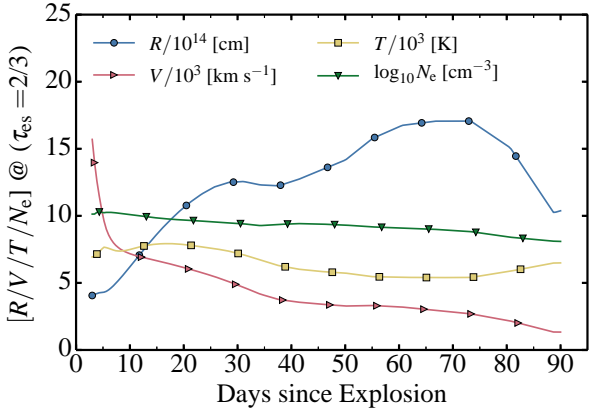,width=8.5cm}
\caption{Evolution of the  ejecta properties at the electron-scattering photosphere
for model 3p65Ax1.
\label{fig_phot_prop_3p65Ax1}
}
\end{figure}

A similar behaviour is seen in SN Ia models (see, e.g., \citealt{blondin_15_02bo})
and in both cases results from a combination of effects. On the rise to maximum, the
ejecta heats up and the emissivity at depth shifts to shorter wavelength --- this
is in part reflected by the behaviour of the electron-scattering photosphere
(Fig.~\ref{fig_phot_prop_3p65Ax1}). The ejecta is thus
not expected to evolve at constant colour. However,
line blanketing causes strong energy redistribution to longer
wavelength. As we approach the bolometric maximum, lines strengthen
(see also Section~\ref{sect_spec_form}), dominating the emissivity and the
opacity over continuum processes. The magnitude of the effect grows with time as
the spectrum forms deeper into more metal-rich regions.

Past maximum, the decline rates are greater in bluer optical bands, and the more
so for lower mass models -- $\Delta M_{15}=$\,1.17\,mag
in the $U$ band for model 3p65Ax1, but only 0.25\,mag for model 6p5Ax1.
In the higher mass model 6p5Ax1, the decline rate over 15\,d
in different bands shows much less scatter than for the lighter model 3p65Ax1, which
reflects the longer diffusion time of the former.

The colour evolution for all three models is qualitatively similar, despite the differences
in mass and composition
(right panels of Figs.~\ref{fig_photometry_3p65ax1}--\ref{fig_photometry_6p5ax1}).
During the first few days after explosion, all models become redder. Unfortunately, although
there may be a physical effect here, this may in a large part stem from the initial relaxation
of the \v1d\ model into \cmfgen. Subsequently, because of the heat supply from decay, they
become bluer on the rise to maximum.
From maximum until 20 (model 3p65Ax1) to 40\,d (model 6p5Ax1) they become
redder, probably because of the increasing effect of blanketing.
Beyond that time, as the ejecta turns progressively nebular, the colours become bluer again.
At nebular times, the colours reflect in part the distribution of  forbidden-line transitions and
thus can show great disparity between filter sets.

In our set of models, the initial colours are systematically red, unlike explosions
occurring in extended stars --- the larger the initial radius, the bluer the colours at early times
and the more delayed the phase of recombination \citep{dessart_etal_13b}.
This is a bias in our sample of progenitors, which is limited
to close-binary systems in which the primary star explodes with a final radius $\lesssim$\,20\,\rsun.

\begin{figure}
\epsfig{file=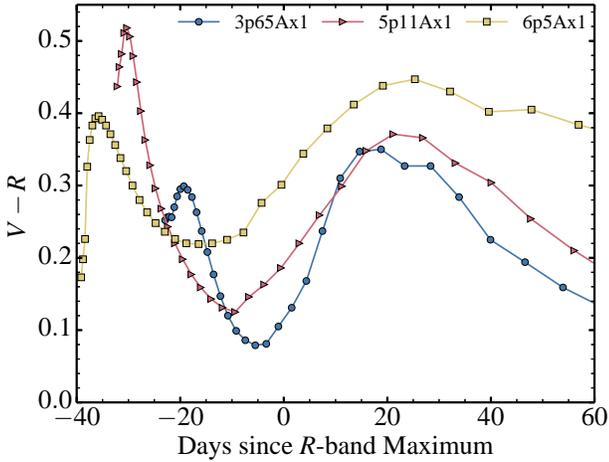,width=8.5cm}
\caption{Comparison of the ($V-R$) light curve for models 3p65Ax1,
5p11Ax1, and 6p5Ax1. At 10\,d after $R$-band maximum, the mean value
is  0.322\,mag and the standard deviation is 0.047\,mag.
\label{fig_vmr}
}
\end{figure}

\citet{drout_11_ibc} reported a small scatter of 0.06 mag about a mean ($V-R$) colour of 0.26\,mag
at 10 days after $V$-band maximum for observed SNe IIb/Ib/Ic with an inferred reddening.
In our models, we also find that the colours early after peak  are comparable. For ($V-R$) at 10\,d after $R$-band maximum,
we obtain a mean of 0.32\,mag, with a standard deviation of 0.04\,mag (Fig.~\ref{fig_vmr}).
The larger sample of models also shows this behaviour, with a similar mean (the standard deviation is reduced if we
limit the sample to lower-mass ejecta; Dessart et al., in prep.).
This uniformity in post-maximum colour can thus be used to constrain the reddening in SNe IIb/Ib/Ic.

\begin{figure}
\epsfig{file=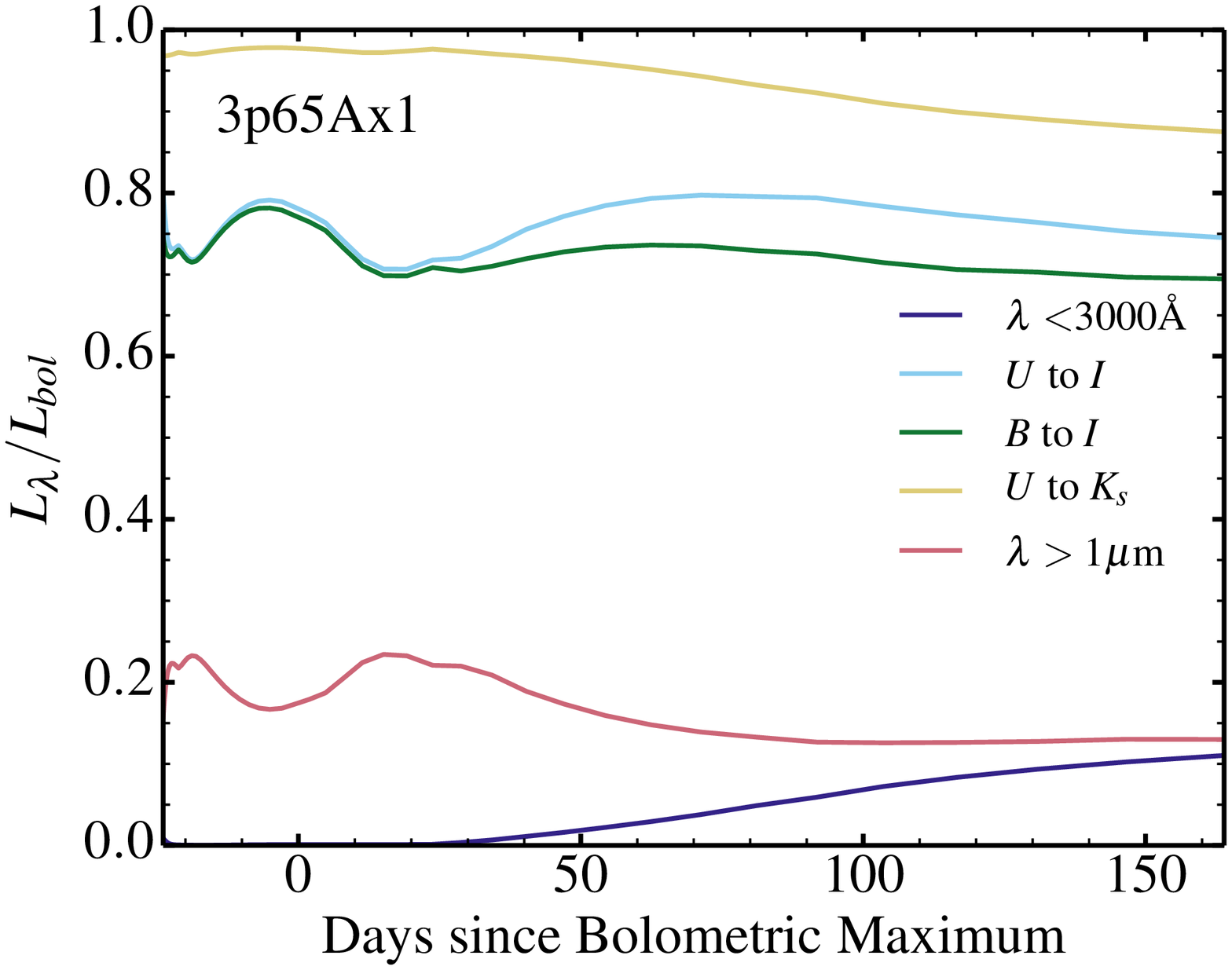,width=8.5cm}
\epsfig{file=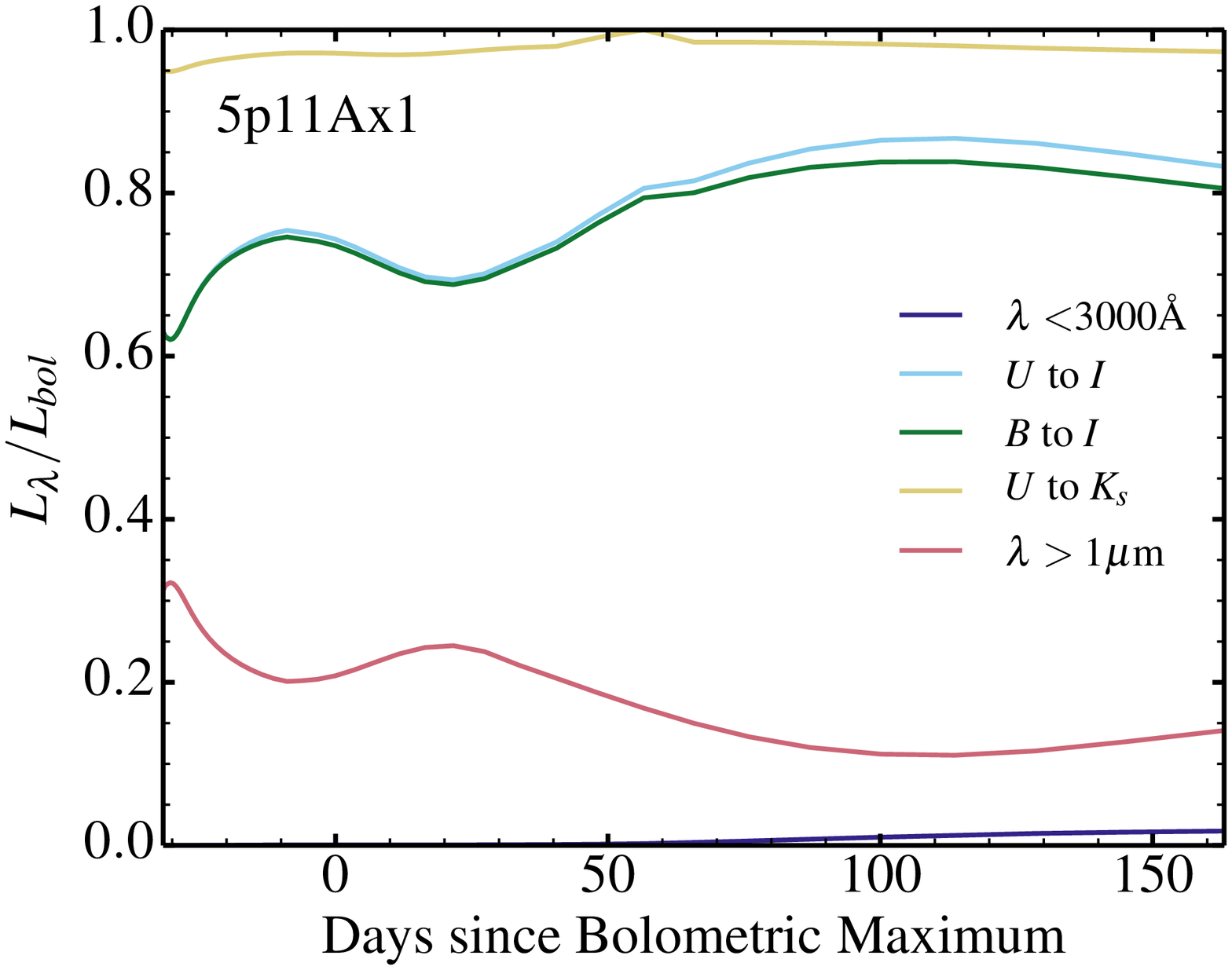,width=8.5cm}
\epsfig{file=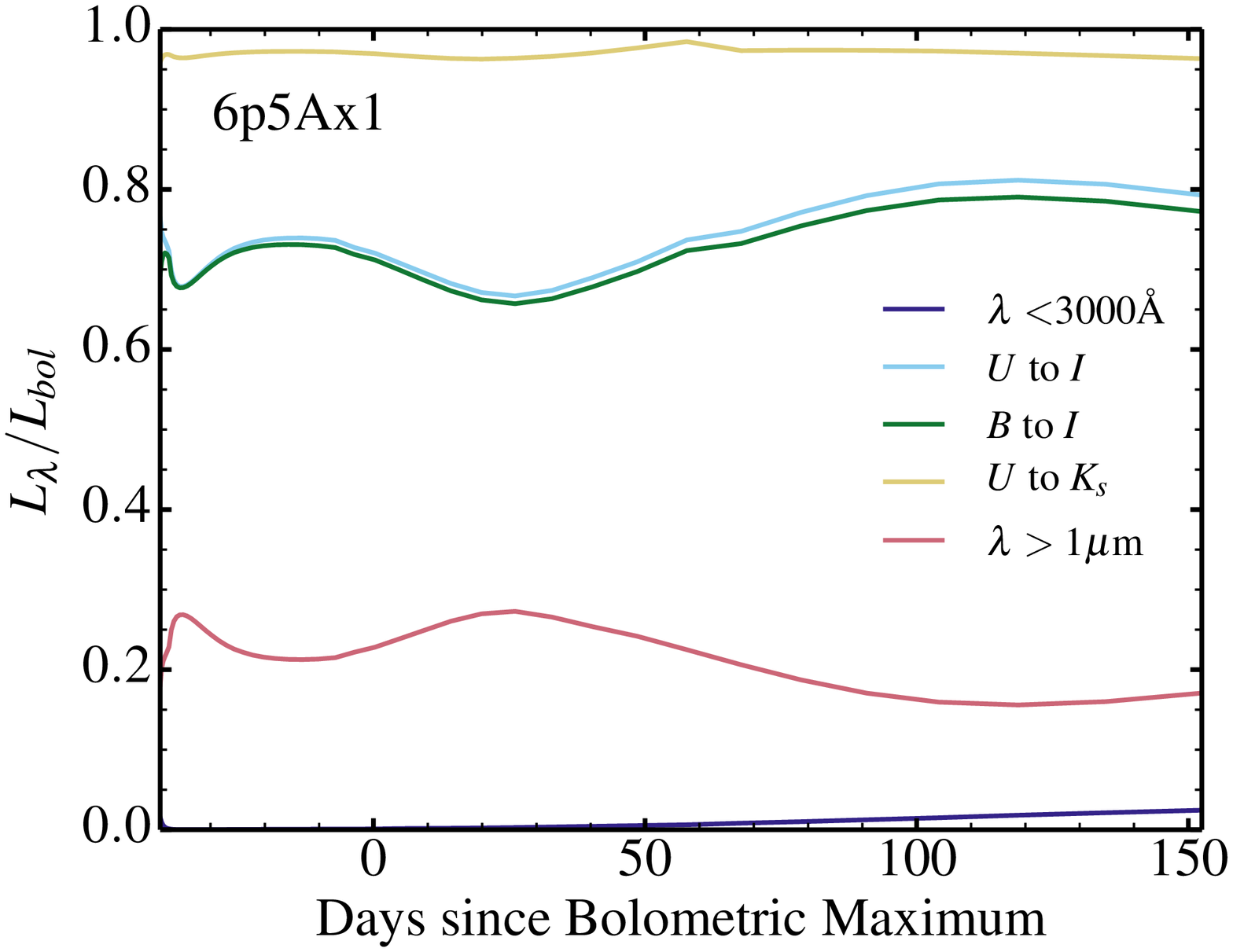,width=8.5cm}
\caption{Fractional flux falling into different spectral regions for
the models 3p65Ax1, 5p11Ax1, and 6p5Ax1 up to about 200\,d after explosion.
\label{fig_frac_flux}
}
\end{figure}

While the bolometric correction can be estimated for our models for  any band, a
more accurate estimate of the bolometric luminosity can be obtained by using multiple passbands.
  Here we extract  the relative fraction of the flux
  that falls within the UV, the optical and the near-IR (Fig.~\ref{fig_frac_flux}).
  Because  of the cool ``photospheric" temperature of our model (Fig.~\ref{fig_phot_prop_3p65Ax1}),
  and the strong blanketing by lines (see below), the UV is subdominant at all times. The UV rises a little at nebular times, in particular
  in the lower mass model, but remains sub-dominant.
  The bulk of the radiation emerges in the optical range ($\approx$\,80\%), with the rest at longer wavelengths
  and primarily in the near-IR --- little flux is emitted beyond 2-3\,$\mu$m.

\begin{figure*}
\epsfig{file=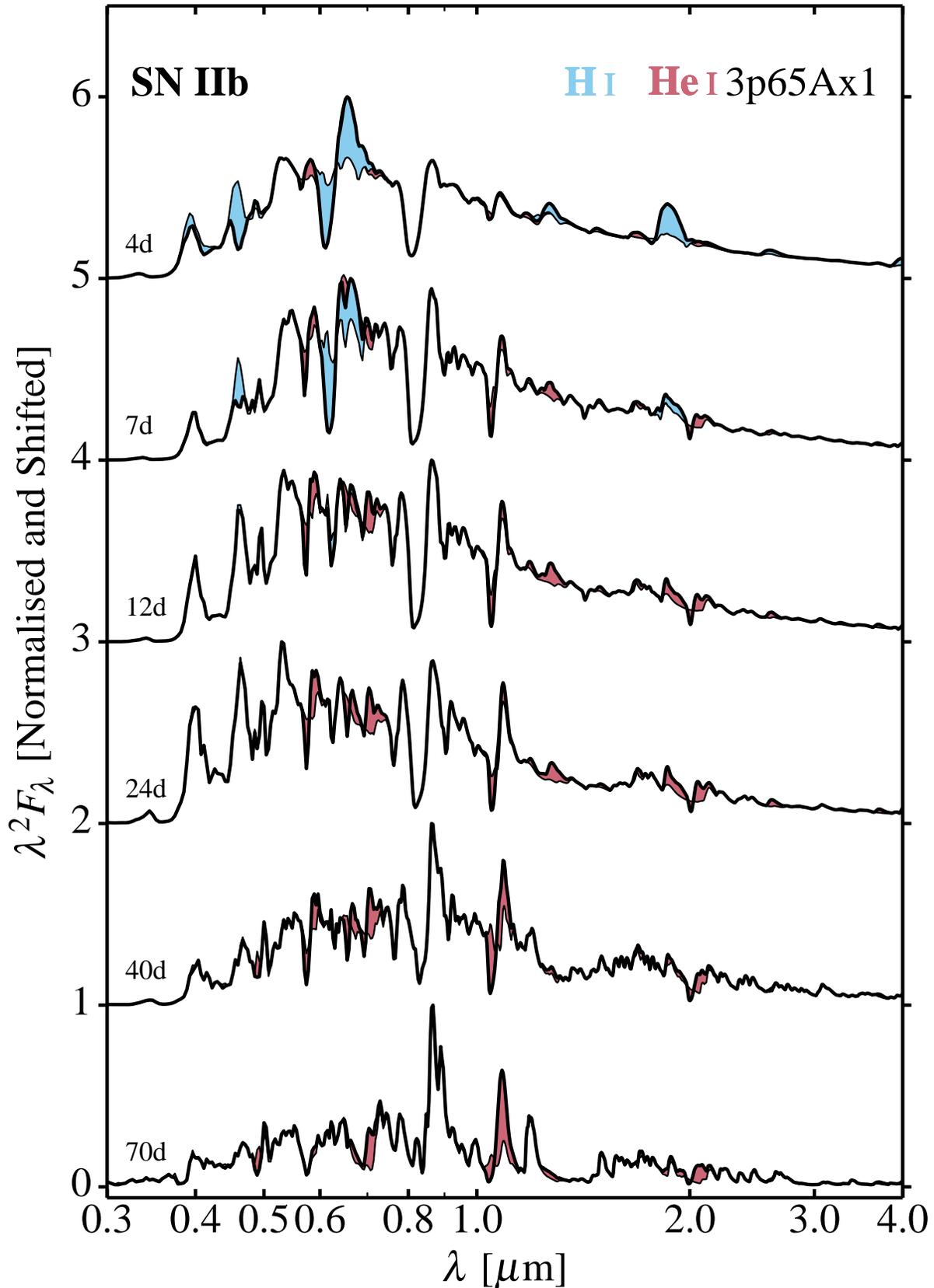,width=17cm}
\vspace{-2.5cm}
\caption{Spectral evolution of model 3p65Ax1 from 0.3 to 4$\mu$m until 70\,d after explosion
(we show $\lambda^2 F_{\lambda}$ for better visibility).
Overlaid on the full synthetic spectrum (black thick line), we hatch the offset in flux obtained
when H\one\ and He\one\ bound-bound transitions are omitted from the formal solution
of the radiative transfer equation. The presence of H\one\ and He\one\ lines at early times,
and the absence of H\one\ at late times, makes this model a type IIb.
\label{fig_3p65Ax1_spec}}
\end{figure*}

\begin{figure*}
\epsfig{file=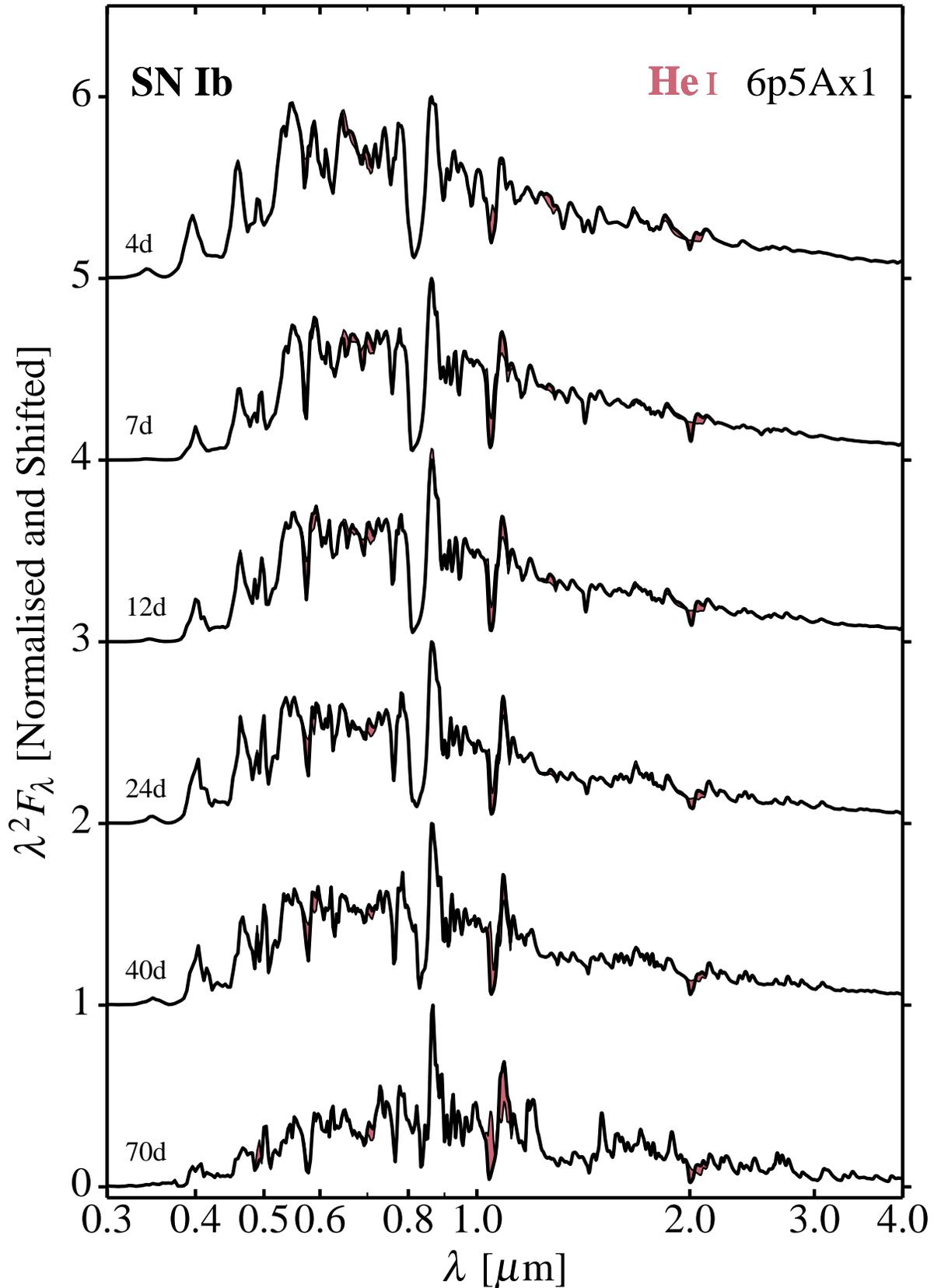,width=17cm}
\vspace{-2.5cm}
\caption{Same as Fig.~\ref{fig_3p65Ax1_spec}, but now for model 6p5Ax1.
H\one\ lines are not plotted since hydrogen is absent in the ejecta.
This model corresponds to a Type Ib SN. Even for this model, with a He mass of 1.67\,\msun,
He\one\ lines (except for $\lambda$10830 and $\lambda$20580) are relatively weak
and/or blended. At 70 days, and unlike earlier epochs, He\one\,10830\,\AA\ is almost
entirely responsible for the P~Cygni profile seen near 10830\,\AA.
\label{fig_6p5Ax1_spec}}
\end{figure*}

\begin{figure*}
\epsfig{file=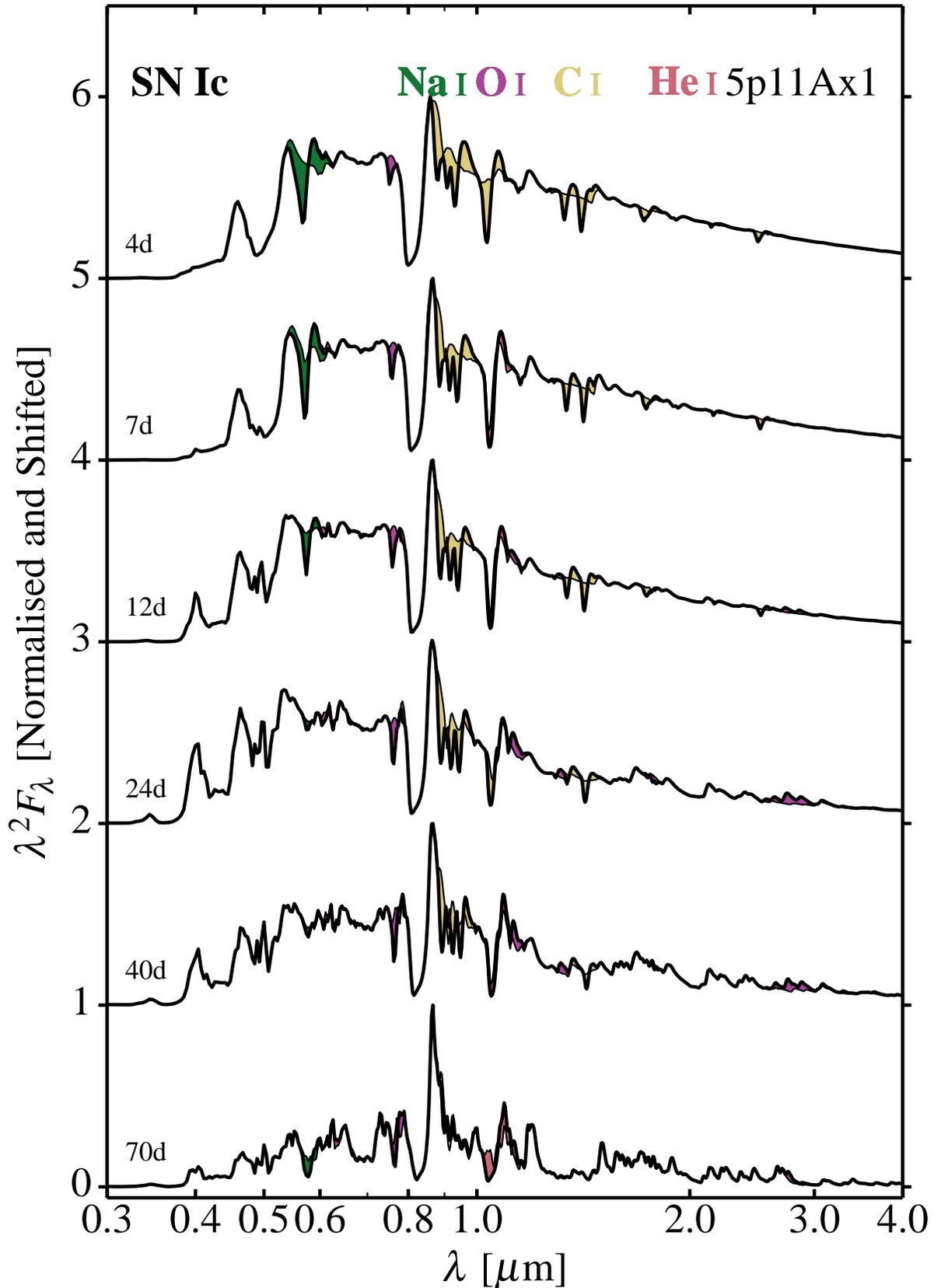,width=17cm}
\vspace{-2.5cm}
\caption{Same as Fig.~\ref{fig_6p5Ax1_spec}, but now for model 5p11Ax1.
Although helium is present in the ejecta, only the He\one\ line at 10830\,\AA\
is visible over the entire optical and near-IR ranges, and only at late times.
This model corresponds to a Type Ic SN. At early epochs Na\one\,D could be erroneously
identified with He\one\ $\lambda$5876.
\label{fig_5p11Ax1_spec}}
\end{figure*}

\section{Spectroscopic properties}
\label{sect_spec}

\subsection{Spectral evolution for our model set}

  The three models 3p65Ax1, 5p11Ax1, and 6p5Ax1 have the same explosion energy ($\approx$\,1.2$\times$\foe),
  a similar \iso{56}Ni mass ($\approx$\,0.09\,\msun),
  similar ejecta masses (2-5\,\msun), and small progenitor radii ($\lesssim$\,20\,\rsun).
 However, a fundamental distinction between these ejecta is their composition. As shown
  in \citet{dessart_11_wr}, even $\sim$\,0.001\,\msun\ of hydrogen can lead to a strong H$\alpha$ line
  at early times, while even a  helium mass fraction of $\sim$30\% may produce no He\,\one\ line \citep{dessart_etal_12}.
  Our three models have composition characteristics (see Tables~\ref{tab_ejecta_mass} and \ref{tab_ejecta_surf},
  and Fig.~\ref{fig_comp_set})  that make them suitable candidates for a type IIb (model 3p65Ax1),
  a type Ib (model 6p5Ax1) and a type Ic (model 5p11Ax1).

%In the preceding section, we have shown that their photometric light curves are qualitatively
% comparable, merely showing moderate variations in rise times, peak magnitudes, post-maximum colour,
 % and decline rates.

  We show the spectral evolution of each model in Figs.~\ref{fig_3p65Ax1_spec}--\ref{fig_5p11Ax1_spec},
  covering from 4 to 70\,d after explosion
  and over the range 0.3--4$\mu$m.   None of our models show blue featureless spectra at early times, likely because the progenitor radii
  are small so that expansion cooling causes a large drain of internal energy in the outer ejecta layers.
 To avoid unnecessary speculations, we delay further discussion to a detailed study focused on that aspect alone.

  There is a great similarity of the spectral energy distribution for the three models at all times shown,
  with the bulk of the flux emerging in the optical range and strong signatures of line blanketing by Fe\two\ and Ti\two\
  (see below).
  However, closer examination reveals significant differences. To highlight these differences, and to
  make more obvious the classification of the model as type IIb, Ib,
  or Ic, we colour the flux associated with specific ions. This contribution is estimated by excluding the bound-bound
  transitions of a given ion in the formal solution of the radiative-transfer equation.

\begin{figure}
\epsfig{file=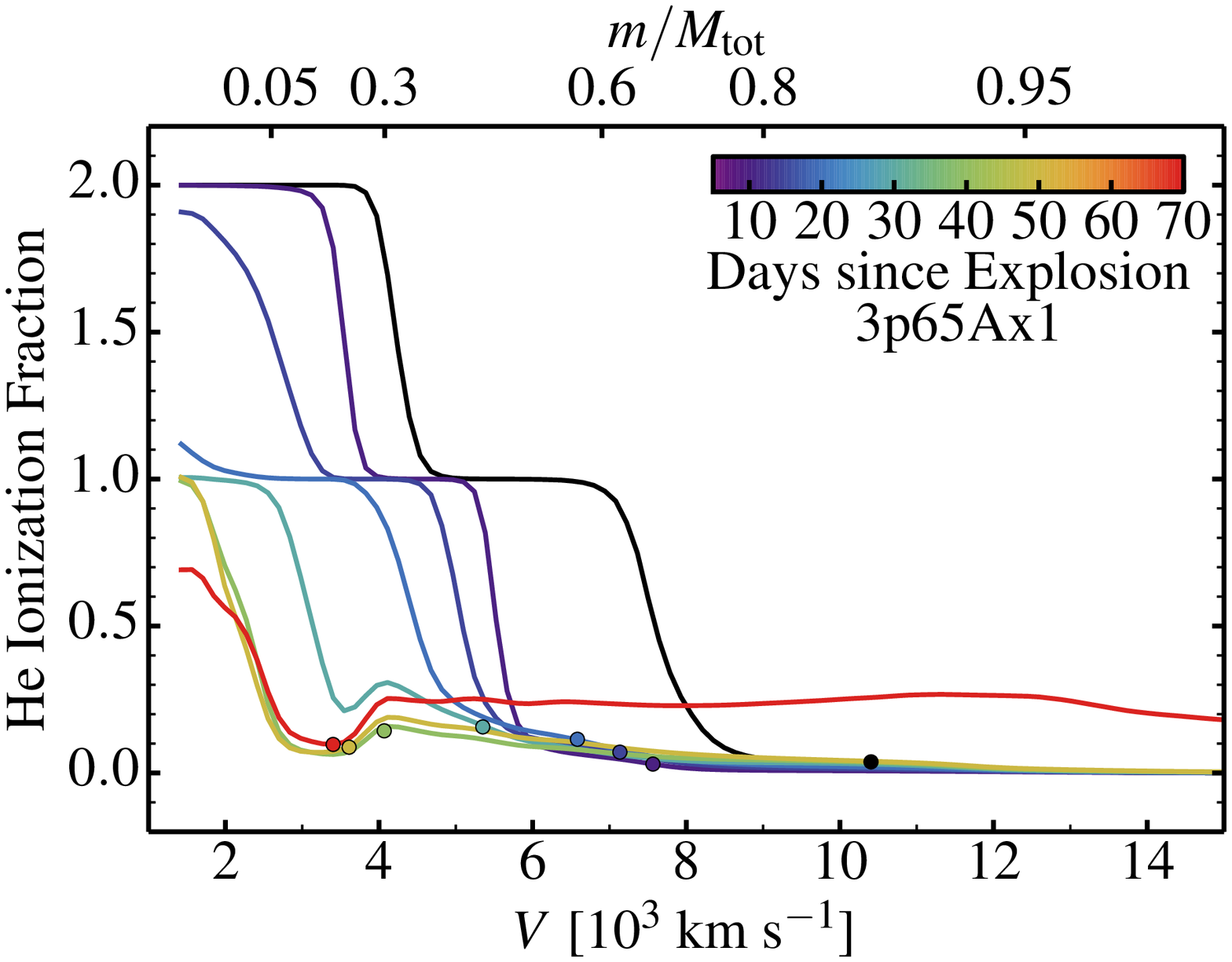,width=8.5cm}
\epsfig{file=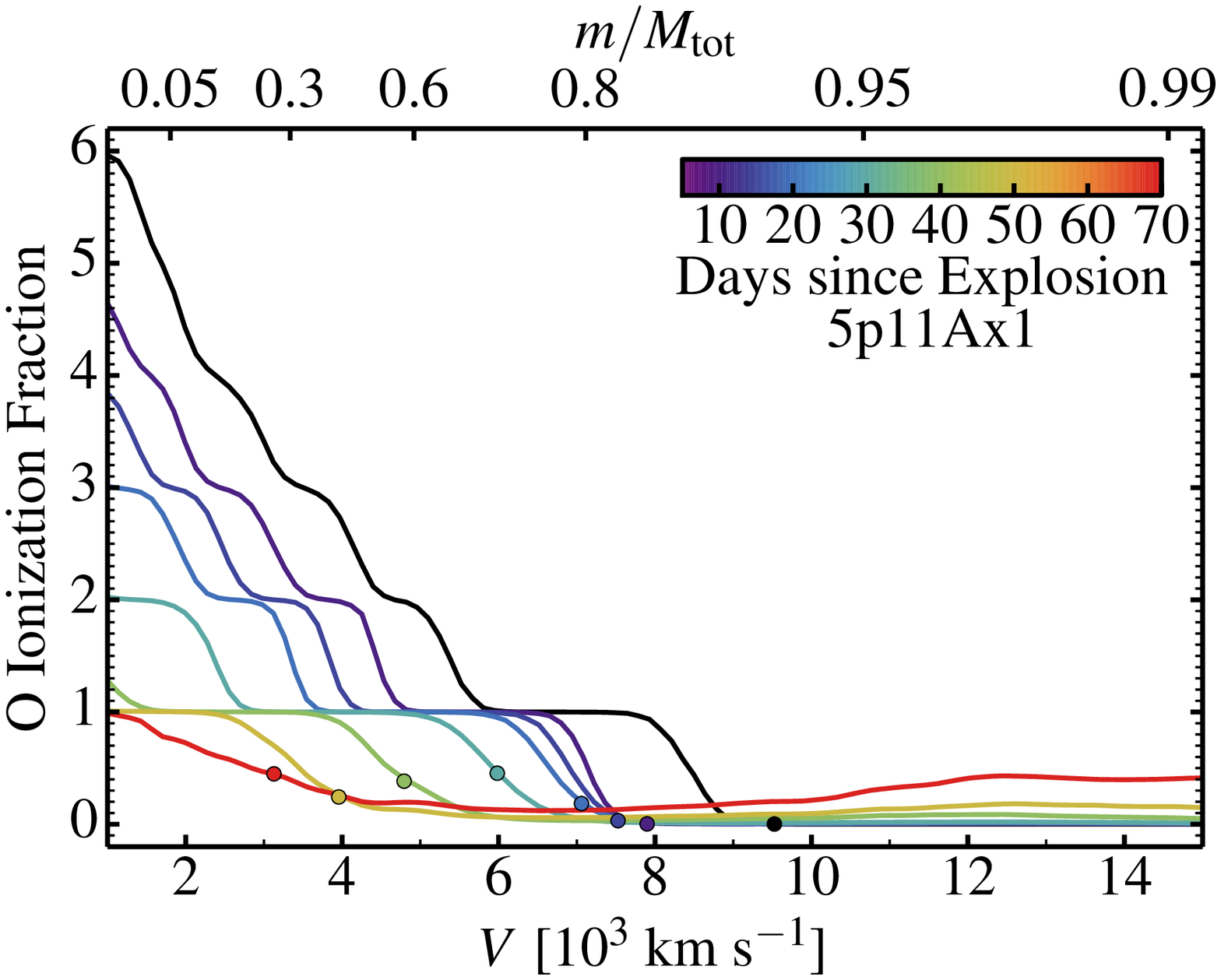,width=8.5cm}
\caption{Evolution of the ionisation fraction of He in model 3p65Ax1 (top) and of O in model 5p11Ax1 (bottom).
Specifically, the quantity plotted is $\sum_i i N_i/N_{\rm tot}$, where $N_i$ is the species number density
when $i$-times ionised and $N_{\rm tot}$ is the species total number density.
Symbols denote the location of the photosphere (using the Rosseland-mean opacity).
The times shown are 4, 6, 8, 10, 15, 20, 30, 40, and 50\,d after explosion for the top panel.
The times shown are 5, 10, 15, 20, 30, 40, 50, and 70\,d after explosion for the bottom panel.
\label{fig_ionfrac}
}
\end{figure}

  Model spectra for  3p65Ax1 exhibit H\one\ lines (primarily H$\alpha$ and H$\beta$
  in the optical, and P$\alpha$ in the near-IR) at early times and up to the peak,
  while He\one\ lines are present at all times, both in the optical and in the near-IR (Fig.~\ref{fig_3p65Ax1_spec}).
  These He\one\ lines may be present early on without the influence of \iso{56}Ni decay \citep{dessart_11_wr},
  but non-thermal excitation and ionisation are essential to produce them after a few days
  \citep{lucy_91, dessart_etal_12, hachinger_13_he, li_etal_12}.
  This is because the temperature in the spectrum formation region is low, so that He\one\ lines
  cannot be thermally excited. The ionisation of He (and of other species) in the spectrum formation region is low (Fig.~\ref{fig_ionfrac}).
  We find He\one\ lines at 5875, 6678, 7068, 10830, 11013.07, 12527.5, 12846.0, 12984.9
  (and numerous other contributions that overlap around 1.25-1.3\,$\mu$m), 15083.6 (weak), 17002.5 (weak),
  18685.4 and 19543.1 (overlap with P$\alpha$), 20581.0 (strong) + 21120.2 (and other similar lines
  that overlap on the red edge of 2.0581$\mu$m), 26881.2 (together with other He\one\ lines that overlap),
  and 37025.6\,\AA.
  With the presence of H\one\ lines early on and the persistence of He\one\ lines, the 3p65Ax1 model corresponds to a SN IIb.

  Model 6p5Ax1 shows a very similar evolution to model 3p65Ax1 but because the ejecta is hydrogen deficient, it shows no H\one\ lines
  (Fig.~\ref{fig_6p5Ax1_spec}). This model corresponds to a type Ib SN. Interestingly, the two sequences for
  the IIb and the Ib models are very similar if we omit the differences made by H\one\ lines.
  Hydrogen is so under-abundant in model 3p65Ax1 that it does not affect significantly the radiative transfer
  solution -- hydrogen acts only as a trace element.

  Model 5p11Ax1 shows the same overall evolution in spectral energy distribution but for this carbon and oxygen
  rich ejecta the spectral signatures are distinct (Fig.~\ref{fig_5p11Ax1_spec}).
  Despite the 0.32\,\msun\ of helium and 30\% surface helium mass fraction,
  He\one\ lines are always absent in the optical. The He\one\ line at 10830\,\AA\ is present at all times, but
  it is badly blended at early times -- absorption that could be erroneously associated with He\one\ alone is also
  due  to C\one\ and Mg\two.
  Paradoxically He\one\ $\lambda$10830 is strongest at late times (see below). This model is significantly
  mixed (just like the other two models) and so the lack of He\one\ lines results from the low helium abundance and the
  large CO core mass \citep{dessart_etal_12}.

\begin{figure*}
\epsfig{file=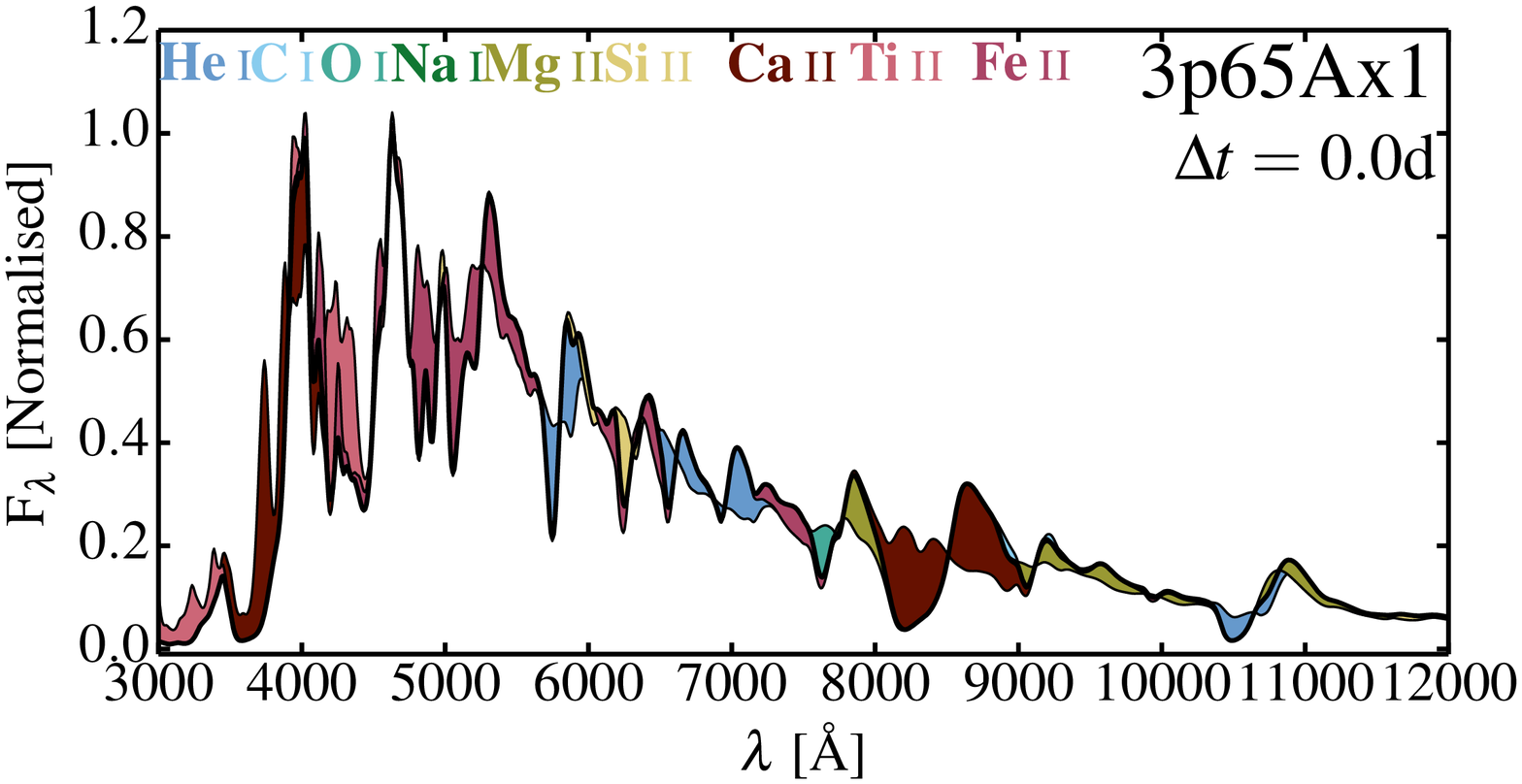,width=15cm}
\epsfig{file=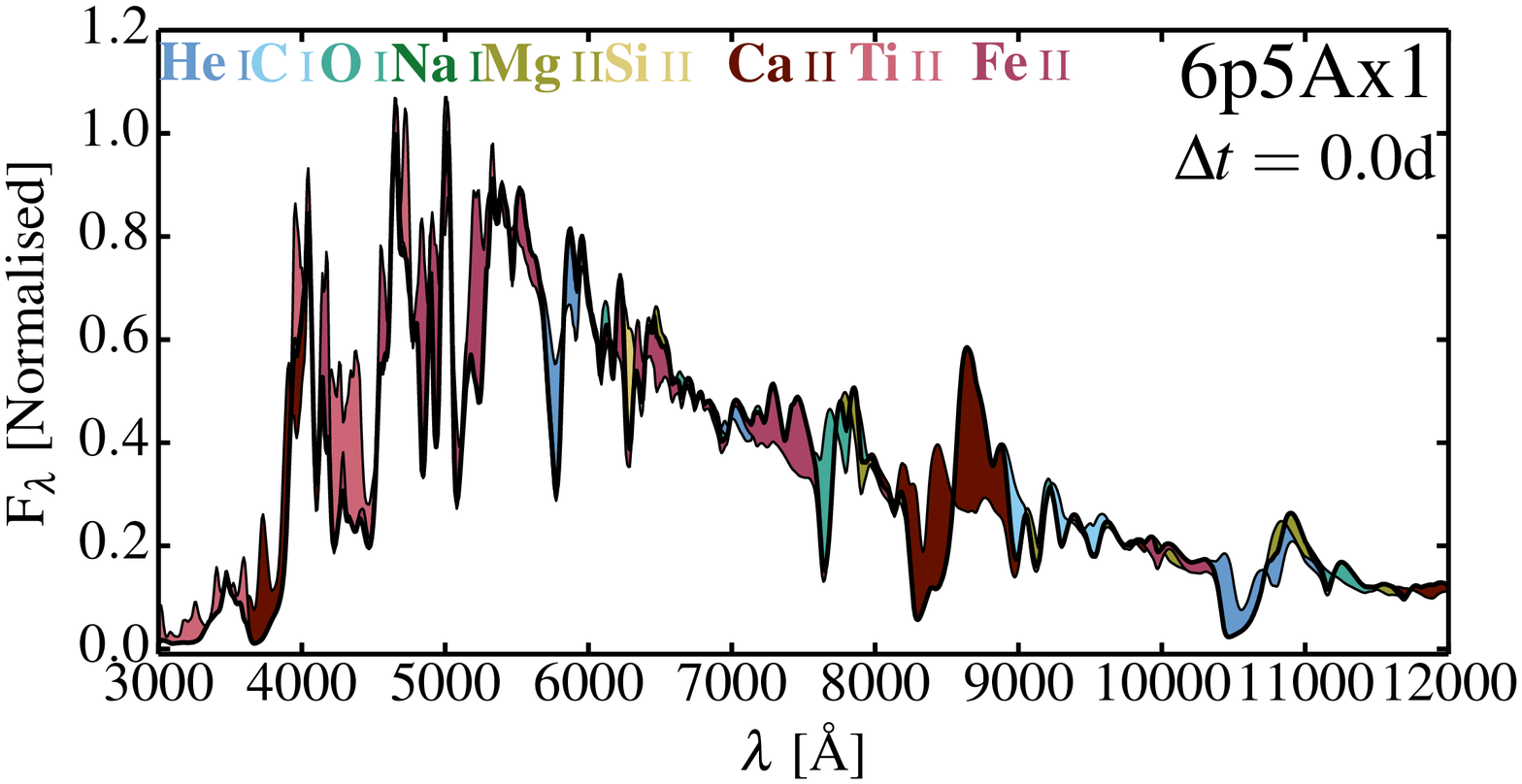,width=15cm}
\epsfig{file=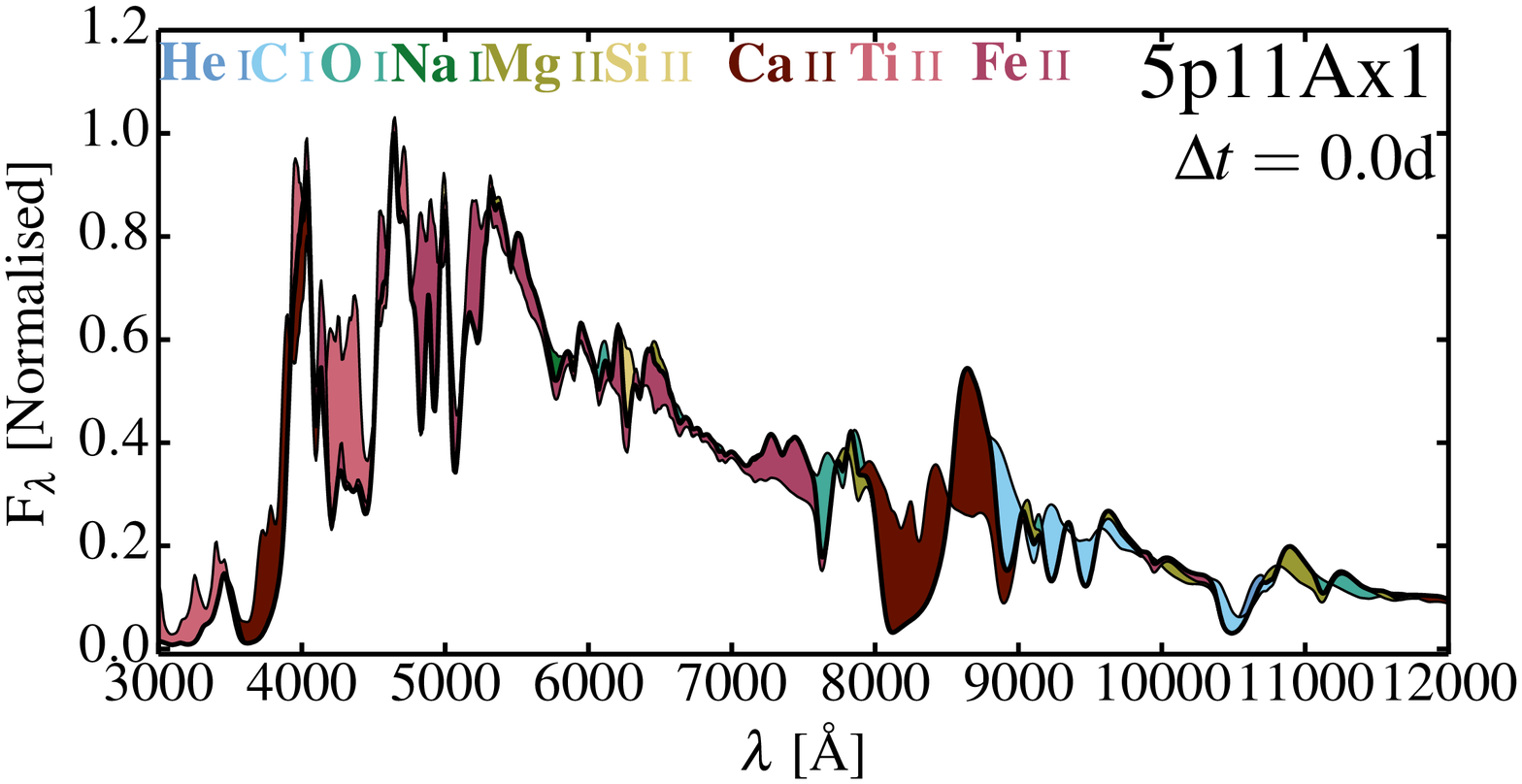,width=15cm}
\caption{Maximum light optical spectra for models 3p65Ax1 (SN IIb), 6p5Bx1 (SN Ib), and
5p11Ax1 (SN Ic).
We illustrate the contributions from lines of different ions by colouring the space that separates
the full spectrum (dark black curve) from that obtained by neglecting a given ion from
the formal solution of the radiative-transfer equation.
\label{fig_mspec_max_optical}
}
\end{figure*}

\begin{figure*}
\epsfig{file=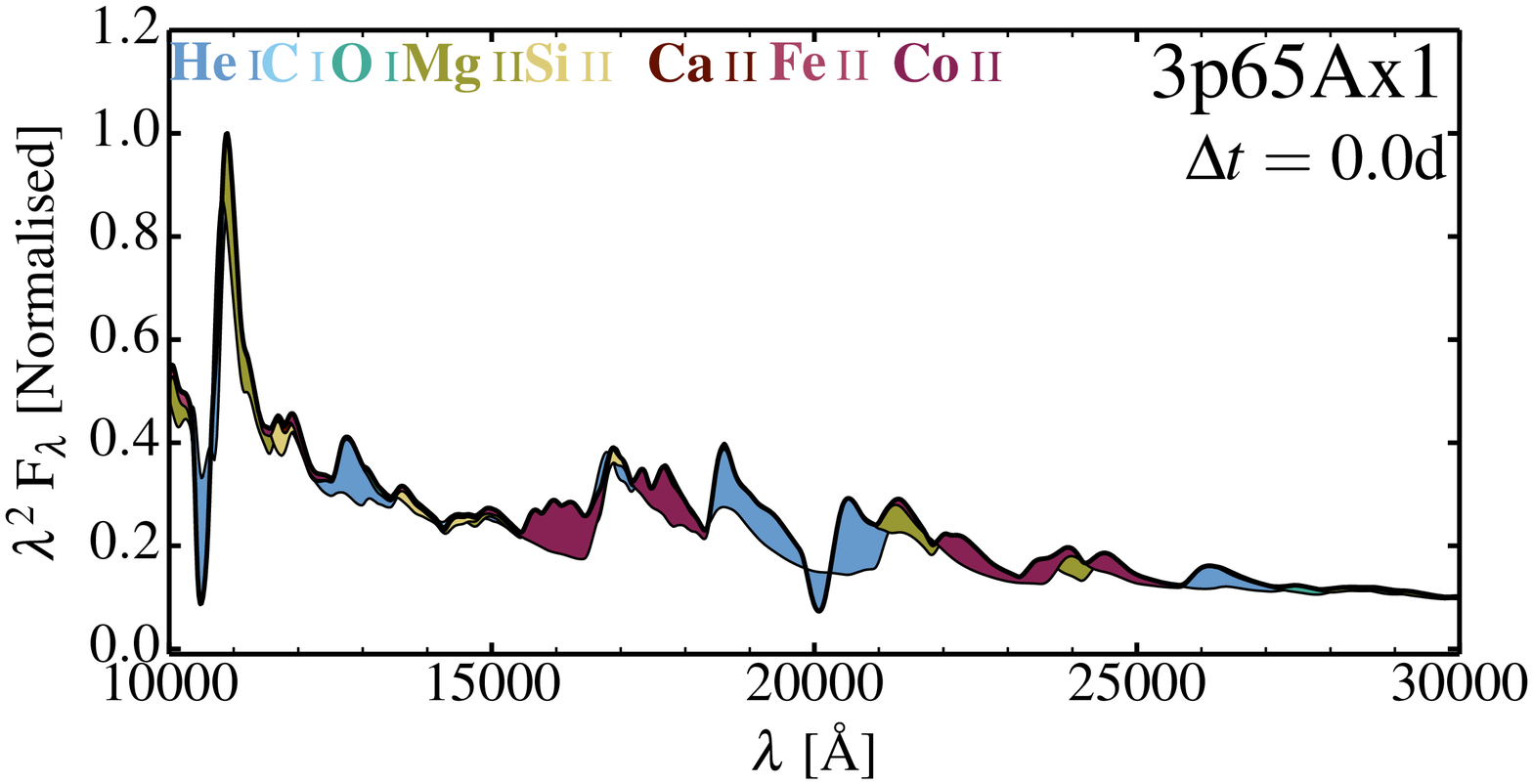,width=15cm}
\epsfig{file=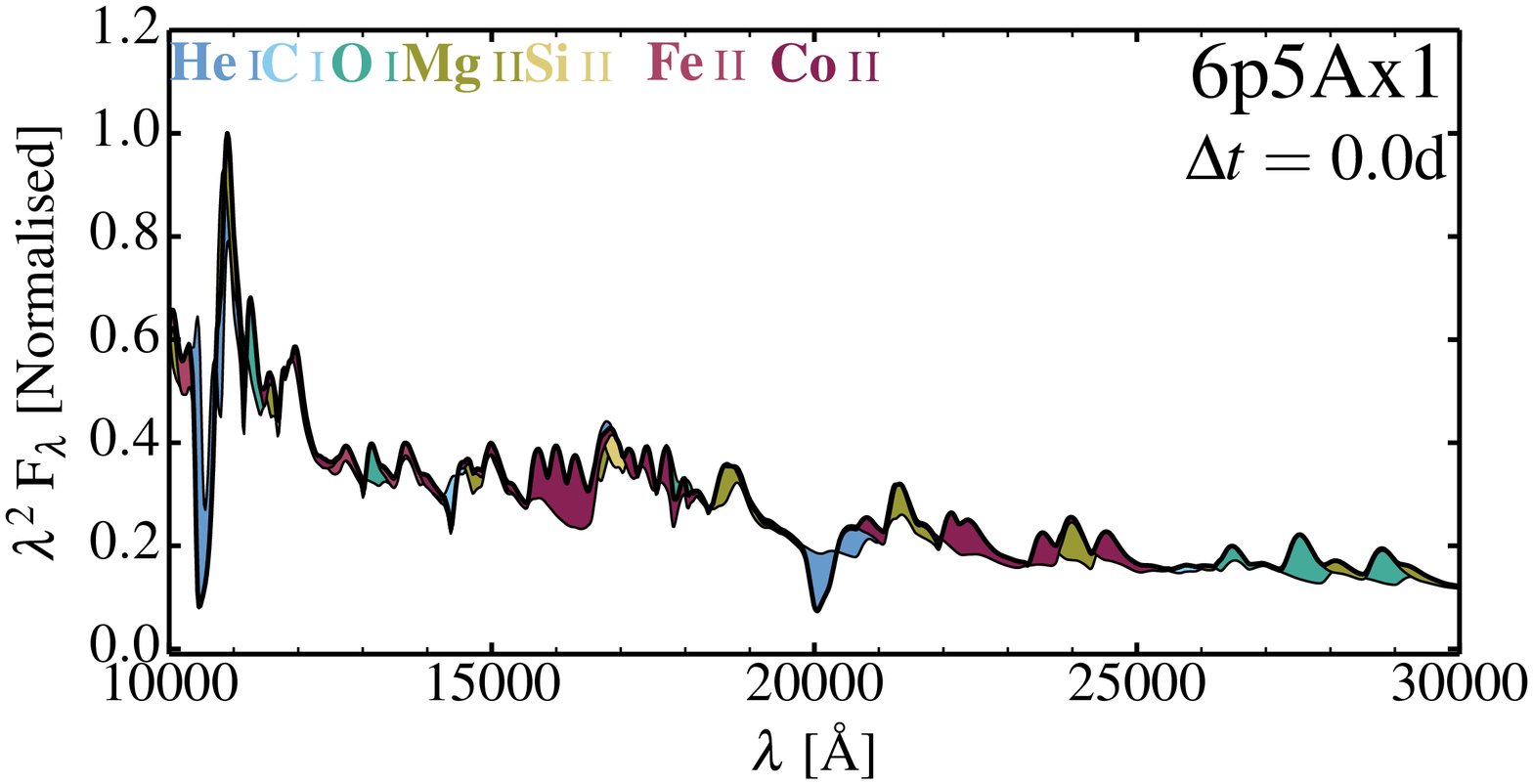,width=15cm}
\epsfig{file=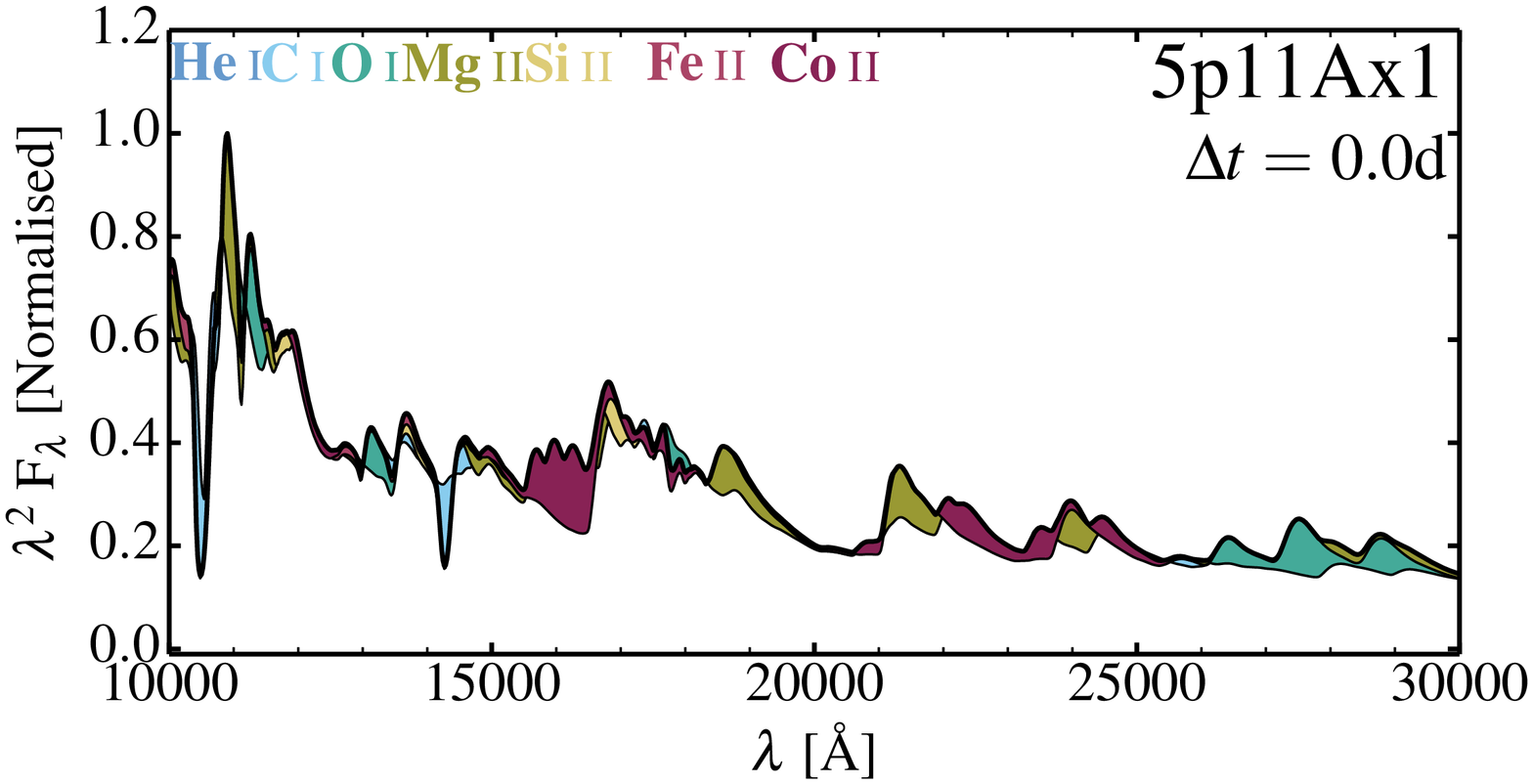,width=15cm}
\caption{
Same as Fig.~\ref{fig_mspec_max_optical}, but now for the near-IR range.
For better visibility, we show the quantity $\lambda^2 F_{\lambda}$.
\label{fig_mspec_max_nir}
}
\end{figure*}

  The buffer of mass lying in between the \iso{56}Ni rich layers and those rich in helium is much larger
  in model 5p11Ax1, preventing efficient
  non-thermal excitation at early times. Model 5p11Ax1 exhibits an extended oxygen-rich shell,
  whose outer part is rich in carbon and poor in helium (mass fraction of 10\%). In contrast, in models 3p65Ax1 and
  6p5Ax1, the oxygen shell is less massive and resides underneath a massive helium-rich shell ($>$90\% mass fraction helium).
  This difference in core structure stems from the difference in main sequence mass. Models 3p65
  and 6p5 stem from stars with initial masses 16 and 25\,\msun, which make a relatively small CO core compared to model 5p11
  which stems from a star with an initial mass of 60\,\msun\ that eventually formed a massive CO core.
  All else being the same (\iso{56}Ni mass, mixing magnitude, explosion energy),
  this difference in core structure causes the lack of He\one\ signatures in model 5p11Ax1. In this SN Ic model,
  the feature at 5900\,\AA\ is solely due to Na\one\,D (strong at early and late times but weak
  around maximum).

  Our result disagrees somewhat with \cite{hachinger_13_he} who
  find in their study that as little as 0.14\,\msun\ of He may suffice for the production of He\one\ lines.
  There is probably no clear threshold because the excitation of He\one\ lines will depend not just on He abundance,
  but also on \iso{56}Ni mass, mixing, the size of the CO core, and the C and O mass fraction in the outer parts
  of the progenitor.

   Iron group elements have a mass fraction equal to the progenitor metallicity value throughout the ejecta
  except in the innermost layers (with the exception of \iso{56}Ni and its daughter nuclei).
  Here, iron-group elements cause blanketing in the optical (Fe\two, Ti\two) and in the UV (Ti\two).
  Numerous lines from Co\two\ produce a broad emission feature at 1.6$\mu$m.
  Because of steady burning within the progenitor helium core,
  intermediate mass elements have abundances that are over- or under-abundant compared to solar,
  with variations between models and also with depth for each model.

  In the type Ic model, most of the features between 9000\,\AA\ and $\approx$\,1$\mu$m are
  due to C\one, while these features
  are very weak in the Type IIb model 3p65Ax1.  In the type Ib model 6p5Ax1, these C\one\ lines
  are visible around maximum and later because of the larger CO-rich shell in this more massive
  progenitor. Hence, C\one\ lines offer a possible diagnostic if the identification of He\one\ lines
  is ambiguous -- a strongly He-rich photosphere would most likely exhibit weak or no C\one\ lines
  in the 9000\,\AA\ and $\approx$\,1$\mu$m region.

  Oxygen shows only one signature in the optical with O\one\,7777\,\AA. This feature is strong at all
  times in model 5p11Ax1 but its presence in the type IIb model 3p65Ax1 likely results from mixing (the
  helium-rich shell is strongly deficient in oxygen in the pre-SN model). Improving our current
  algorithm for mixing is work in progress --- while \iso{56}Ni may be mixed far out in the ejecta,
  intermediate mass elements may not \citep{wongwathanarat_15_3d}.

  Na\one\,D is conspicuous only in the type Ic model, and could be confused with He\one\,5876\,\AA.
  Neon lines are absent at all times and in all models.
  Magnesium lines stem primarily from Mg\two. In the optical, we predict (usually blended) lines at 7790,
  8213--8234, 9218--9244, 10910, 21368--21432, 24042--24124\,\AA. In model 3p65Ax1, these lines
  are probably present even at early times  because we mix magnesium out in velocity space
  (O, Ne, Mg are originally under-abundant in the He-rich shell).

  Lines from Ca\two\ (H\&K lines and the triplet at 8500\,\AA) are strong at all times, and of comparable strength
  in the three models. Being closely tied to the ground state (directly or through Ca\two\,H\&K for the triplet), these
  lines are strong even at solar abundance so they provide little constraint to help distinguish different ejecta.

  All these signatures are illustrated at the time of bolometric maximum for the optical range
  in Fig.~\ref{fig_mspec_max_optical} and for the near-IR range in Fig.~\ref{fig_mspec_max_nir}.
  These figures are merely for illustration, not quantitative estimates, since excluding a given ion
  in the computation of the spectrum breaks the consistency of the calculation.

\section{Near-IR H\lowercase{e}\one\ lines}
\label{sect_hei}

   The identification of He\one\ lines in SN Ib/c spectra is a recurring issue in the SN
   community. The problem is overemphasised since helium is likely present in
   all core-collapse SNe, simply in different abundances. For example, optical He\one\ lines
   are seen in Type II SN spectra only prior to hydrogen recombination. During the
   rest of the photospheric phase, optical He\one\ lines are absent although about 30\% of
   the total hydrogen-envelope mass is helium \citep{dessart_etal_13}. In that case, there is little
   debate about the presence of helium because hydrogen is present, although the difficulty in
   producing He\one\ lines in such cool atmospheres is already evident.

   In hydrogen poor ejecta, the same issue returns, but here the problem is whether the SN
   atmosphere lies within the He-rich shell ($\gtrsim$\,90\% helium mass fraction) or within
   the CO core (typically $\lesssim$\,10\% helium mass fraction).
   Hence, the production of He\one\ lines depends on the ejecta composition. Furthermore,
   the need for non-thermal excitation/ionisation represents a major hurdle. Any ejecta with
   a massive CO core will probably never show strong He\one\ lines because of the large
   mass buffer between the \iso{56}Ni rich regions and the outer He-rich regions (whose
   presence is not even guaranteed because of mass loss). This suggests that SNe IIb/Ib progenitors
   preferentially arise from lower mass progenitors (which have small CO cores), where even
   moderate mixing can foster the production of He\one\ lines \citep{dessart_etal_12}.

\begin{figure*}
\epsfig{file=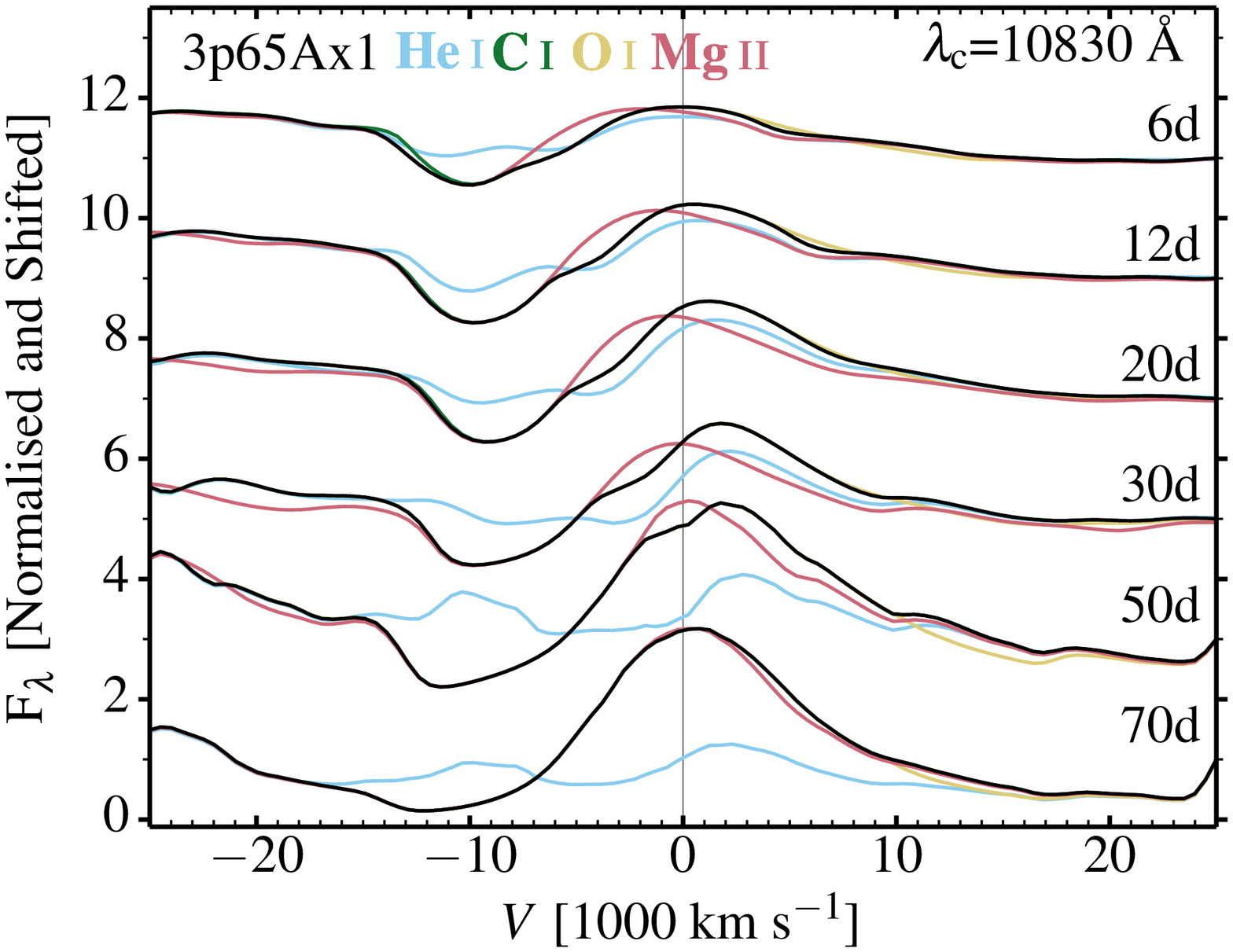,width=8cm}
\epsfig{file=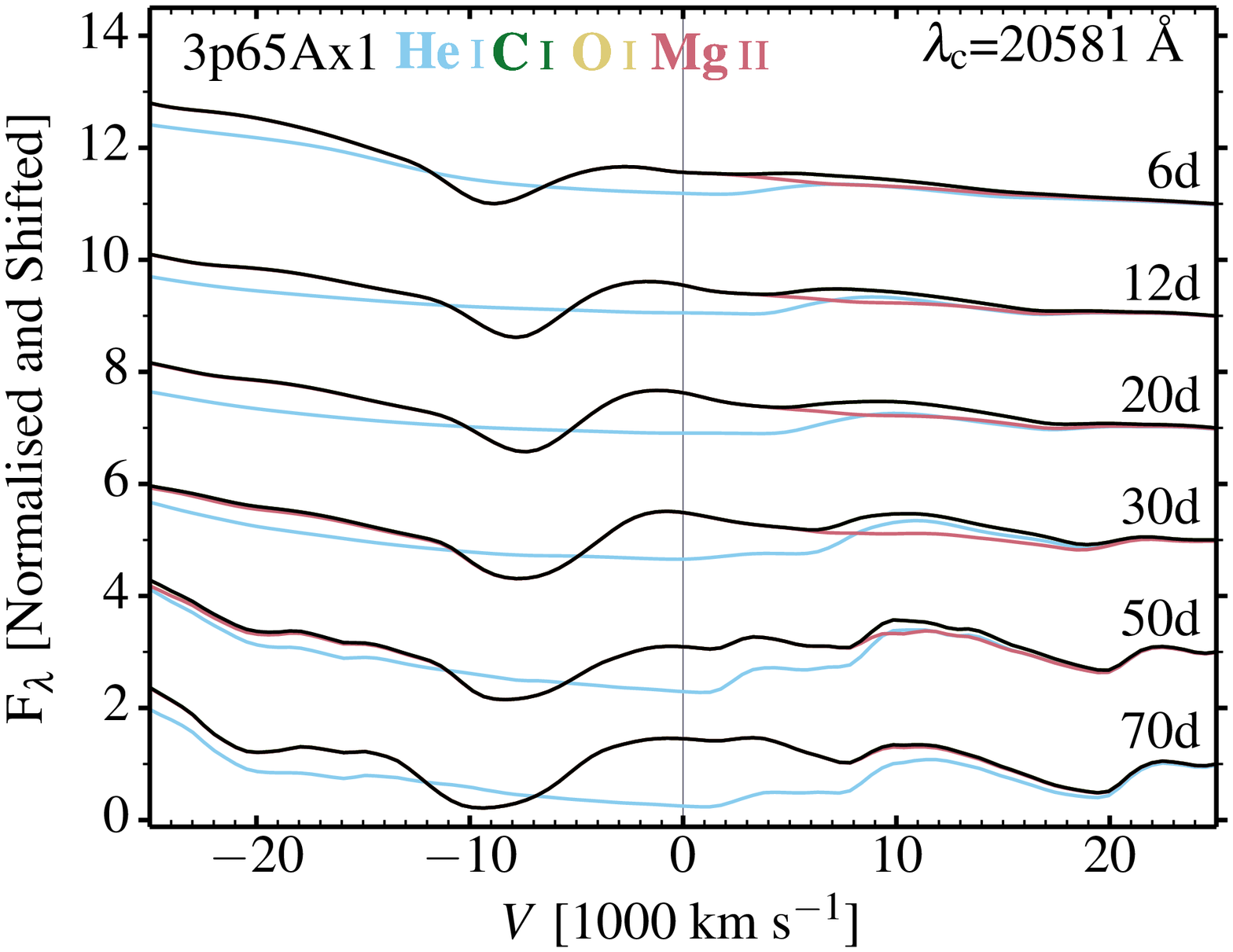,width=8cm}
\caption{Illustration of the 10830 and 20581\,\AA\ regions for model 3p65Ax1 (type IIb).
We show the contributions from He\one, C\one, O\one, Mg\two\ lines  to highlight the fractional
importance of each species/ion. In these IIb spectra, high velocity absorption (around $-$10,000\,\kms) associated
with both 10830 and 20581 provides unambiguous evidence for the presence of He. However, even in this
case Mg\two\ has a significant influence on the emission component of 10830\,\AA.
\label{fig_3p65Ax1_hei}}
\end{figure*}

\begin{figure*}
\epsfig{file=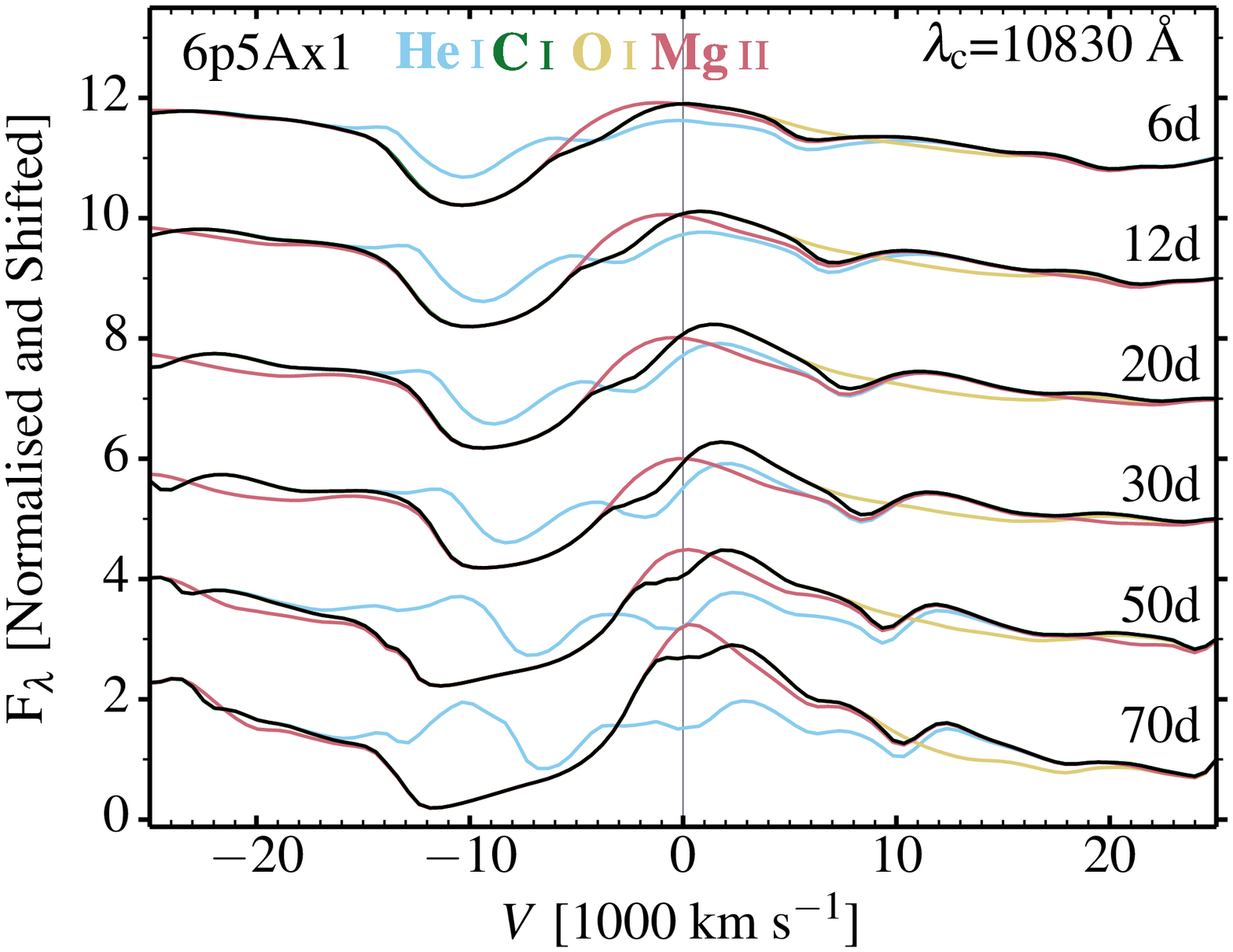,width=8cm}
\epsfig{file=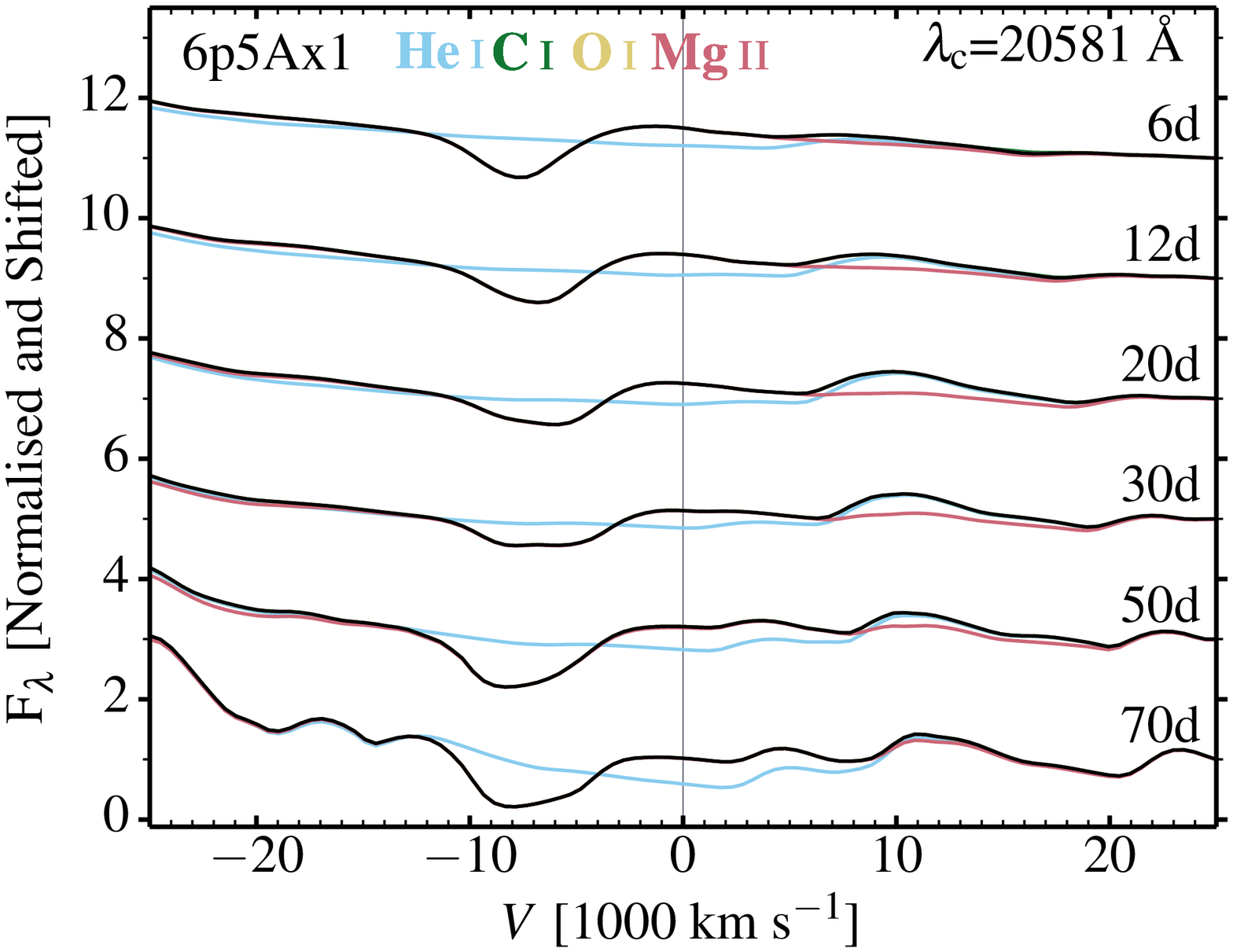,width=8cm}
\caption{
Same as Fig.~\ref{fig_3p65Ax1_hei}, but now for model 6p5Ax1 (type Ib).
He\one\,20581\,\AA\ is present at all times, and hence provides a very
useful diagnostic of the presence of helium.
\label{fig_6p5Ax1_hei}}
\end{figure*}

\begin{figure*}
\epsfig{file=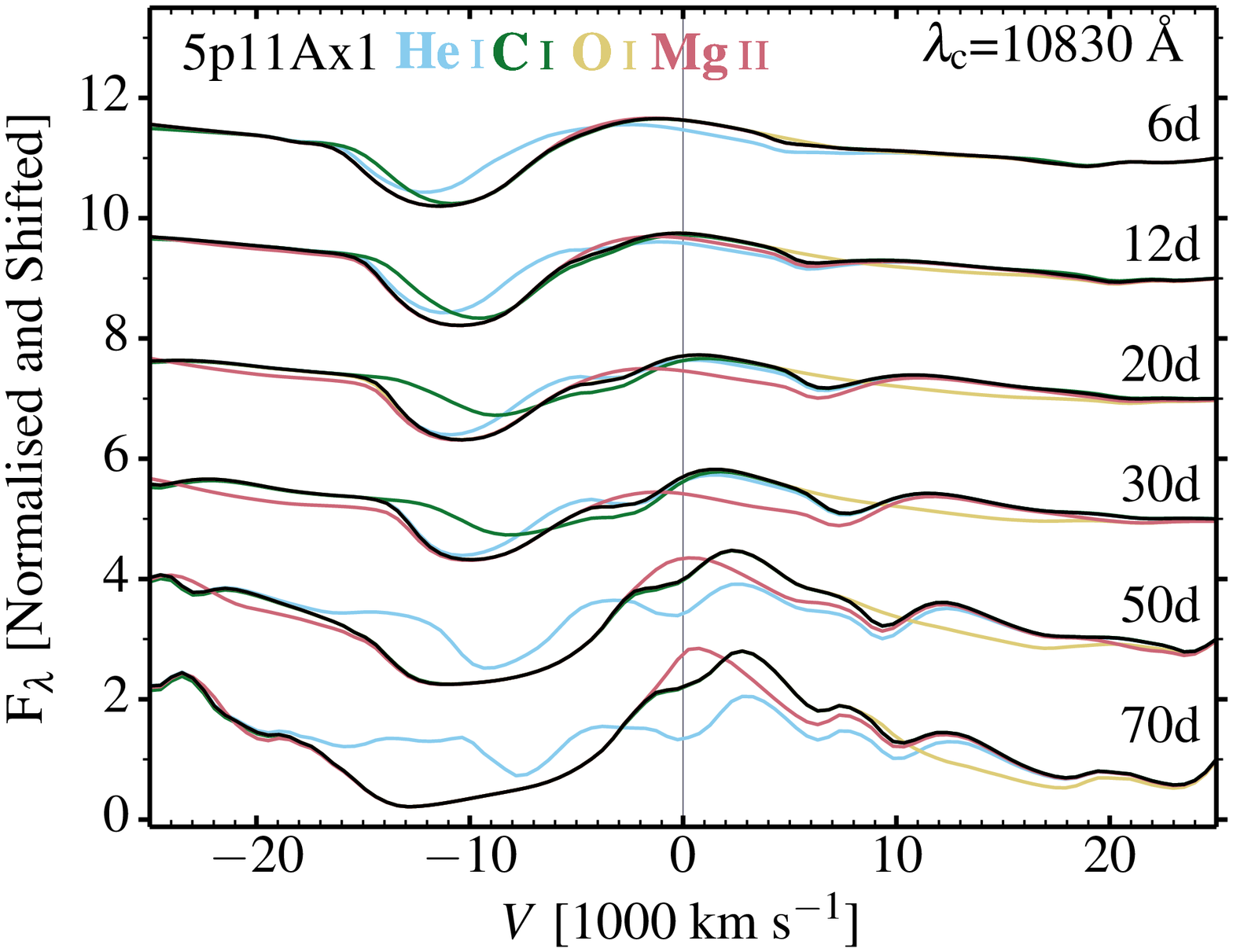,width=8cm}
\epsfig{file=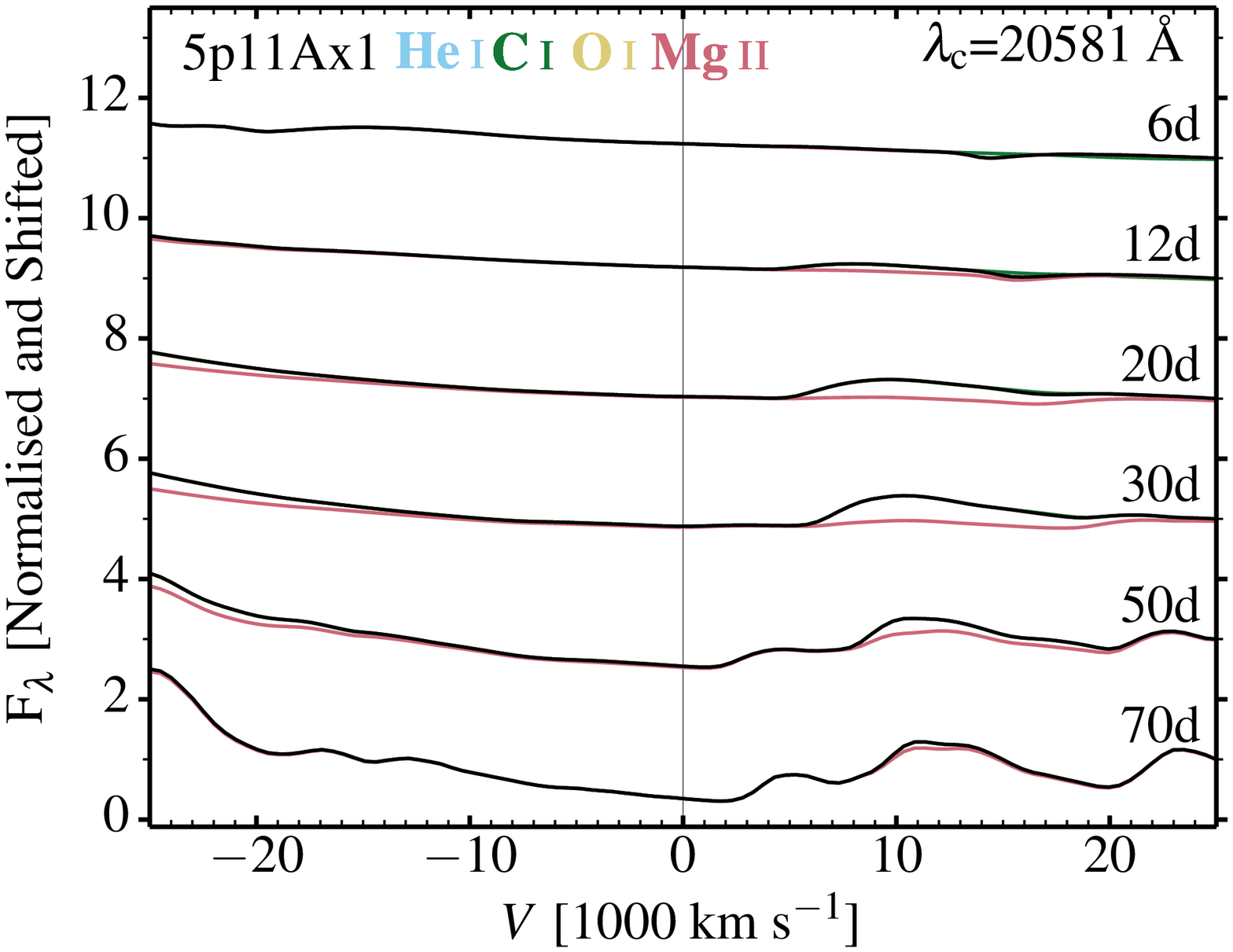,width=8cm}
\caption{Same as Fig.~\ref{fig_6p5Ax1_hei}, but now for model 5p11Ax1 (type Ic).
Notice the complicated blend associated with the P~Cygni profile at 10830\,\AA\
while there is a firm non-detection of He\one\ at 20581\,\AA. At late times the 10800\,\AA\
feature shows high velocity absorption (beyond $-$10,000\,\kms) that is only associated with
He\one\,10830\,\AA.
\label{fig_5p11Ax1_hei}}
\end{figure*}

   In this section, we document the properties of two spectral regions notorious
   for their possible He\one\ signatures, at 10830\,\AA\ and 20581\,\AA\
   (Figs.~\ref{fig_3p65Ax1_hei}--\ref{fig_5p11Ax1_hei}).
   He\one\,10830\,\AA\ is the strongest line, and has the potential to confirm or deny the presence of
   He in SN Ic. \citet{filippenko_etal_95} used the presence of strong absorption near 10830\,\AA\ to argue
   for the presence of He in the Type Ic SN\,1994I.
   Later work by \citet{millard_etal_99} found, using the parameterised synthesis code {\sc synow}, that the feature
   could not be fitted by He\one\ alone. They suggested that it was a possible blend with C\one.
   Later work by \citet{sauer_etal_06} confirmed this suggestion.
   Other studies \citep[e.g.,][]{meikle6_10830,hachinger_13_he} show that Mg\two\ is also a possible contributor.

   In type IIb/Ib models 3p65Ax1 and 6p5Ax1, the 10830\,\AA\ region is dominated by
   the He\one\ line, with a modest contribution of Mg\two\,10914\,\AA\ (multiple transitions),
   and O\one\,11300\,\AA\ in the red wing (multiple transitions).
   At late times, numerous lines from Ca\two, Fe\two, and Co\two\ appear around 1$\mu$m
   and affect the blue wing of the He\one\ line.
   In model 5p11Ax1, similar features are present, but with the addition of a strong overlapping
   component from numerous C\one\ transitions around 10700\,\AA. Around bolometric
   maximum, the He\one\ contribution to the 10900\,\AA\ feature is weak relative to that of C\one\ and Mg\two.
   However, it is still important. To remove the feature seen near 10900\,\AA\ we would need to simultaneously
   omit He\one, C\one, and Mg\two\ lines from the spectrum calculation.
   The relative strength of these different contributions varies with time, and He\one\ is a significant
   component at all times only in the type IIb/Ib models.

    In the 20581\,\AA\ region, the dichotomy between SNe IIb/Ib and SNe Ic is more evident.
    Indeed, He\one\,20581\,\AA\ is present in our type IIb/Ib models but completely absent
    in the type Ic model. Because there are no other lines at that location, the lack of a feature
    there is clear evidence that there is no He\one\ line. \citet{hachinger_13_he} also highlight
    the importance of the  He\one\,20581\,\AA\ line in assessing the He abundance of SN Ibc ejecta.
    Note that in our SNe IIb/Ib models,
    there are He\one\ lines around 1.9\,$\mu$m (e.g., 19543.1\,\AA), which spread the influence of He\one\ far to
    the blue of the 2.0581\,$\mu$m feature.

    An interesting feature of our simulations, also observed (see, e.g., \citealt{ergon_14_11dh}
    for SN 2011dh), is the migration of the location of maximum absorption (or more
    generally the broadening) of the He\one\ lines at 10830 and 20581\,\AA. This is the spectral
    counterpart of the rapid photometric decline at late times. Indeed, before $\gamma$ rays start to escape,
    they first deposit their energy non locally within the ejecta, increasingly influencing the outer layers
    after bolometric maximum. When this effect arises, non-thermal excitation and ionisation processes
    are boosted in the outer ejecta layers (which are poor in \iso{56}Ni initially), increasing the optical
    depth in these strong He\one\ lines. In type Ic model 5p11Ax1, only the He\one\ contribution to the
    10900\,\AA\ feature shows this behaviour, which may serve as a tracer of helium in the ejecta.

\begin{figure}
\epsfig{file=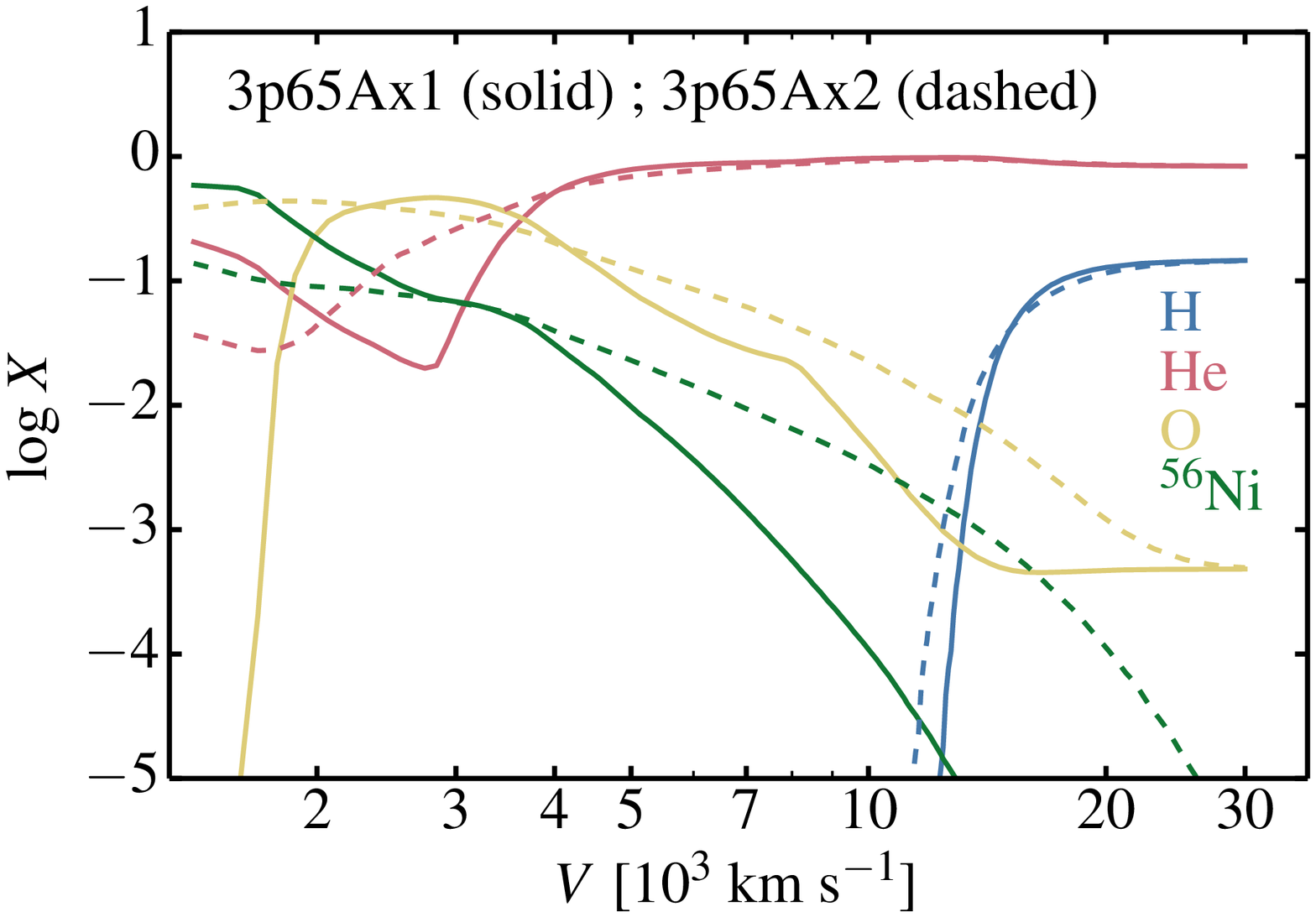,width=8.5cm}
\epsfig{file=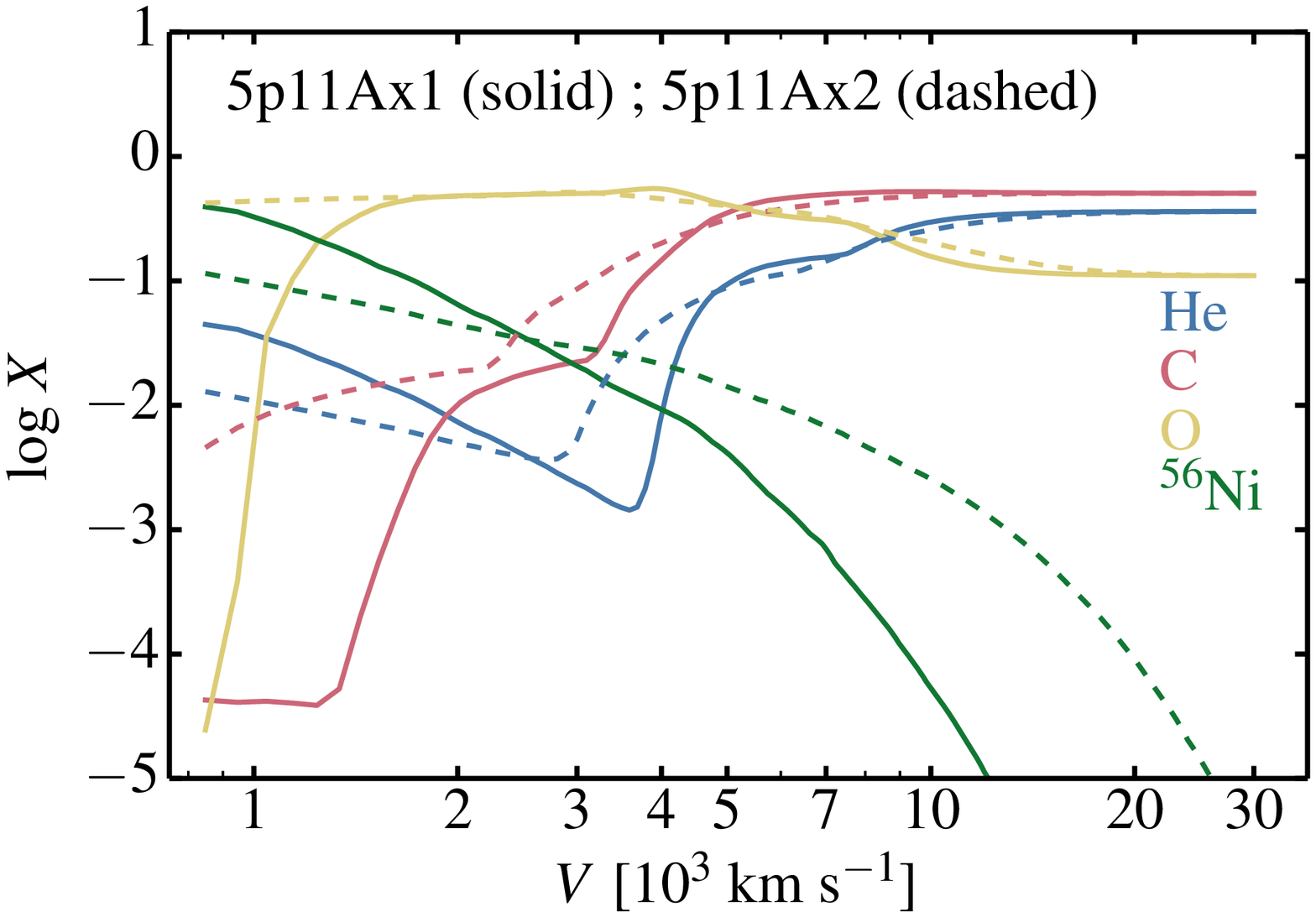,width=8.5cm}
\caption{Ejecta composition profiles for models 3p65A (top) and 5p11A (bottom)
with moderate (x1) and strong (x2) mixing.
[See Sections~\ref{sect_setup} and \ref{sect_mixing} for discussion.]
\label{fig_comp_mix}}
\end{figure}

\begin{figure}
\epsfig{file=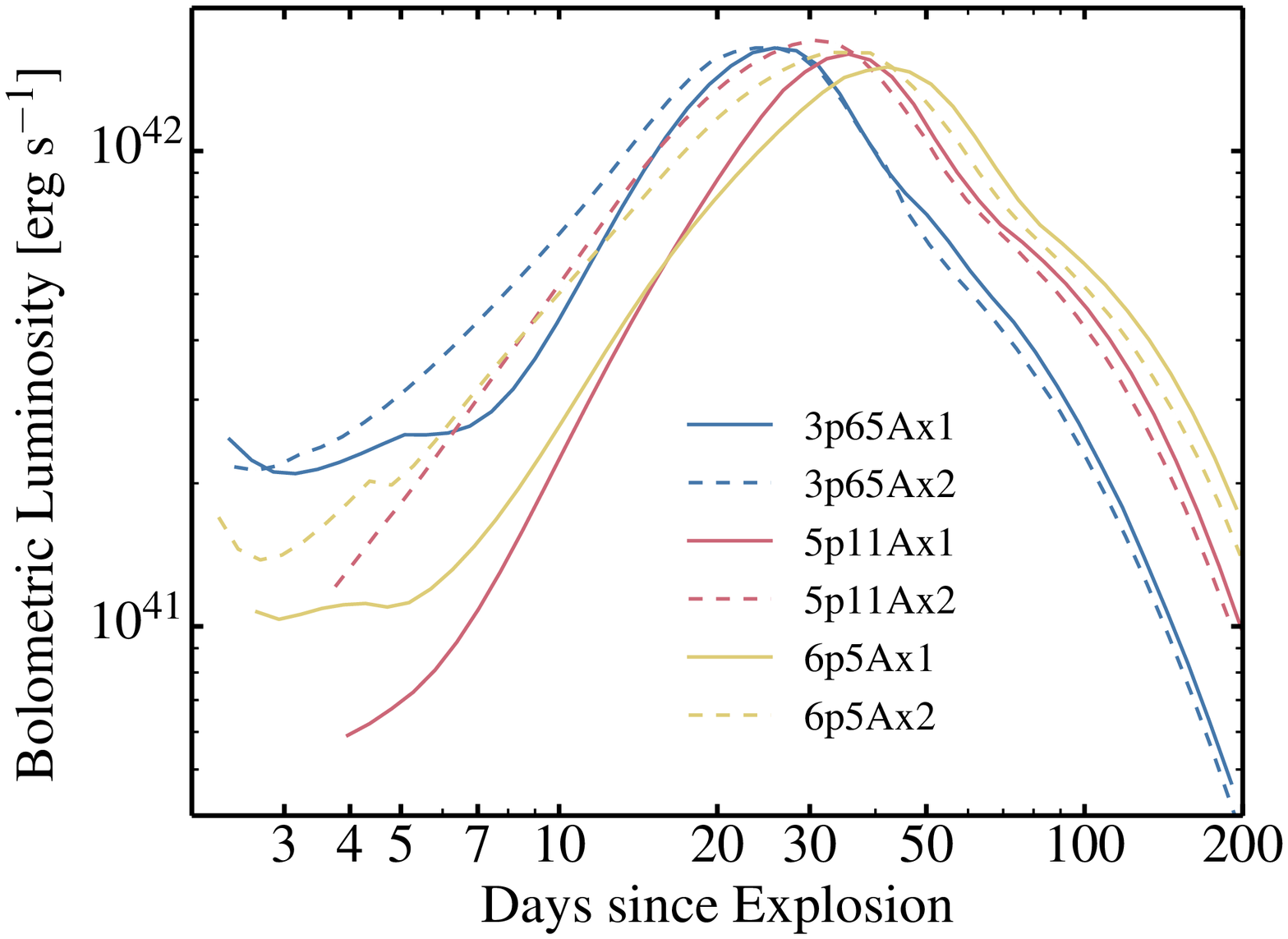,width=8.5cm}
\epsfig{file=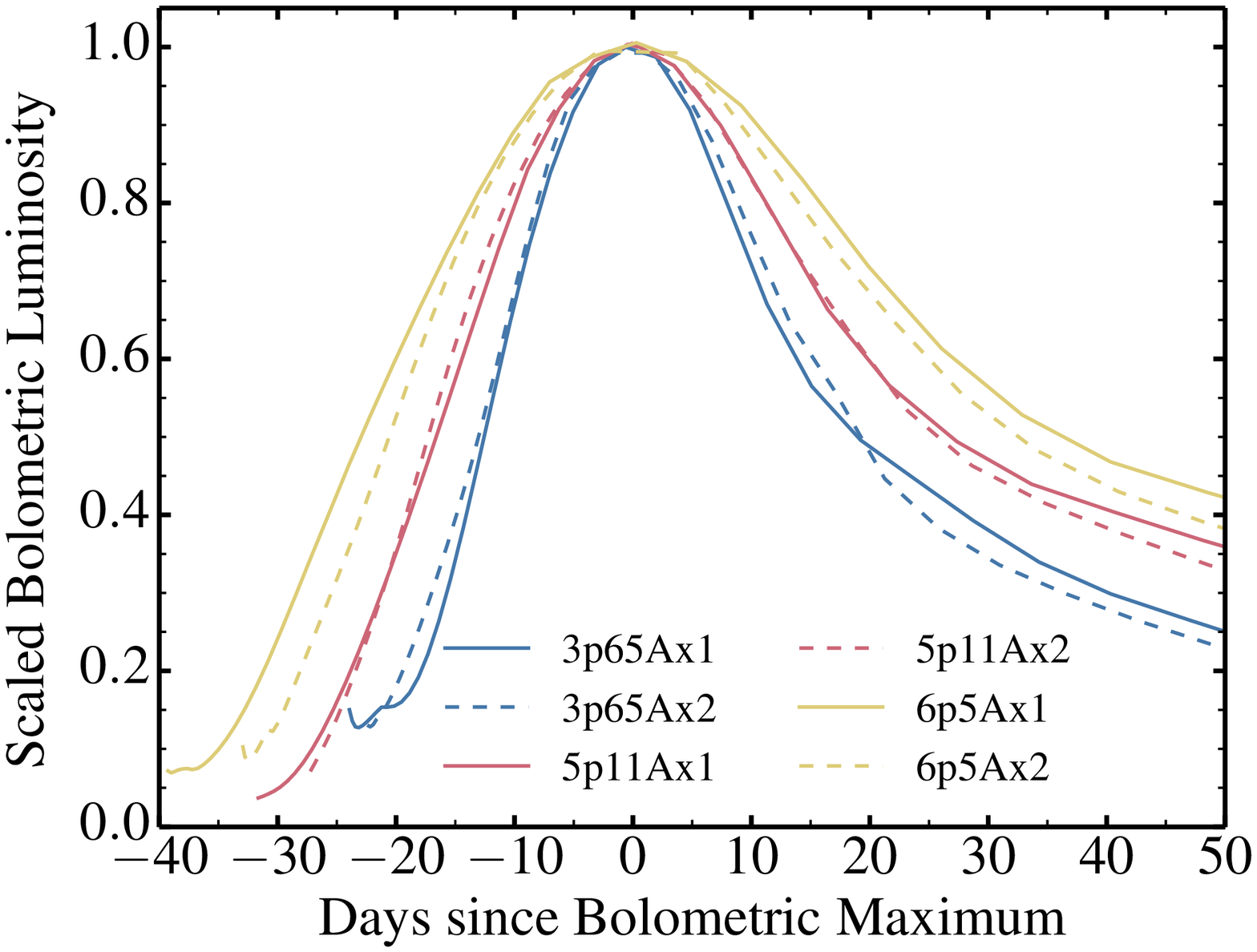,width=8.5cm}
\caption{
{\it Top:} Illustration of the effect of a variation in the magnitude of mixing
(x1 versus x2, i.e., moderate versus strong mixing) on the bolometric
luminosity light curve for our three ejecta models.
{\it Bottom:} Same as top, but now showing the bolometric light curves normalised to their
maximum value and shifted so that the origin is the time of bolometric maximum.
\label{fig_lbol_mix}}
\end{figure}

\begin{figure}
\epsfig{file=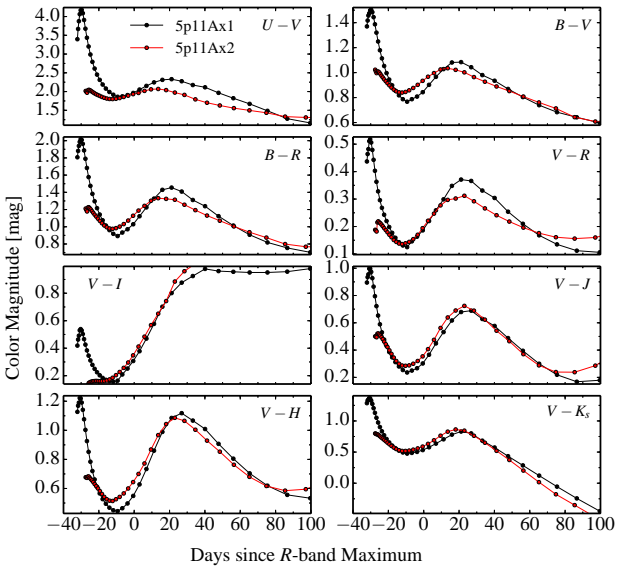,width=8.5cm}
\caption{Illustration of the impact of mixing on the colour evolution of model 5p11A.
Time is given in days since $R$-band maximum. Model with suffix
x1 has weaker mixing than model with suffix x2 (see Fig.~\ref{fig_comp_mix} for the composition profiles).
\label{fig_color_mixing}}
\end{figure}

\section{Influence of mixing}
 \label{sect_mixing}

    In \citet{dessart_etal_12}, we performed controlled experiments on the effect
    of mixing in He-rich low-mass ejecta models. This study emphasised how
    the distribution of \iso{56}Ni is critical for the production of He\one\ lines.
    Indeed, even in ejecta abundant in helium, weak mixing would lead to a type Ic SN.
    The existence of SNe IIb/Ib is a robust indication that extensive (macroscopic)
    mixing takes place in these explosions, and by extension, is a likely feature of all
    core-collapse SNe.\footnote{We note that, theoretically, He\one\, lines may be present at the earliest
    times without the influence of non-thermal effects, but their presence in this case is very short lived
    and limited to progenitors with a He-rich shell (for a discussion, see \citealt{dessart_11_wr}).}

    The production of He\one\ lines and a type Ib SN requires both the presence of a large helium
    mass and a strong mixing of \iso{56}Ni into the He-rich layers. This implies that
    any progenitor star characterised by a large CO core would tend to form a SN Ic
    because the He-rich regions are too remote from the inner ejecta where \iso{56}Ni
    is produced.
    It is likely that mixing is a ubiquitous feature of core-collapse SNe. The details and degree
    of mixing as a function of progenitor and explosion properties are, however, uncertain.

    In the present work, all models have therefore been mixed.
    From our set of models,
    we indeed produce type IIb/Ib and Ic SNe for the reasons described above.
    But to quantify the uncertainties associated with mixing, we run strongly-mixed
    counterparts to models 3p65Ax1, 5p11Ax1, and 6p5Ax1 --- these strongly mixed
    models are given the suffix x2.  Our mixing procedure at present applies to
    all species and is thus indiscriminate (Fig.~\ref{fig_comp_mix}).
   In contrast, models without mixing would show shells with nearly 100\% \iso{56}Ni
   mass fraction in the innermost layers of the ejecta and none above, as in the (unmixed) models studied
   in \citet{dessart_etal_12}.

\begin{figure*}
\epsfig{file=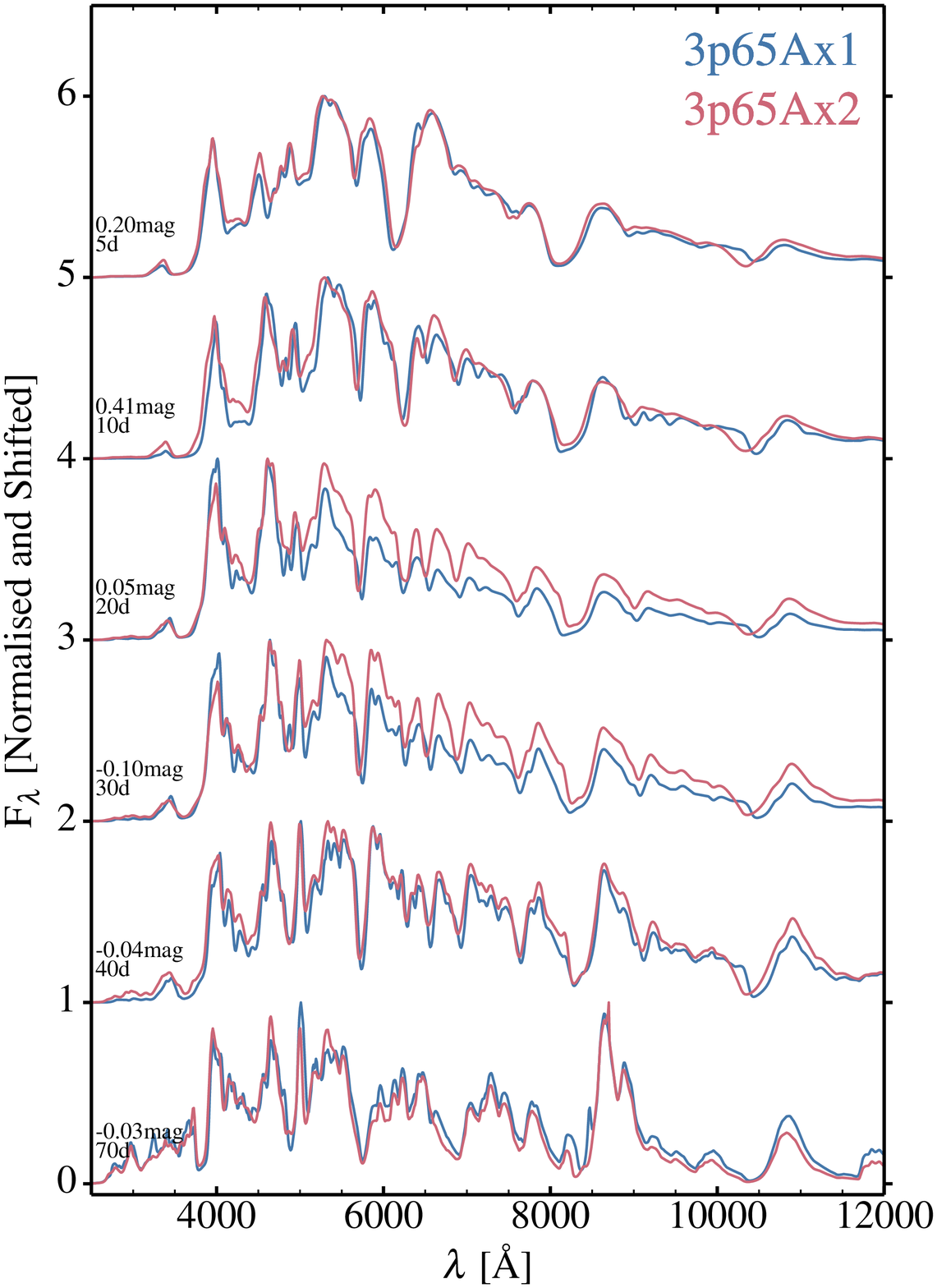,width=5.8cm}
\epsfig{file=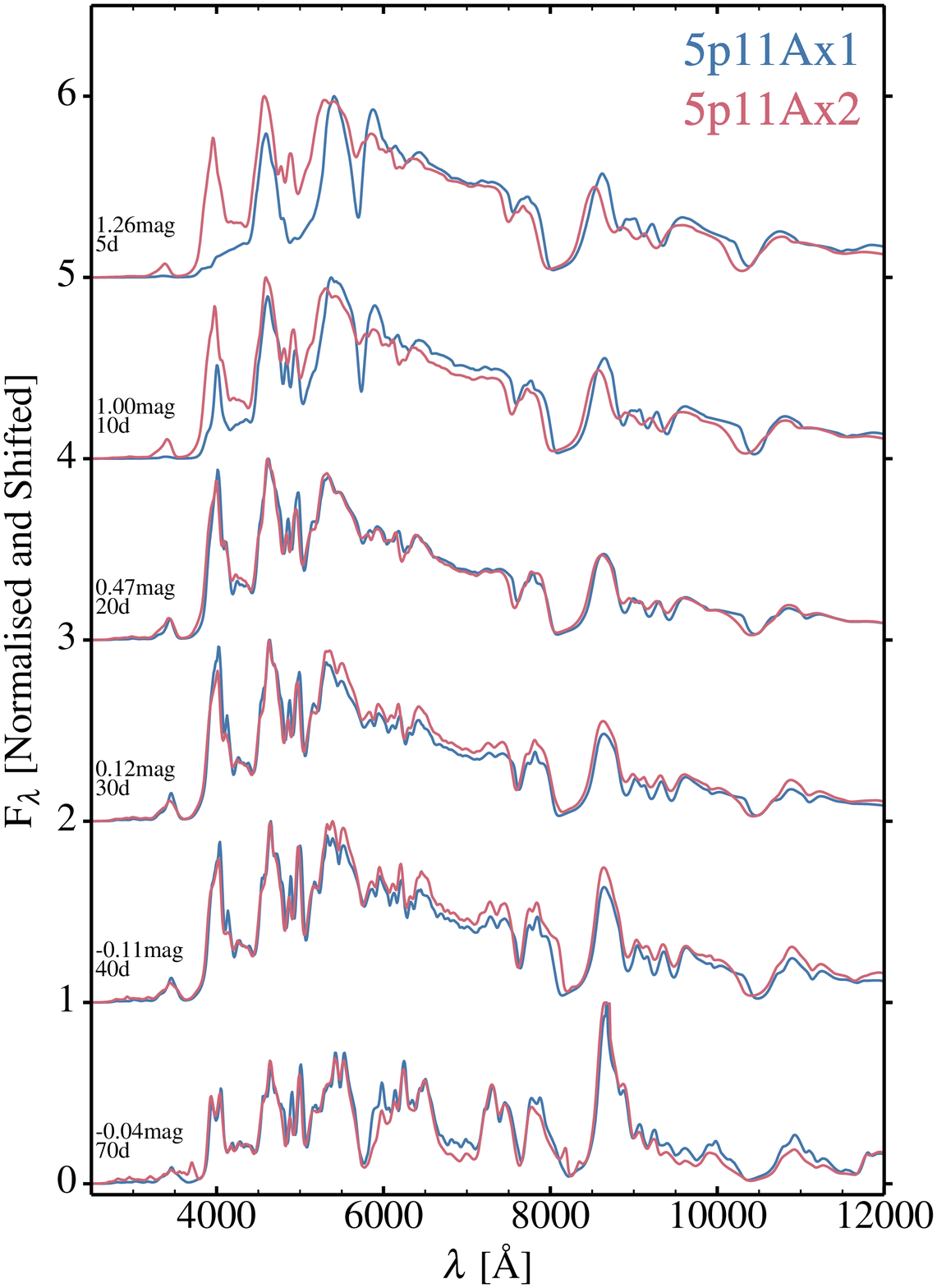,width=5.8cm}
\epsfig{file=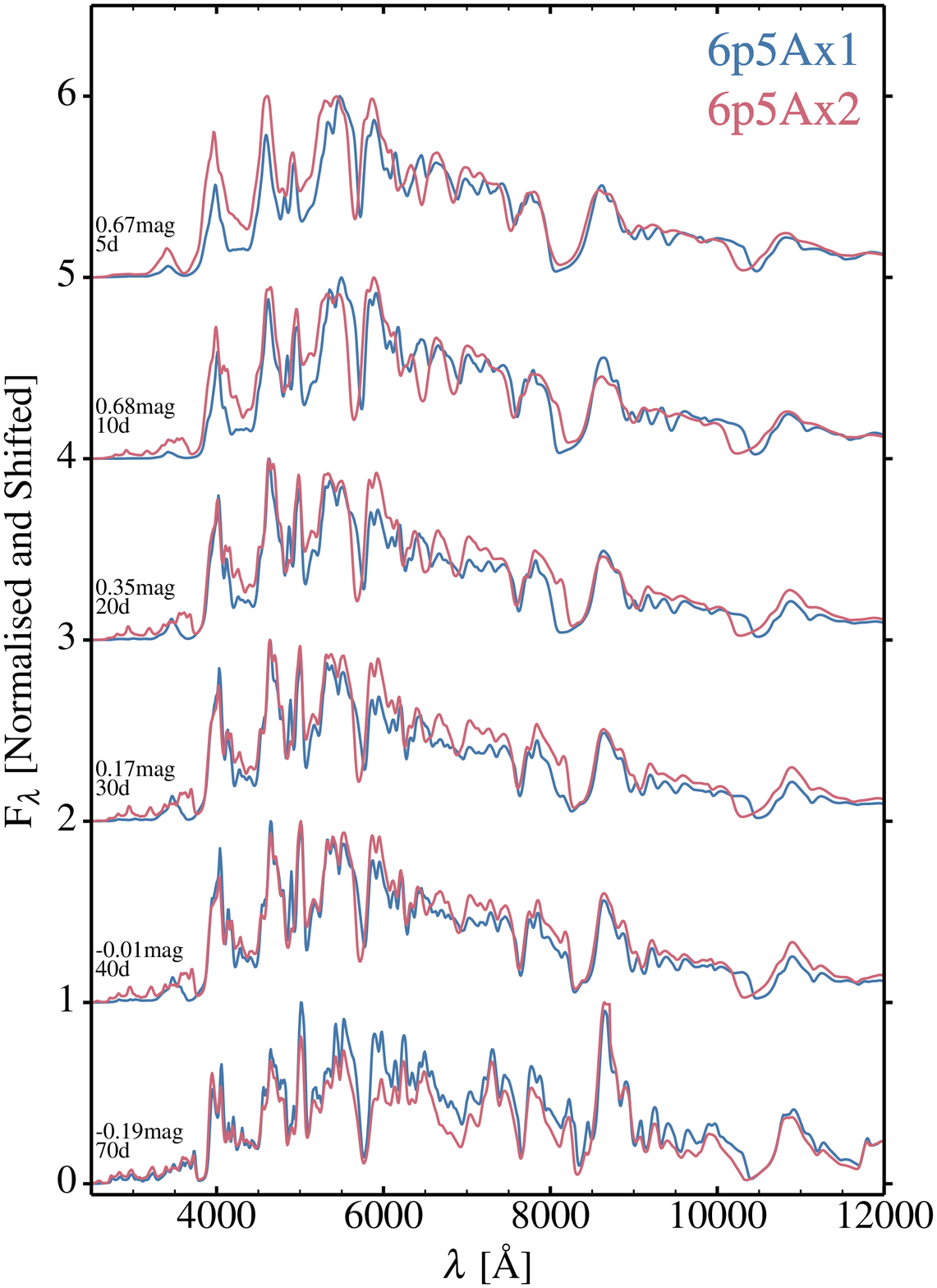,width=5.8cm}
\vspace{-1cm}
\caption{Illustration of the impact of mixing on the spectral evolution of our three models.
Each pair corresponds to ejecta with a moderate (x1) and a strong (x2) mixing.
Labels provide the time since explosion as well as the magnitude offset between the two models
at each date.
Besides the impact on the bolometric luminosity and colour, stronger mixing boosts
non-thermal processes and leads to broader lines (at fixed $E_{\rm kin}$ and M$_{\rm e}$).
The effect is therefore more evident on He\one\ lines.
\label{fig_spec_mixing}}
\end{figure*}

    As is well known, stronger mixing causes a higher luminosity  \cite[e.g.,][]{bersten_etal_12_11dh} and bluer
    colours \citep{dessart_etal_12} at early times.
 The larger amount of \iso{56}Ni at larger velocities causes enhanced escape as time progresses,
 causing a lower luminosity after maximum.
 In all three models, mixing causes the maximum to shift to
 earlier times, with the largest shift occurring for the highest mass ejecta  (Fig.~\ref{fig_lbol_mix}).
 For 3p65Ax1/x2 the shift is 1.4\,d, for 5p11Ax1/x2 it is 5.0\,d, and for 6p5Ax1/x2 it is 6.7\,d.
 For model 6p5Ax2 the maximum bolometric luminosity
 is about 10\% larger compared to the model with the lower mixing; for the other models the change is even smaller.
 When we use the time of bolometric maximum as the time origin, the curves for the moderately mixed
 and strongly mixed models show
 a remarkable similarity for the two models with the lowest ejecta mass -- the bolometric light curve for 6p5Ax2 is somewhat
 broader (by about a week at 50\% maximum) than the light curve for 6p5Ax1 (Fig.~\ref{fig_lbol_mix}).
 This morphological similarity for different mixing means that the explosion date could be in error by as much as a week
 if only the bolometric light curve was used.

   In our procedure, mixing enhances the mean atomic weight of the outer ejecta layers.
While this should strengthen line blanketing, we observe bluer colours
in strongly mixed models because of the extra heating from \iso{56}Ni decay (Fig.~\ref{fig_color_mixing}).
After maximum, the colour offset due to different mixing is much weaker. For ($V-R$) at 10\,d
past $R$-band maximum, the offset is negligible.

   Synthetic spectra corroborate these properties but also illustrate how stronger mixing leads
   to broader lines (see also \citealt{dessart_etal_12}), even at fixed ejecta mass and kinetic energy 
   (Fig.~\ref{fig_spec_mixing}).
   The effect is particularly visible on He\one\ lines and directly stems from the modulation of
   the magnitude of non-thermal processes --- SN Ic spectral line profiles are less sensitive
   to mixing in that respect.
   Uncertainties in the mixing magnitude in SNe Ib introduces
   an uncertainty in the inferences on the explosion energy, ejecta mass etc.

   The impact of mixing is greater in higher mass ejecta models like 5p11Ax1 and 6p5Ax1 because
   mixing is the only way the outer ejecta can be influenced by non-thermal processes. In lower
   mass ejecta, like model 3p65Ax1, strong mixing is less critical because the outer ejecta
   (basically the He-rich shell) is close to the \iso{56}Ni rich layer -- it is only separated by a thin O-rich layer.

\section{The ambiguous notion of a photosphere}
\label{sect_spec_form}

\begin{figure}
\epsfig{file=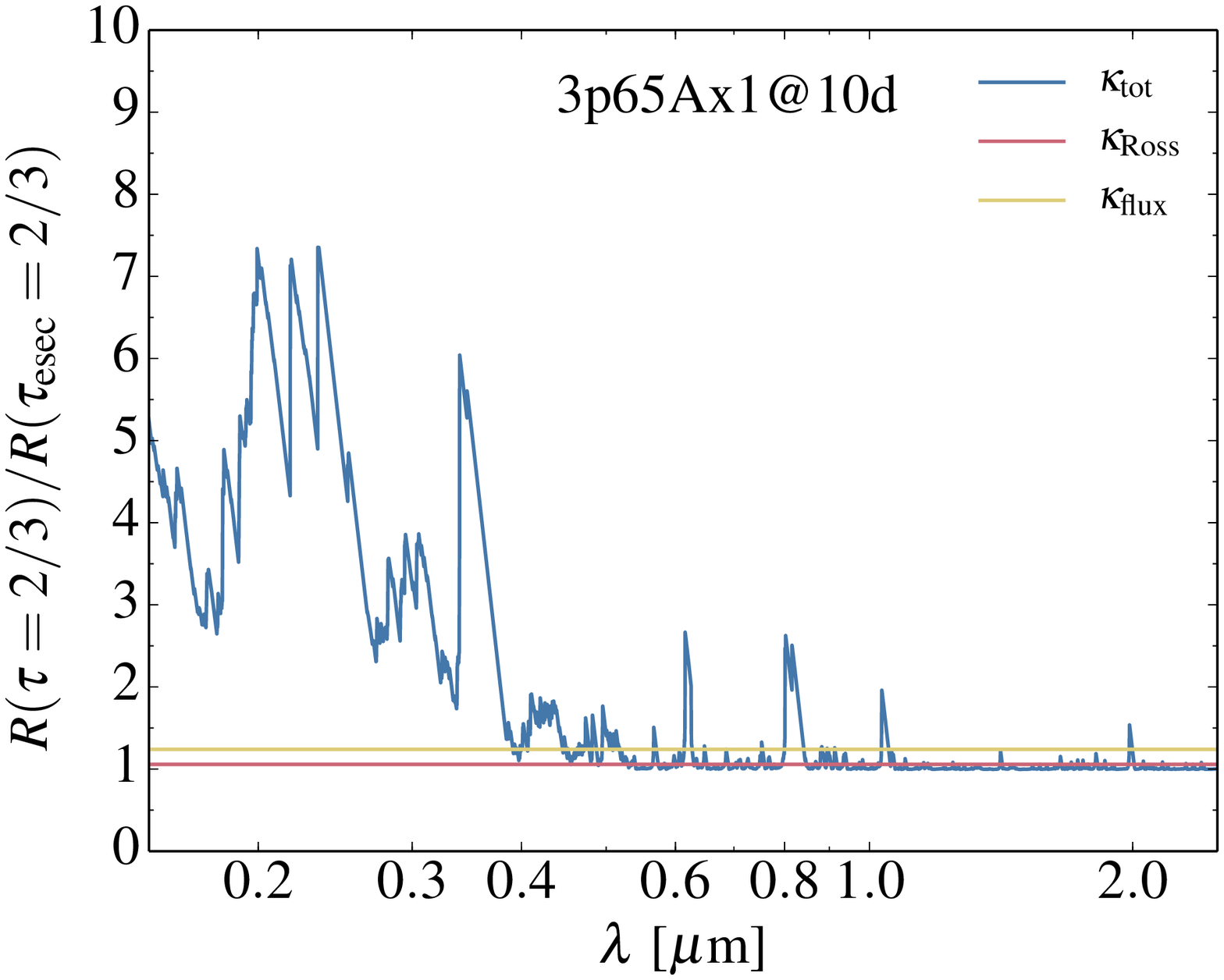 ,width=8.5cm}
\epsfig{file=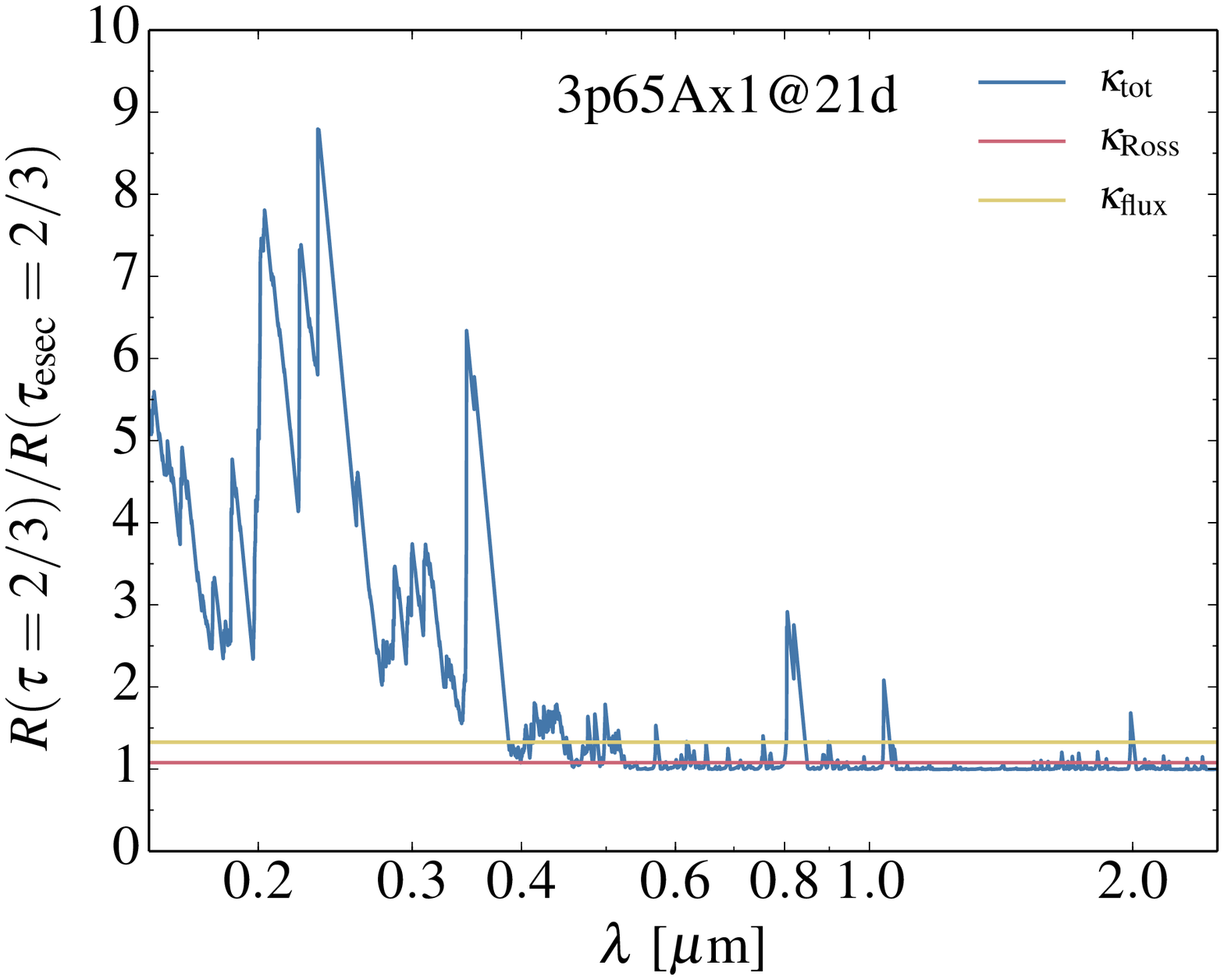,width=8.5cm}
\epsfig{file=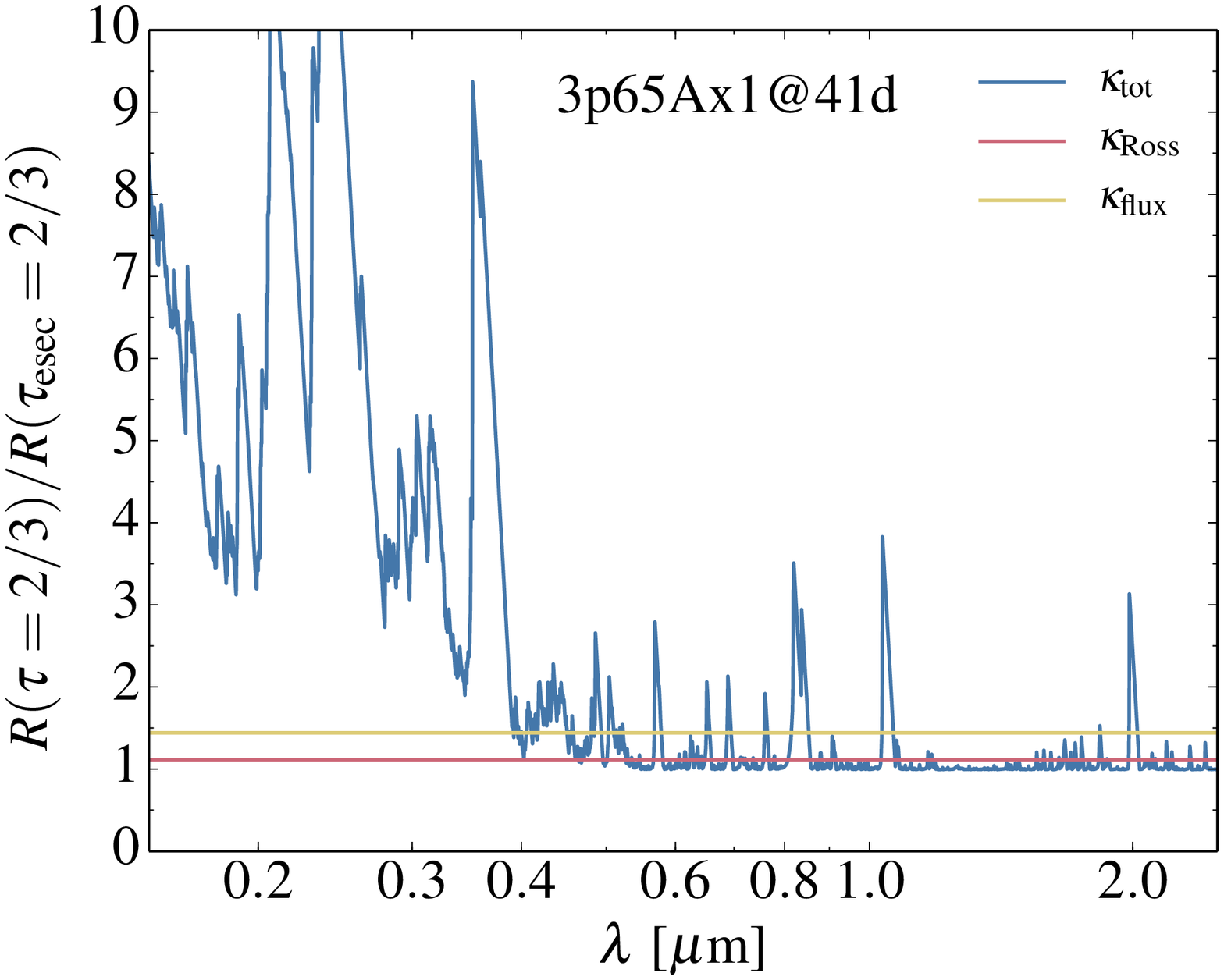,width=8.5cm}
\caption{Wavelength variation of the photospheric radius $R(\tau=2/3)$, normalised to the
radius of the electron-scattering photosphere $R(\tau_{\rm esec}=2/3)$, when all opacity sources are included,
and when we use the Rosseland-mean opacity $\kappa_{\rm Ross}$ or the flux-mean opacity
$\kappa_{\rm flux}$.
From top to bottom, we show the results for model 3p65Ax1 at 10, 21, and 41\,d after explosion.
The distinct locations traced by each curve emphasise the ambiguity
of the notion of a photosphere in these Type IIb/Ib/Ic SNe.
\label{fig_rtau}
}
\end{figure}

For stellar work, the photosphere is usually defined to be the location at which the Rosseland-mean optical depth is 2/3,
a definition which follows from the Eddington-Barbier relation for a plane-parallel atmosphere
\citep[e.g.,][]{mihalas_78}. For extended atmospheres a more practical relation might be the atmospheric
location where the probability of photon escape is 50\%, or alternatively, that location beyond which 50\% of the observed flux
originates. Unfortunately, because the opacity has a strong wavelength-dependence, and because of the
rapidly expanding ejecta, the photosphere is not uniquely defined,
and even at a single wavelength photons will escape over a region which is spatially extended.
The strong effect of line blanketing in the UV increases the effective photospheric
radius for bluer photons, while redder photons in the optical, decouple from the gas at a radius
comparable to the Rosseland-mean or flux-mean photosphere (Fig.~\ref{fig_rtau}).
Still, even within the optical,  this decoupling radius can vary by a factor of 2--3.

\begin{figure}
\epsfig{file=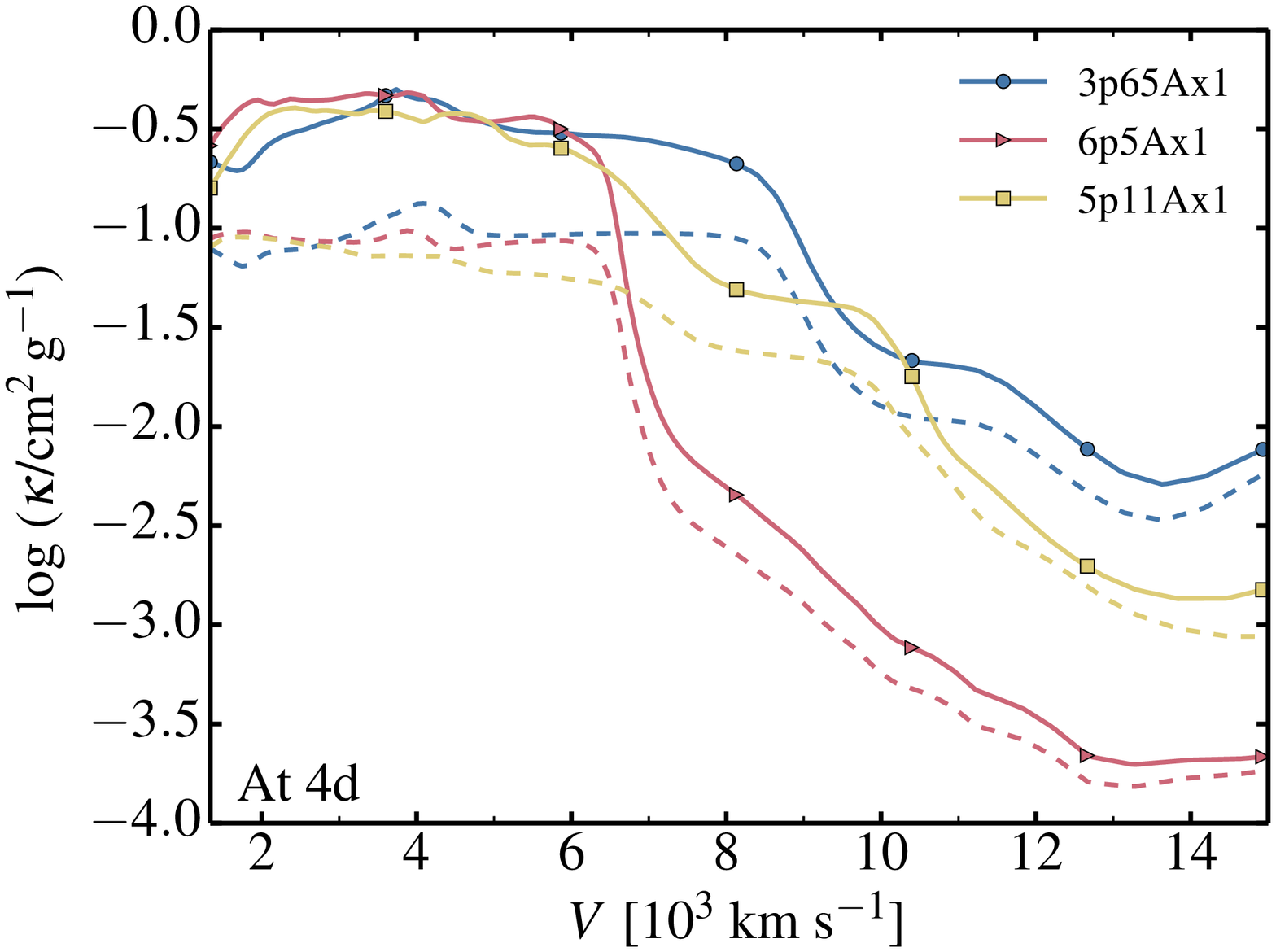,width=8.5cm}
\epsfig{file=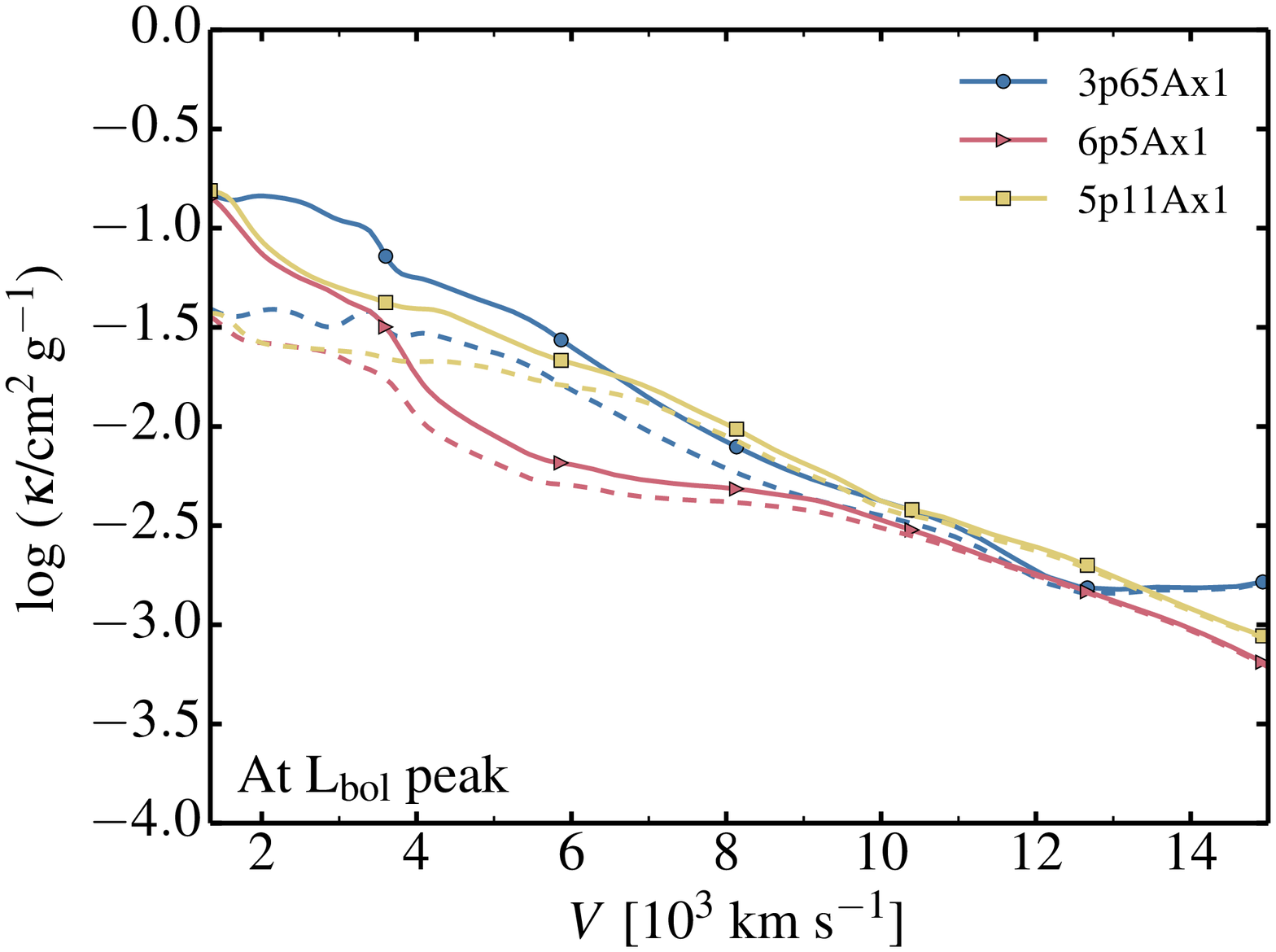,width=8.5cm}
\caption{Variation of the mass absorption coefficient associated with the Rosseland-mean opacity $\kappa_{\rm Ross}$ (solid)
and the electron-scattering opacity $\kappa_{\rm esec}$ (dashed) in the ejecta of models 3p65Ax1 (SN IIb),
6p5Ax1 (SN Ib), and 5p11Ax1 (SN Ic)
at $\approx$\,4\,d after explosion (top) and at bolometric maximum (bottom).
The same ordinate range is used in each panel.
All opacities show a large variation with ejecta location, reflecting variations both
in composition and ionisation (see also Fig~\ref{fig_ionfrac}). The range in mass absorption
coefficient is significantly greater than for Type Ia SNe, which possess a more uniform composition
(see, e.g., Fig.~12 in \citealt{DH15}).
\label{fig_kappa}
}
\end{figure}

In type II-Plateau SNe, the continuum opacity dominates at early times throughout most of the optical,
and even at late times, there are fairly clean ``continuum" regions devoid of line opacity and emissivity,
for example, between H$\alpha$ and the Ca\two\ triplet at 8500\,\AA\ \citep{DH05b}.
In type IIb/Ib/Ic, the large abundance of He (and its low ionisation in the photospheric layers)
and intermediate mass elements reduces the
importance of continuum opacity sources and  favours the role of
lines (Fig.~\ref{fig_kappa}). Electron scattering is sub-dominant and may represent only a third
of the total Rosseland-mean opacity, even at bolometric maximum.

\begin{figure*}
\epsfig{file=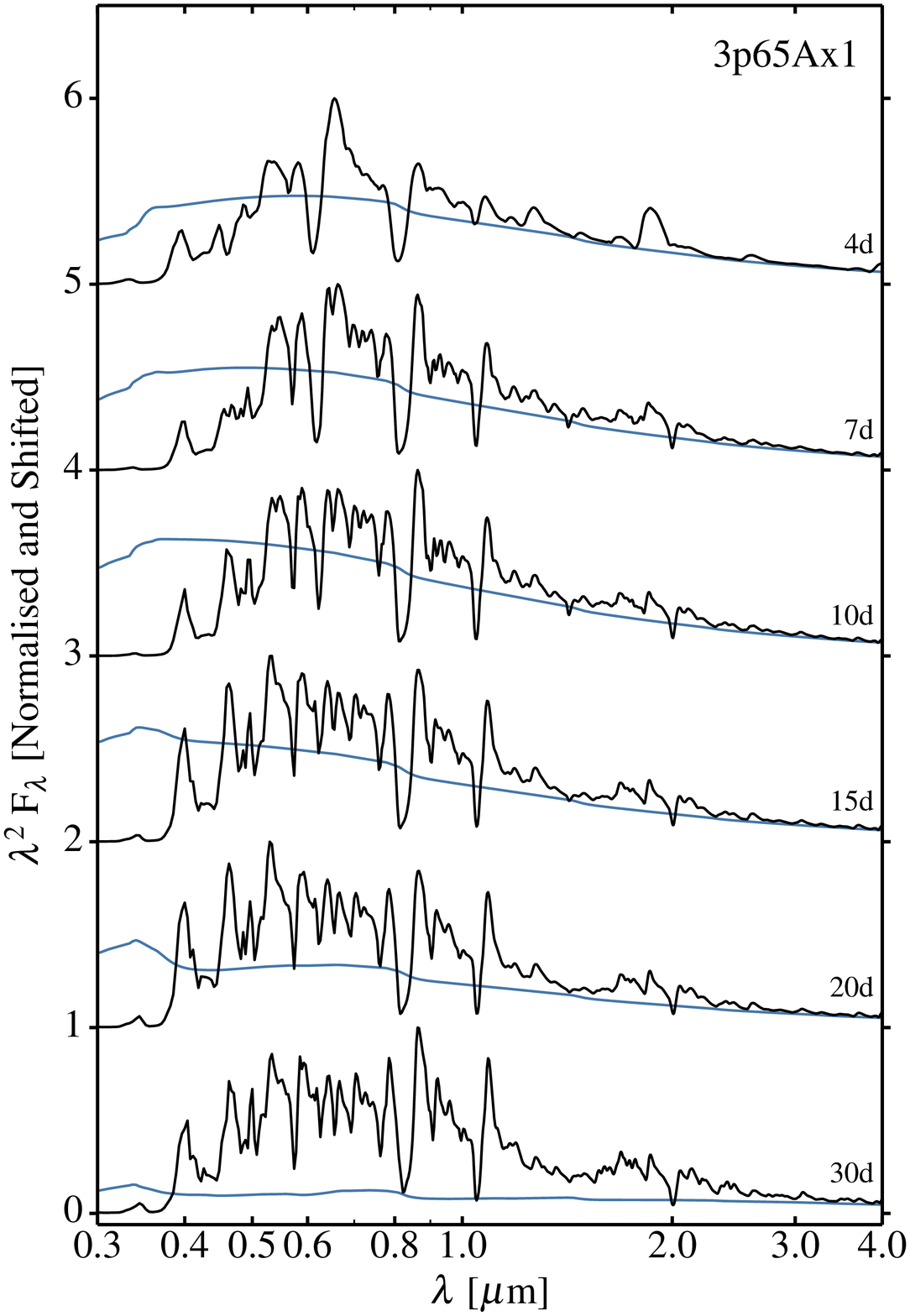,width=5.5cm}
\epsfig{file=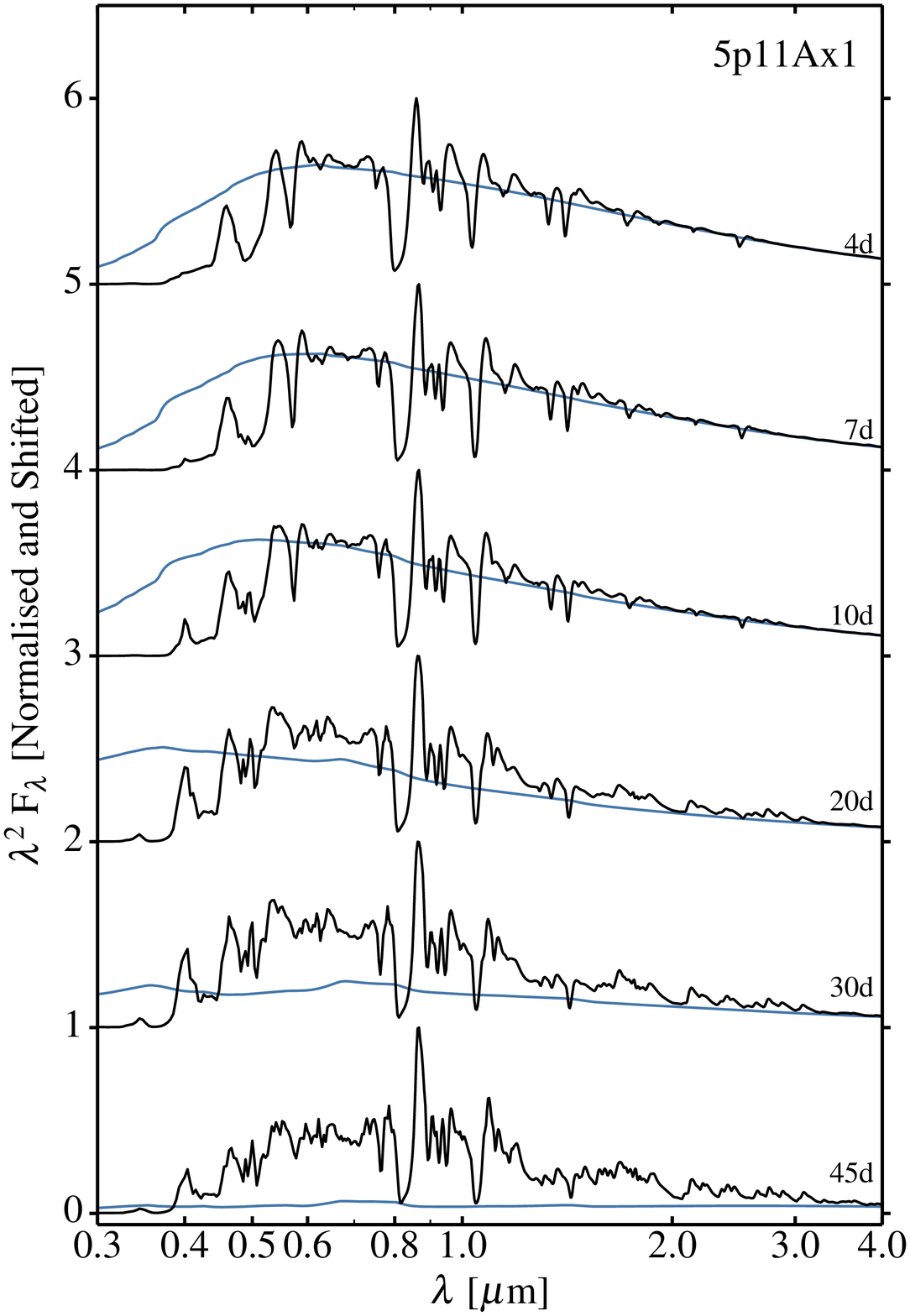,width=5.5cm}
\epsfig{file=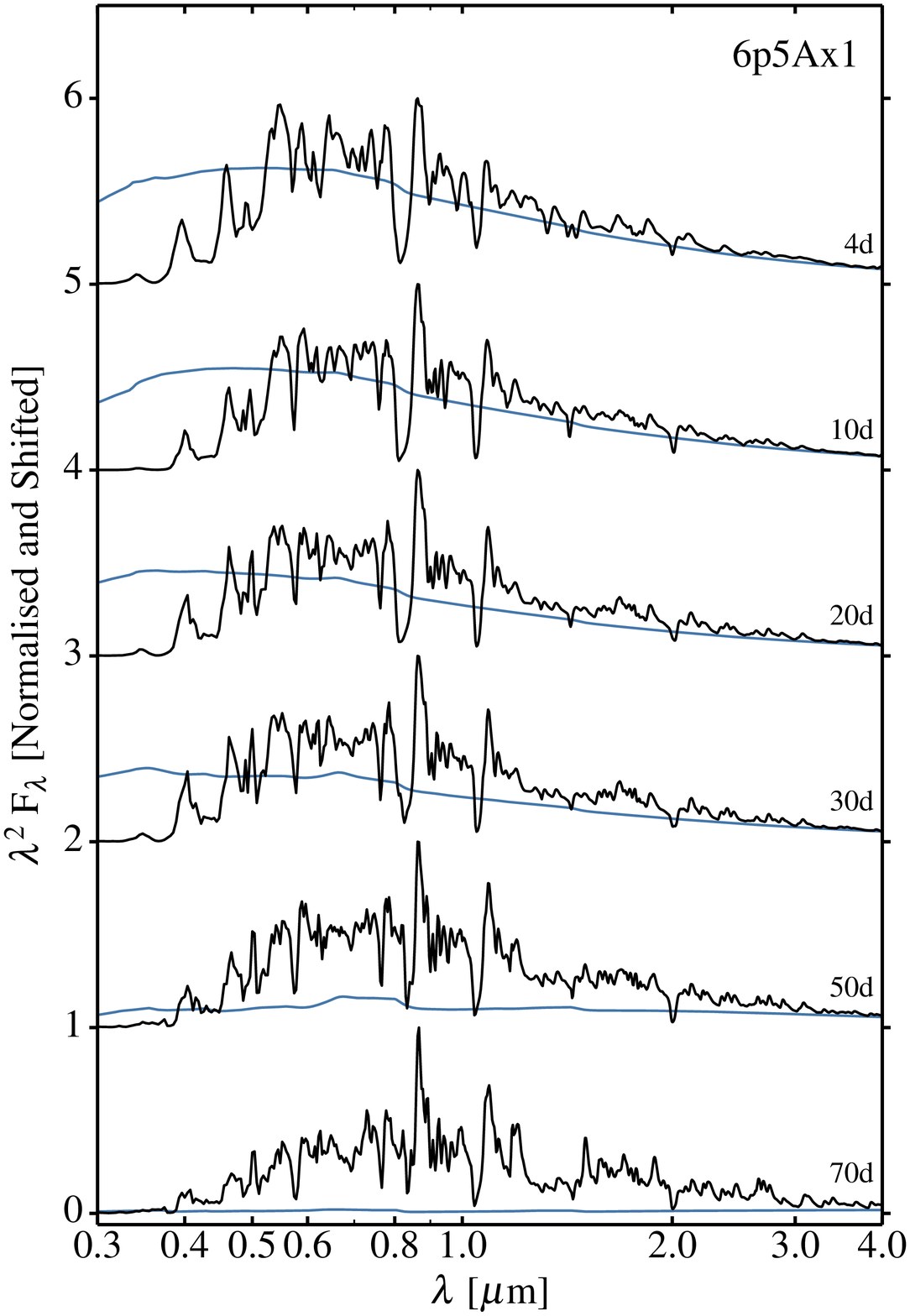,width=5.5cm}
\caption{Evolution of the total flux (black) and the continuum flux (blue) in the optical and near-IR
for the models 3p65Ax1, 5p11Ax1, and 6p5Ax1 during the photospheric phase.
For better visibility of the near-IR, we show the quantity $\lambda^2 F_{\lambda}$.
For the continuum flux calculation, only bound-free and free-free processes are included in
the formal solution of the radiative-transfer equation.
Note the growing separation between the total flux and the continuum flux as time progresses, caused by the increasing role
played by lines over pure continuum processes. At many epochs no pure continuum region
can be seen. In the optical, lines act through both emission and absorption, and in a very tangled way,
while at early times in the UV lines cause a flux deficit through a blanketing effect.
\label{fig_spec_cont}
}
\end{figure*}

As time progresses and the ejecta turns nebular, the influence of lines grows. In our simulations,
this causes the steep decline of the continuum flux below the representative flux level of the model,
which only shows a quasi-continuum (Fig.~\ref{fig_spec_cont}).
The emergent spectrum even at the peak is a collection of overlapping lines, causing simultaneous
absorption, scattering, and emission.
P-Cygni profile formation is altered from the general conception since it is not continuum photons
that interact with the line but instead the background radiation arising from overlapping lines.

\section{Inference of the ejecta expansion rate}
\label{sect_vm}

One important characteristics of SN ejecta is their kinetic energy. It represents
the left-over explosion energy after unbinding the progenitor envelope.
To infer the ejecta kinetic energy, we need to infer its expansion rate. It is customary
in the community to do this by constraining the location of the photosphere.

However, it is clear from the preceding section that photons subject to continuum opacity
will decouple much deeper in the ejecta than photons overlapping with a strong
line, such as H$\alpha$ in a SN IIb at early time. To different lines correspond a whole range
of intermediate heights and legitimate but distinct photospheres.

\begin{figure}
\epsfig{file=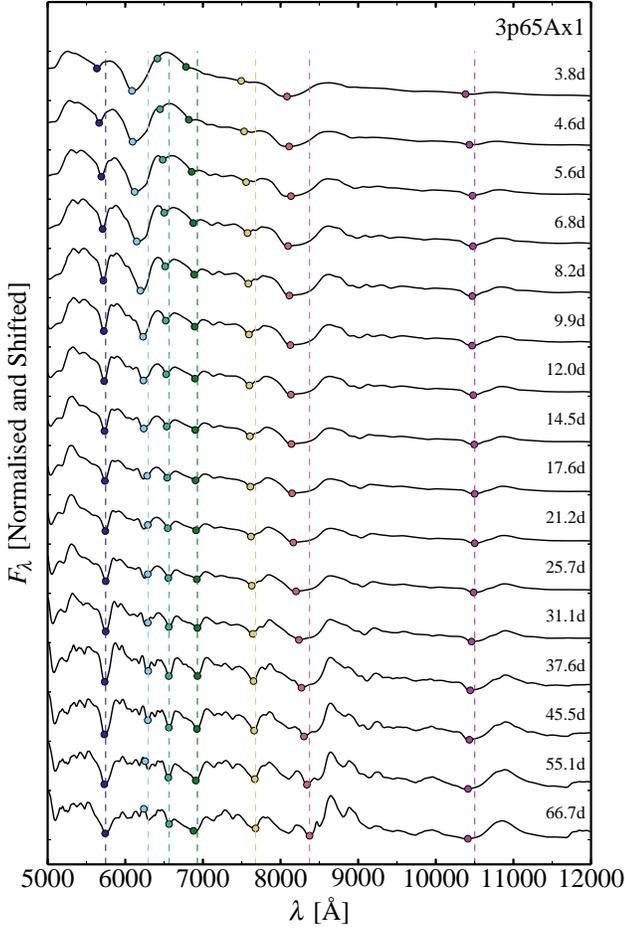,width=8.5cm}
\caption{Spectral sequence for model 3p65Ax1 during the photospheric phase.
We overlay the location at maximum absorption in
H$\alpha$, in the He\one\ lines at 5875.7\AA,  6678.1\,\AA, 7065.2\,\AA, and 10830.2\,\AA,
in O\one\,7773.0\,\AA,   and in the Ca\two\, triplet at 8569.0\,\AA.
To better reveal the evolving velocity shifts for each transition, we also draw a vertical line
(dashed and coloured)  through the maximum absorption closest to the rest wavelength.
With most features, the location of maximum absorption shifts to smaller velocities
with time. However for He\,\one\,10830\,\AA\ this is only true initially -- at late times
the maximum absorption shifts to larger velocities due to the influence
$\gamma$-rays and non-thermal effects in the outer ejecta.
\label{fig_vabs_spec}}
\end{figure}

Determining the location of these photospheres is nonetheless an interesting exercise. Using
the location of maximum absorption, one can constrain the chemical stratification
of an ejecta, bearing in mind that a line profile is also subject to multiple processes
and uncertainties (e.g., ionisation, details of atomic physics, blending). In Fig.~\ref{fig_vabs_spec},
we show a spectral sequence in the optical for model 3p65Ax1. Using synthetic
spectra for individual species, we can unambiguously determine the location of
maximum absorption in lines of sufficient strength and not overlapping with other
lines of the same ion. We show the results (one red mark for each line and each epoch)
for H$\alpha$, He\one\ lines at 5875.7\AA,  6678.1\,\AA, 7065.2\,\AA, and 10830.2\,\AA,
O\one\,7773.0\,\AA,   and the Ca\two\, triplet
at 8569.0\,\AA.
We omit the lines from Fe\two\ or Ti\two\ in the blue part of the optical since their
contributions overlap --- we see broad absorptions with internal dips which prevent
the unambiguous identification of a specific line.

The guidance of the model helps for the velocity measurement on the weak optical He\one\ lines
at early times, or the vanishingly small line strength of H$\alpha$ at and beyond bolometric
maximum. It helps also with ambiguous dips. The depression at 8000\,\AA\ in the spectrum
at 60.6\,d is not caused by Ca\two\ absorption at high velocity but corresponds to the blue edge of its emission.
It also allows to establish with certainty that the progressive migration of the 1.05\,$\mu$m absorption
to larger velocity is associated with the broadening of the He\one\,10830\,\AA\ line, following the
growth of the $\gamma$-ray mean free path and the non-thermal effects on He\one\ in the outer
ejecta.

Repeating the process for the other two models, we illustrate the trajectory of the velocity
at maximum absorption for representative lines in Fig.~\ref{fig_vabs_spec_traj}
(we use the $gf$-weighted mean wavelength for multiplets).
In the type IIb model, $v_{\rm abs}$(H$\alpha$) converges to $\approx$\,13000\,\kms,
which indeed corresponds to the base of the H-rich shell in the ejecta (Table~\ref{tab_ejecta_glob}).
Some sudden variations can occur when line profiles have double dips of comparable
strength or when the absorption is an extended trough. The jumps at early times may
also arise when we switch on the treatment of non-thermal processes.
However, overall, the evolution of all lines is to exhibit a maximum absorption that recedes
in velocity space as we approach the time of bolometric maximum.

After maximum, non-local energy deposition and non-thermal processes alter the
line opacity in the outer ejecta and lead to the broadening of the absorption troughs.
The effect is strong in He\one\,10830\,\AA, whose maximum absorption occurs
at 12000\,\kms\ at late times. The chemical stratification matters here since He is systematically
more abundant in the outer ejecta of these models --- intermediate mass elements are less likely
to exhibit this behaviour because their abundance peaks in the inner ejecta.

While the photosphere, and its associated velocity, are poorly defined, the representative  ejecta velocity
$v_m \equiv \sqrt{2E_{\rm kin}/M_{\rm e}}$ is a fundamental
characteristic of the ejecta and an important parameter of analytical models for SN light curves.
In our three simulations, we find that around bolometric maximum,
this quantity is matched to within $\approx$\,10\% by  $v_{\rm abs}$(He\one\,5875\,\AA)
in model 3p65Ax1 ($v_m=$\,7476\,\kms), or $v_{\rm abs}$(O\one\,7773\AA) in models 6p5Ax1
($v_m=$\,5047\,\kms) and 5p11Ax1  ($v_m=$\, 5959\,\kms).
 This correspondence is useful to constrain the ejecta kinetic energy and mass and
drops the additional step, with its associated ambiguities, to locate ``the'' photosphere.
In Paper\two, we will study how the quantity $v_m$ correlates with various lines in the full
grid of models. This exploration also needs to be extended to progenitors with large surface radii.

\section{Correction factors}
\label{sect_corfac}

   Comparing a SN spectrum to a blackbody can give some information
on the colour temperature. In type II SNe, the comparison is meaningful at
early post-explosion times because there are extended continuum regions
where the flux resembles a blackbody, merely ``diluted'' through  the effect
of scattering in the atmosphere \citep{mihalas_78,E96,DH05b}.

One must apply a correction to this blackbody spectrum in order to obtain the
total luminosity and to infer the distance to a type II SN.
Unfortunately, in practice, one is led not so much to correct for the effect
of scattering but instead for the effect of lines (see discussion in \citealt{DH05b}).
With type IIb/Ib/Ic SNe, it is also possible to produce such correction factors,
but the correction arises primarily from the effect of line emission/absorption.
The second problem, as discussed above, is that there is no unambiguous
photosphere, since it depends much on the opacity considered.

\begin{figure}
\epsfig{file=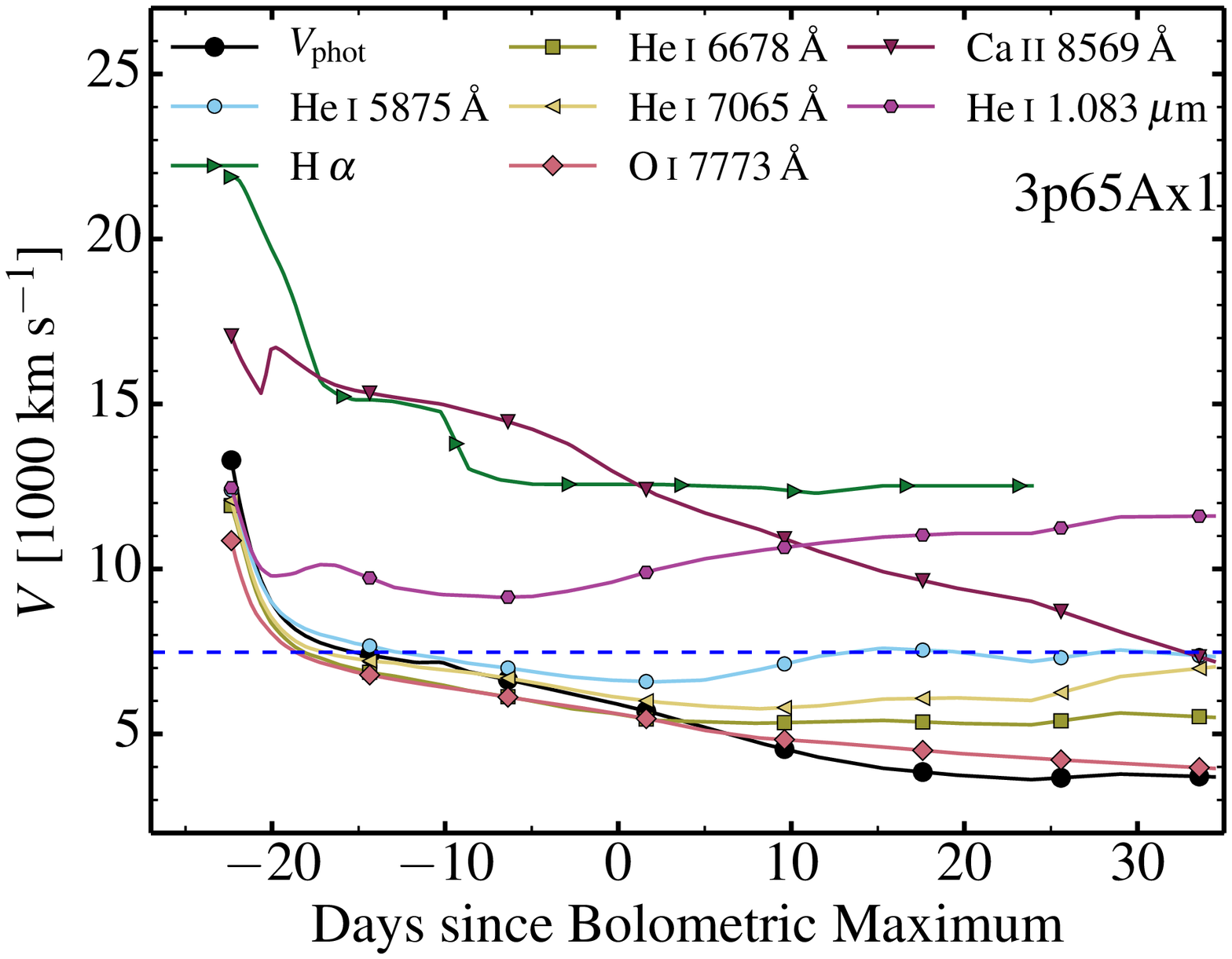,width=8.5cm}
\epsfig{file=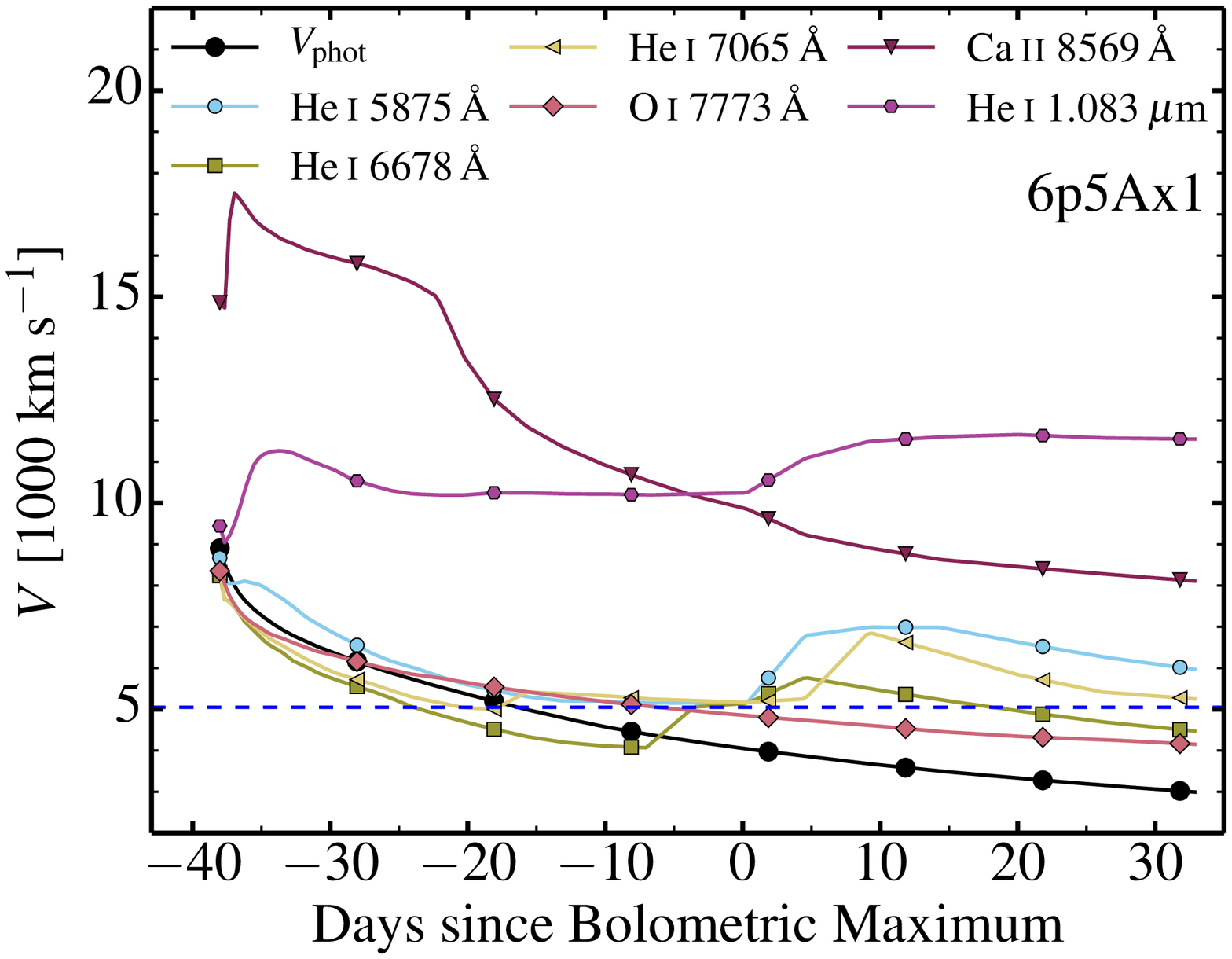,width=8.5cm}
\epsfig{file=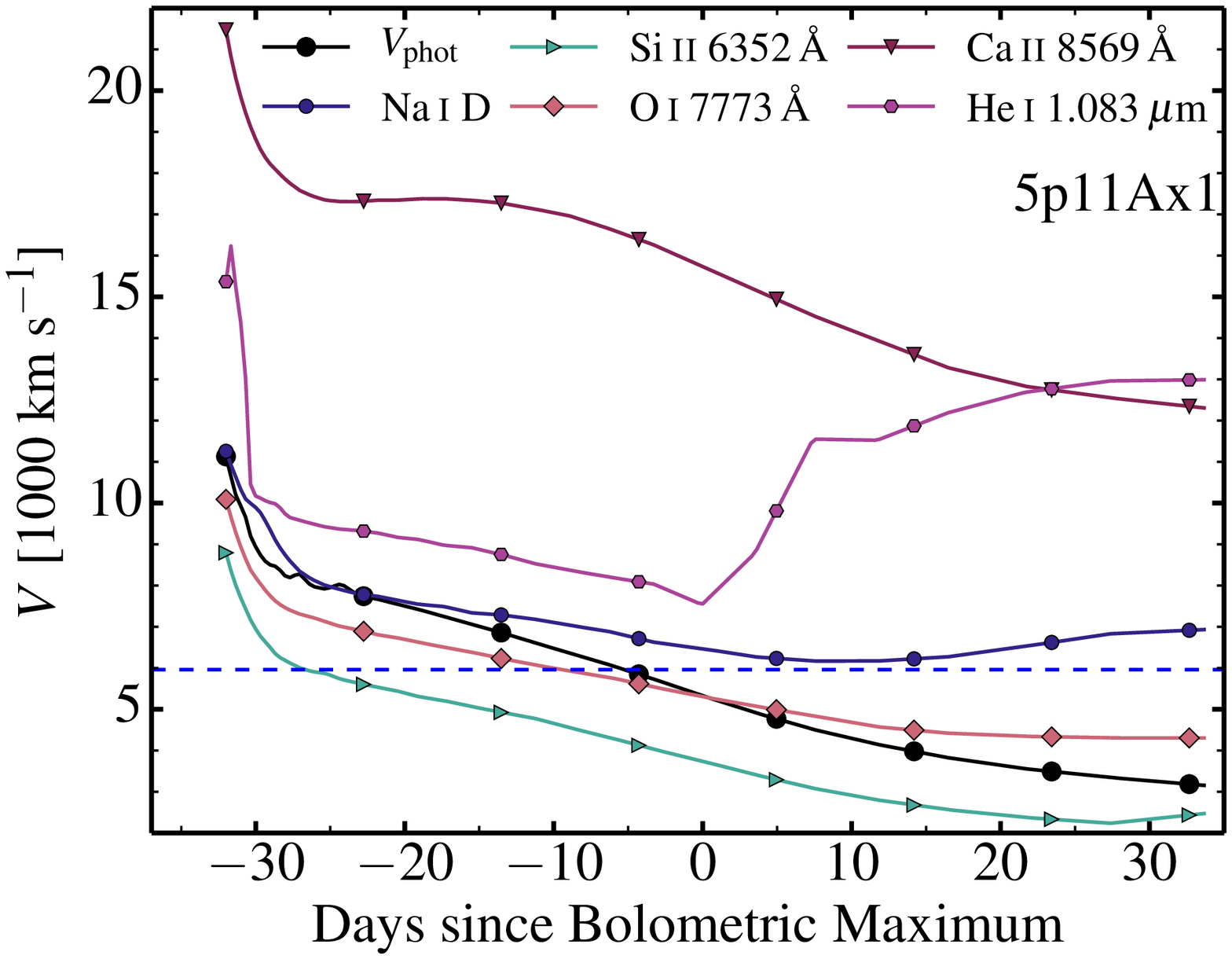,width=8.5cm}
\caption{
Evolution of the velocity at maximum absorption in representative lines for
the three models 3p65Ax1 (SN IIb), 6p5Ax1 (SN Ib), and 5p11Ax1 (SN Ic).
We use the $gf$-weighted mean wavelength for multiplets.
We also add the photospheric velocity (for which we employ the Rosseland-mean opacity).
The dashed line corresponds to the representative ejecta velocity given by $\sqrt{2E_{\rm kin}/M_{\rm e}}$.
\label{fig_vabs_spec_traj}
}
\end{figure}

Nonetheless, we have repeated the calculations of \citet{DH05b} (this study was
based on steady state models; correction factors for time-dependent  simulations
of type II SNe was discussed in \citealt{DH08}) on the present set of models.
We follow the procedure described in Section~2 of \citet{DH05b}, and specifically
compute the correction factors through a minimisation of their Equation~9.
Since electron scattering is not a good representation of the total opacity, we employ
the Rosseland mean instead and focus on times when the ejecta Rosseland-mean
optical depth is $\gtrsim$\,10 (i.e., the photospheric phase). Velocities associated
with the photosphere, and how to infer them, can be gleaned from Fig.~\ref{fig_vabs_spec_traj}.

\begin{figure*}
\epsfig{file=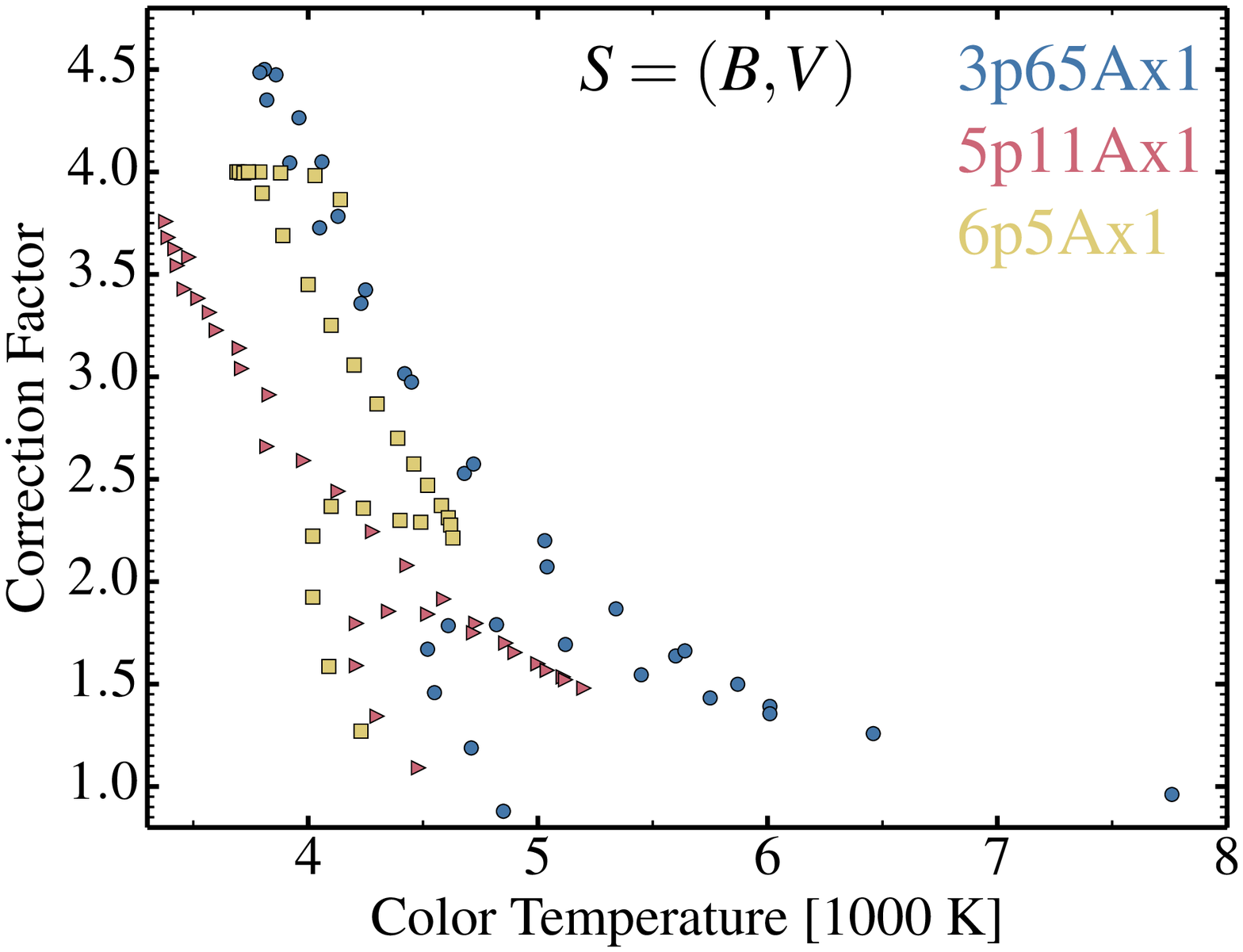,width=5.8cm}
\epsfig{file=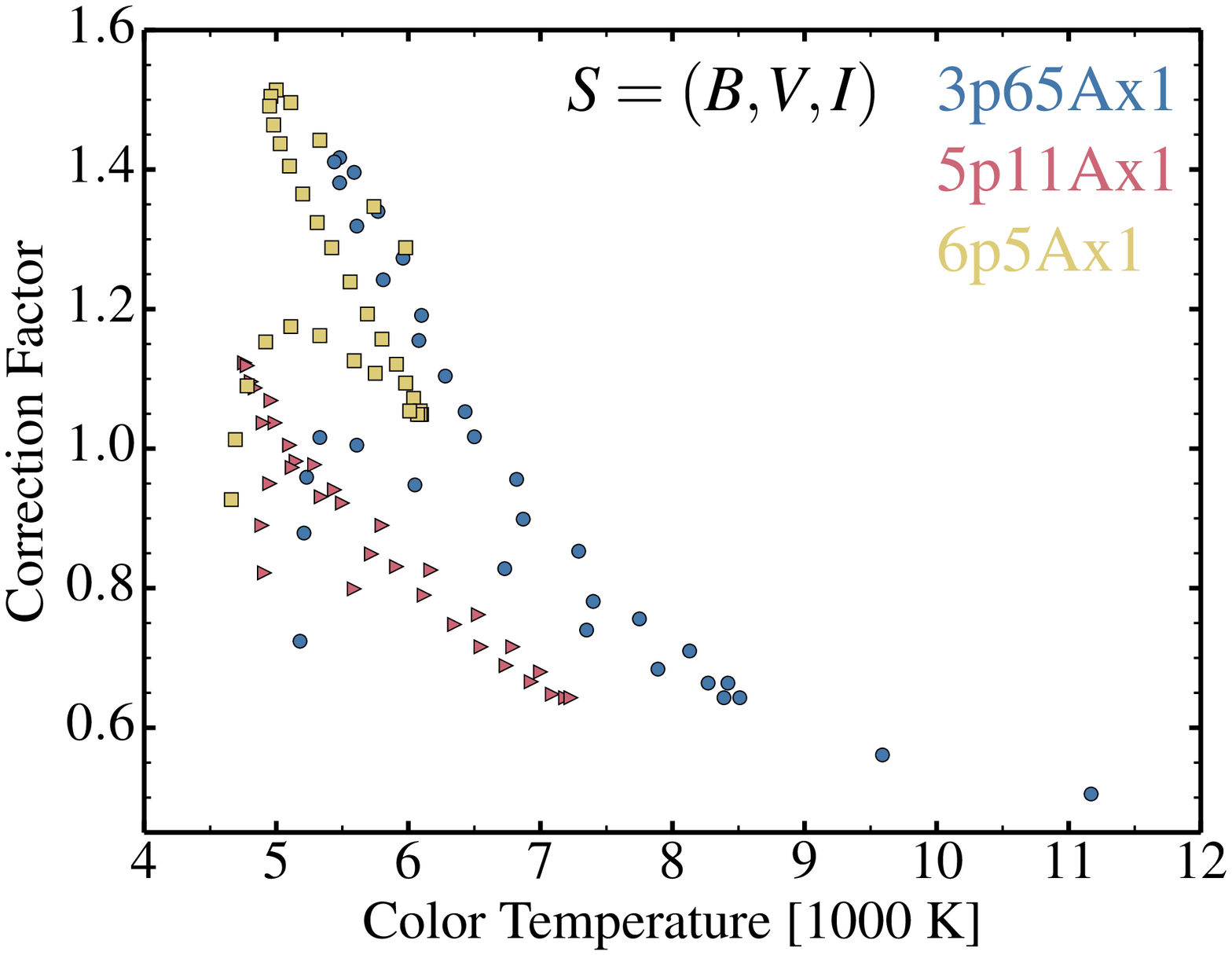,width=5.8cm}
\epsfig{file=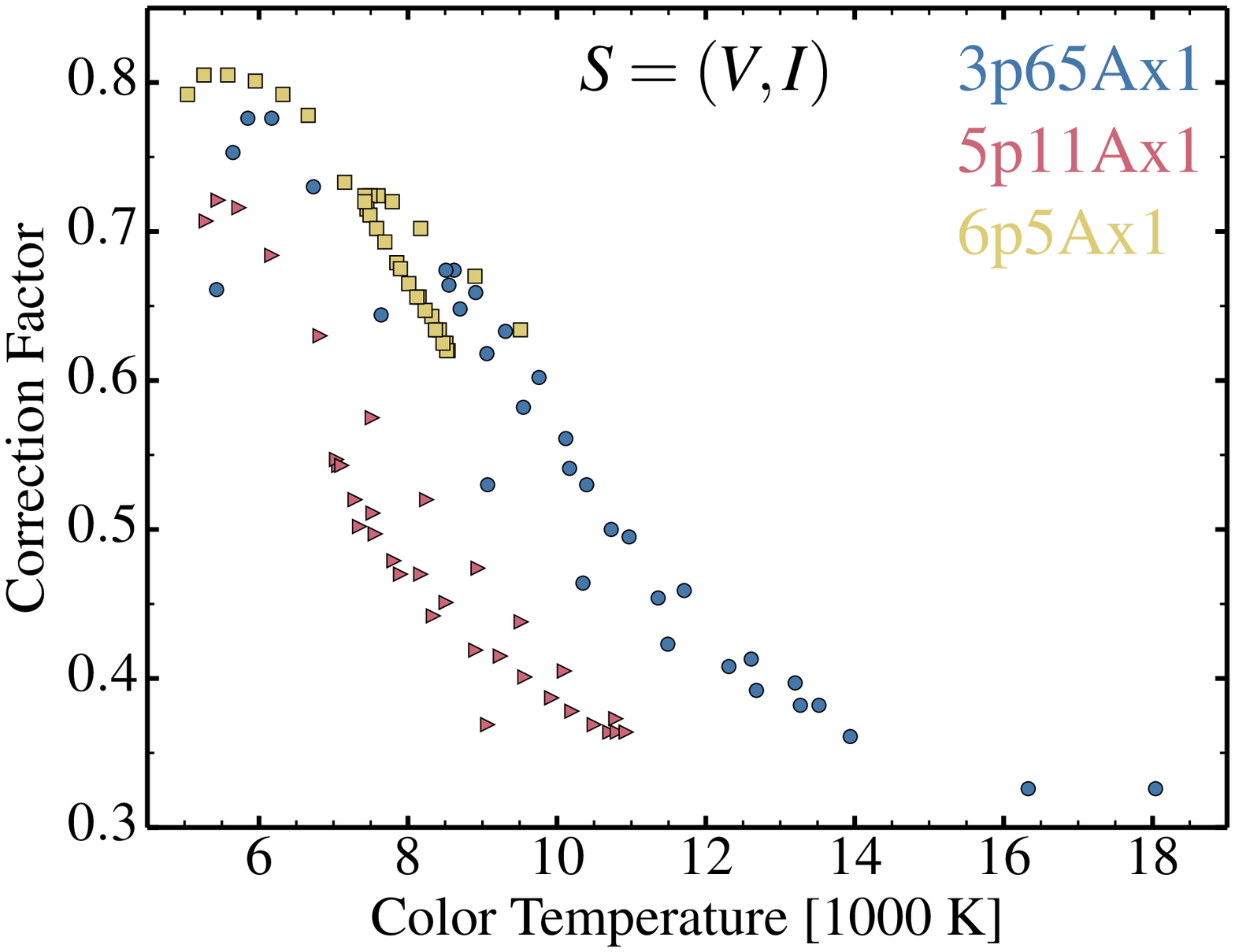,width=5.8cm}
\epsfig{file=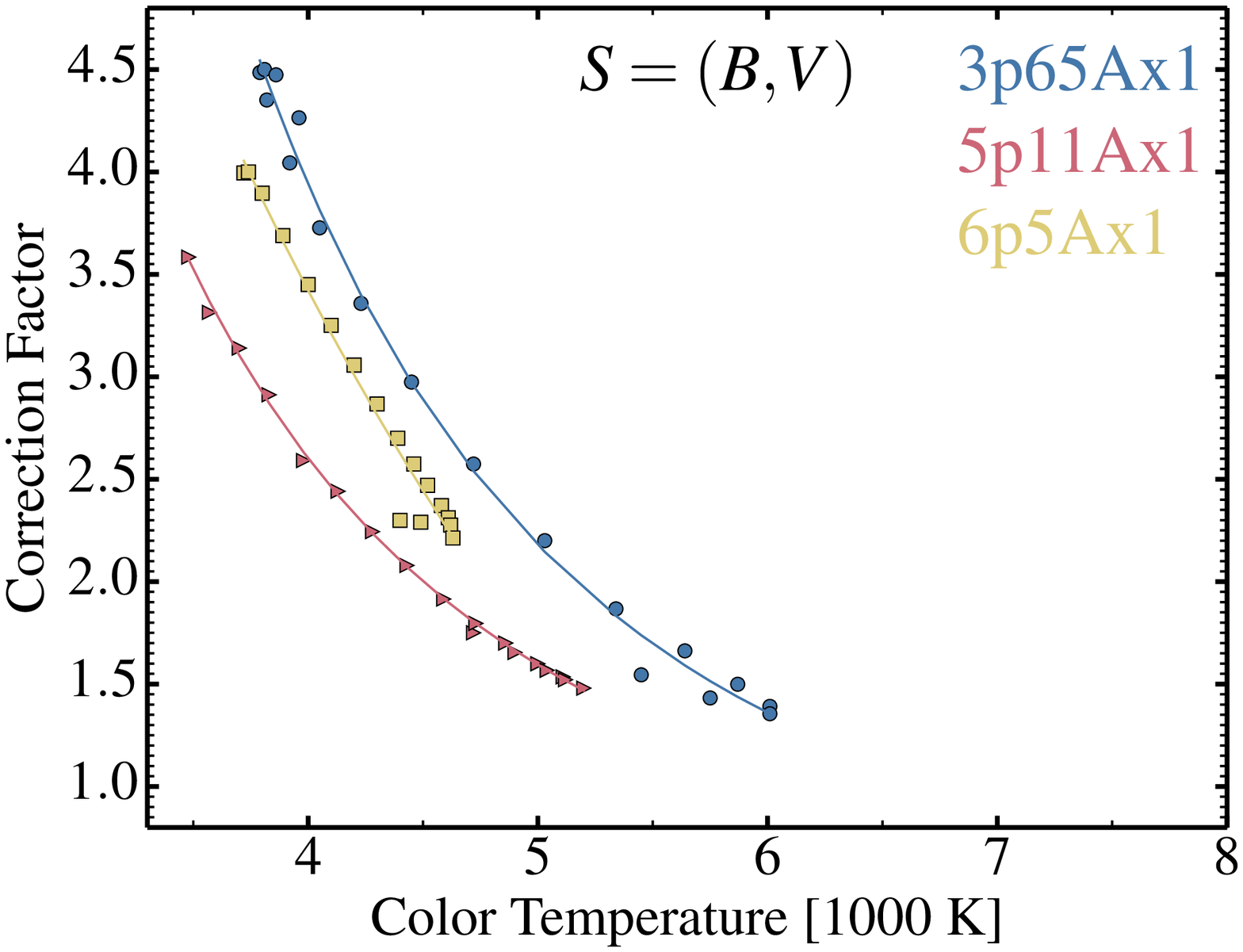,width=5.8cm}
\epsfig{file=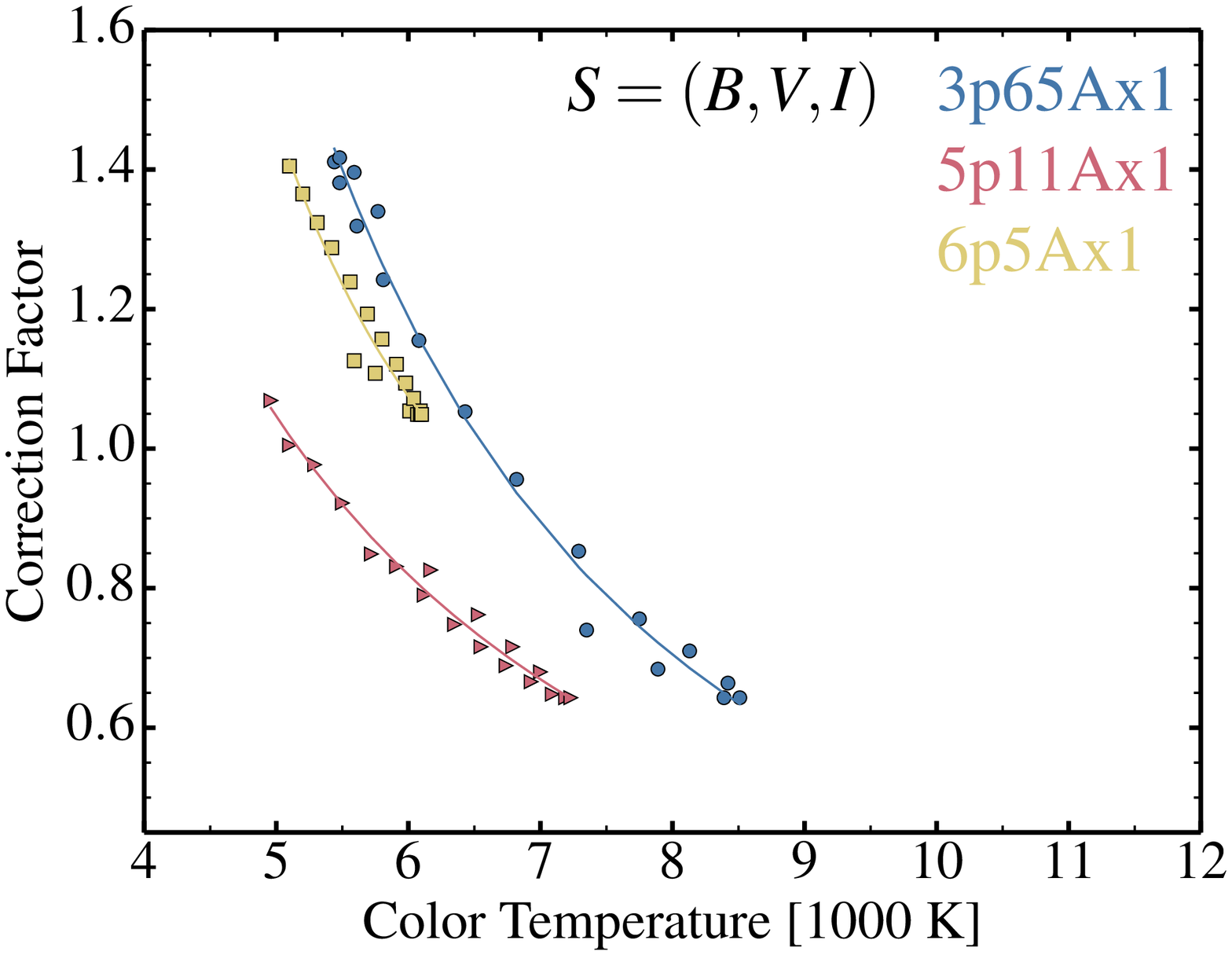,width=5.8cm}
\epsfig{file=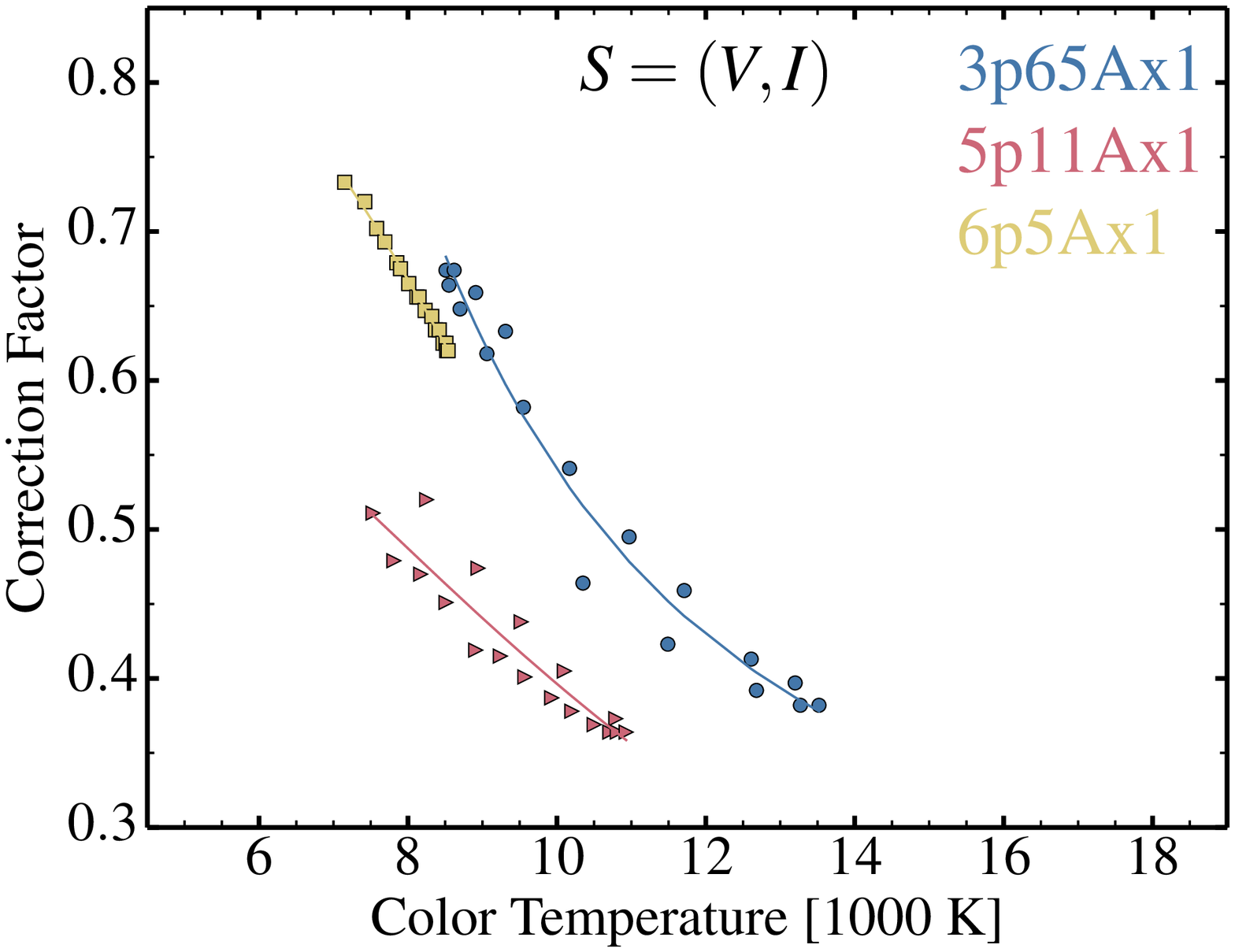,width=5.8cm}
\caption{{\it Top row:}
Distribution of correction factors for models 3p65Ax1, 5p11Ax1, and 6p5Ax1 for optical
band combinations $S=(B,V)$ (left), $(B,V,I)$ (middle), $(VI)$ (right), and based on photospheric-phase epochs.
Each set $S$ must contain at least two bands to constrain the colour temperature.
{\it Bottom row:}
Same as top, but now limiting the epochs to the range [$t_{\rm max}$/5,1.1$\times t_{\rm max}$], where $t_{\rm max}$
is the time of bolometric maximum.
Also plotted is a fit to the distribution of correction factors using a second-order polynomial.
The coefficients of the fitted polynomial are given in Table~\ref{tab_corfac_fit}.
The correction factors and color temperatures shown here are tabulated in the appendix.
\label{fig_dist_corfac}
}
\end{figure*}

Correction factors show a complex behaviour (Fig.~\ref{fig_dist_corfac}).
Unlike type II SNe during the photospheric phase, which show a monotonic evolution towards
cooler temperature at fixed photospheric composition \citep{DH11},
type IIb/Ib/Ic SNe exhibit a non-monotonic colour evolution
(Fig.~\ref{fig_photometry_3p65ax1}--\ref{fig_photometry_6p5ax1})
with a photospheric composition that becomes progressively metal dominated.
This is the reason for the bifurcation in the distribution of correction factors
(compare the top row panels with the bottom row counterparts).

When we limit the time span to times around bolometric maximum, the distribution
of points shows more uniformity (bottom row of Fig.~\ref{fig_dist_corfac}). For
convenience we have fitted the correction factors using a second-order polynomial in the inferred
colour temperature. The polynomial coefficients are provided in Table~\ref{tab_corfac_fit}.
We also note that the correction factors are systematically offset between models, likely
arising from the distinct composition in the atmospheres of each model.

\begin{figure*}
\epsfig{file=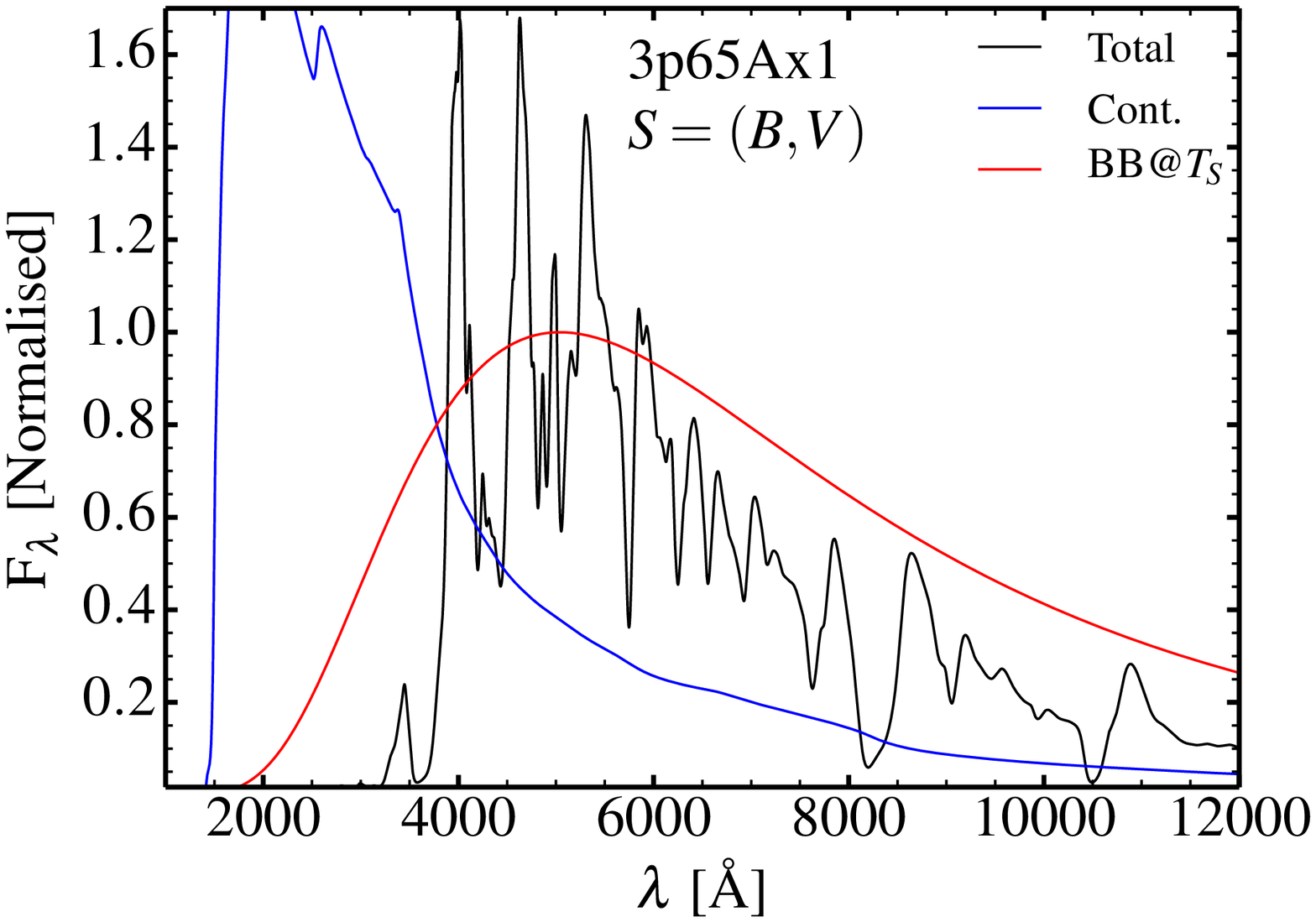,width=5.8cm}
\epsfig{file=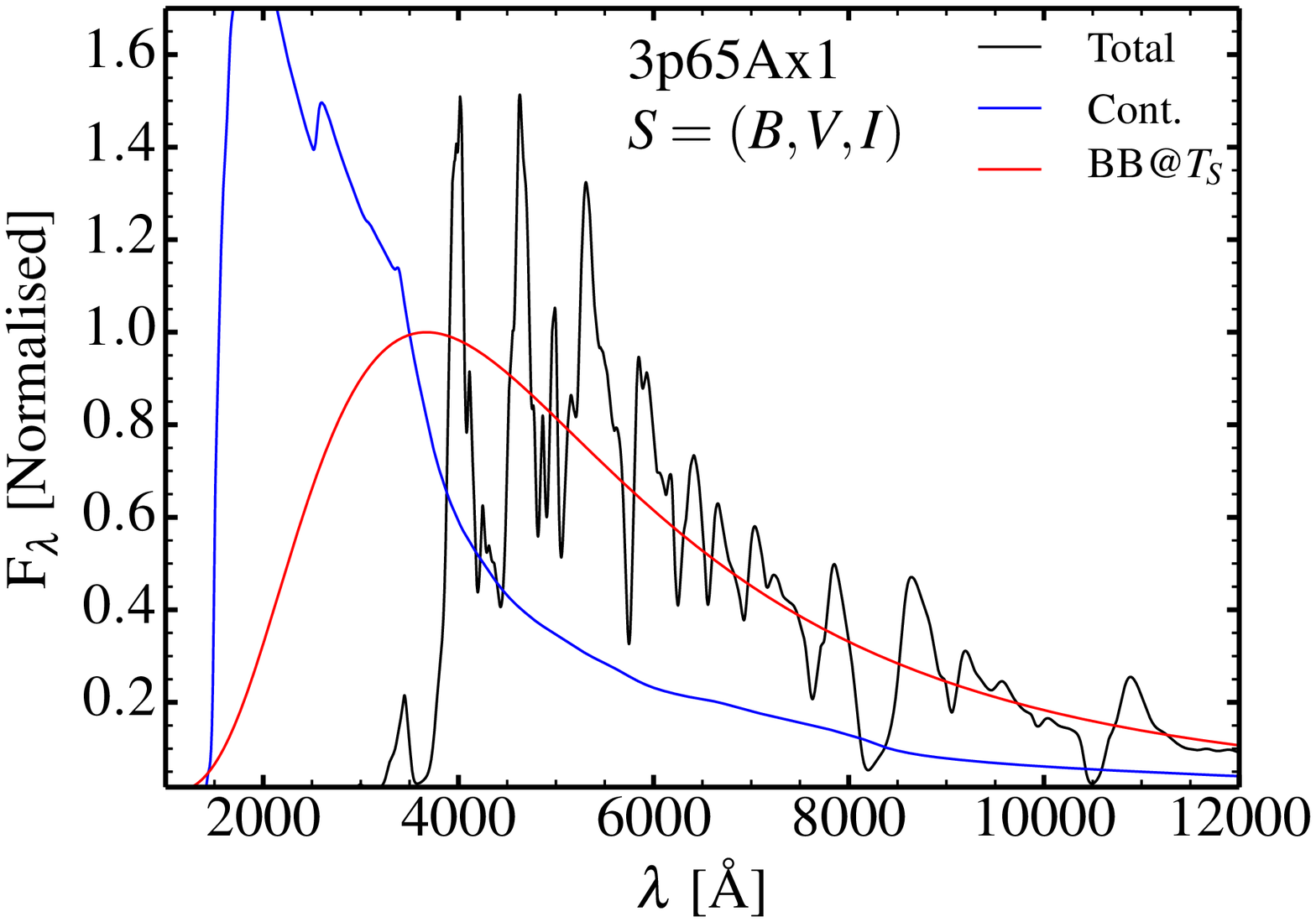,width=5.8cm}
\epsfig{file=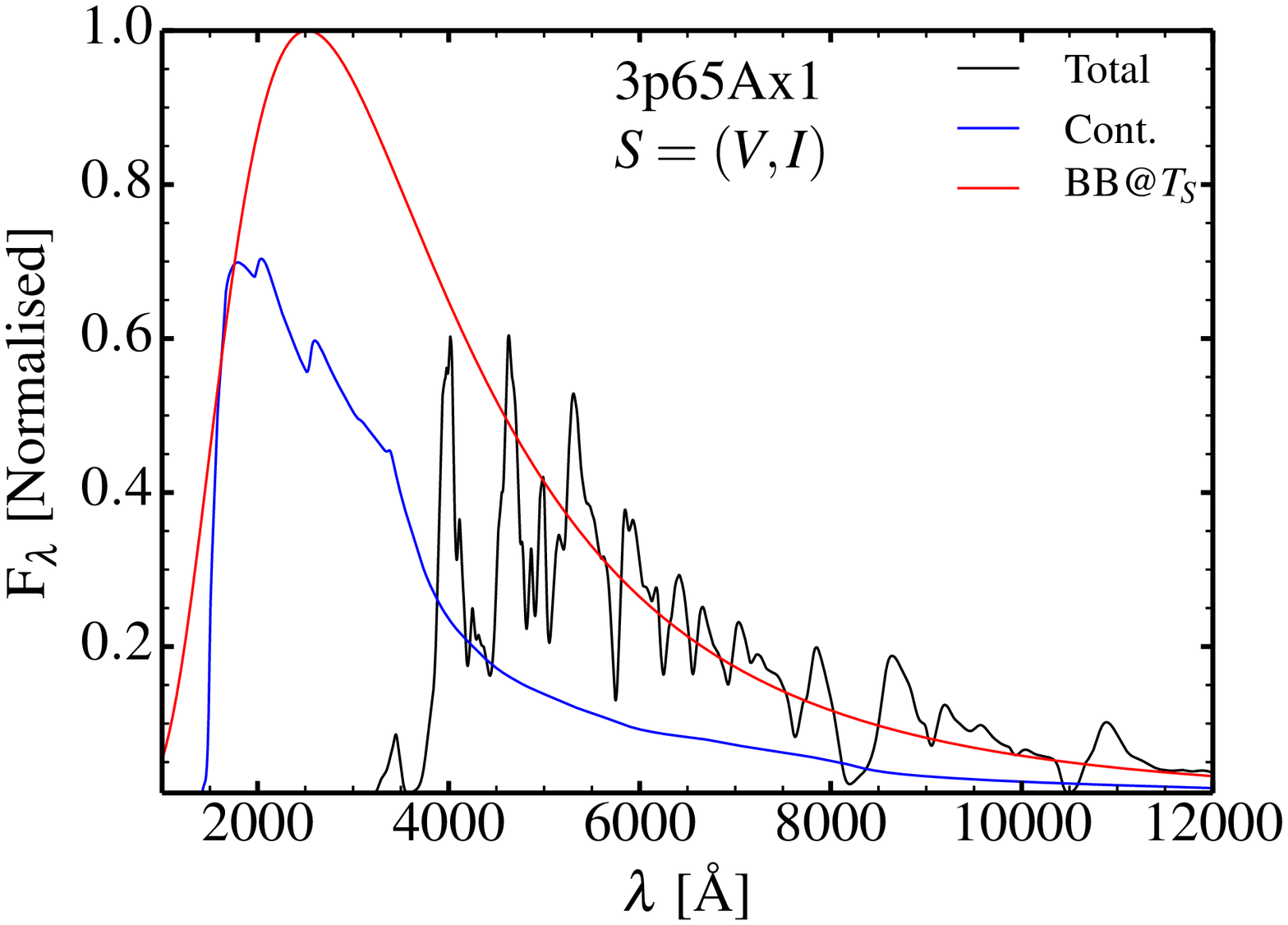,width=5.8cm}
\caption{Comparison, near the time of bolometric maximum, of the total and continuum fluxes for model 3p65Ax1
with the blackbody flux computed with the associated correction factor and colour temperature $T_S$
(see Table~\ref{tab_corfac_365Ax1} in the appendix). From left to right, we show the results
for the sets $S=(B,V)$, $(B,V,I)$, and $(V,I)$.
\label{fig_bb_fits}}
\end{figure*}

These correction factors have to be used with circumspection. In Fig~\ref{fig_bb_fits}, we show the blackbody fits to
the different optical bands for model 3p65Ax1 around bolometric maximum.
The present context is far from the original notion of flux dilution in scattering atmospheres. The correction is instead used to
recover from the questionable approximation of the model flux as a blackbody.

\section{Conclusions}
\label{sect_conc}

  This paper is the first of a series describing a set of radiative transfer simulations for SNe IIb, Ib, and Ic from
  binary star progenitors.  In this paper we focused our attention on three models that were selected for their distinct composition.
  Model 3p65A has a sizeable He-rich shell and some residual H in the outermost ejecta layers. It produces H\one\ lines
  early on, and He\one\ lines at all times, making it a Type IIb SN. Although more massive, the ejecta associated
  with model 6p5A  is hydrogen deficient and has 1.65\,\msun\ of He ($\approx$\,35\% by mass). It produces He\one\ lines and is of Type Ib.
  Model 5p11A stems from a typical WC star composition, with a mixture of He, C, and O in the outermost ejecta.
  Even with strong mixing, it shows no He\one\ line in the optical and is thus of Type Ic.

  Despite these differences in composition, which produce different SN types, the qualitative properties of the
  bolometric light curves, multi-band light curves, and colours are comparable for all models. Having the same
  ejecta kinetic energy, the differences stem primarily from the different ejecta masses. As expected, the higher
  the ejecta mass, the longer the rise time, the broader the light curve, and the slower the post-maximum decline
  rate. These properties stem from the greater trapping efficiency of low-energy photons (which in number make the bulk
  of the ejecta internal energy) and of high-energy photons (which originate from radioactive decay).
  About 80\% of the total flux falls within the optical range, with the rest in the near-IR --- little flux comes out in the UV
  except at very late times. At early times this holds because our progenitor models have small radii --- it is not a fundamental property
  of all SNe IIb/Ib/Ic.

  Quantitatively, Arnett's rule does not produce a very accurate prediction of  the \iso{56}Ni mass -- applied to our models
  it leads to an overestimate of the  \iso{56}Ni mass by about 50\%.
  In contrast, the method of \citet{katz_13_56ni} works very accurately with our models. Using an expression based
  on a solution of the energy equation, we can recover the original \iso{56}Ni to within 5\% by integrating the bolometric
  luminosity over time starting at 3\,d after explosion (this offset can be reduced to $\lesssim$2\% if we take into
  account the trapped radiation energy at 3\,d). The challenge with this method is the need to monitor the multi-band brightness
  from very early times (the later we start, the more inaccurate is the method), while a better accuracy requires the knowledge
  of the trapped radiation energy at the start of the integration.

  As  observed, we find that our models exhibit an earlier light-curve peak for bluer bands. Despite the uncertainties
  in mixing and ejecta mass/composition, the early post-maximum colours are comparable. The ($V-R$) colour shows a good
  uniformity with a value of 0.32\,mag (standard deviation of 0.04\,mag).
  The effect of mixing is to enhance the brightness and make the SN bluer at early times. This leads to a flatter slope
  of the bolometric light curve as it approaches maximum. The effect of mixing on the multi-band light curves
 is only of minor importance after maximum.

  We emphasise the spectral differences between our
  three models. We find that H$\alpha$ is eventually a good tracer of the velocity of the outer H-rich shell
  in a SN IIb. We also reproduce the non-monotonic behaviour of He\one\ lines with time, which tend to become narrower
  from early time to bolometric maximum, before becoming broader again after maximum. This property is a vivid illustration
  of the combined effects of non-local (decay) energy deposition and non-thermal processes. This effect is primarily
  seen in He\one\ lines, which are the most sensitive to non-thermal processes in these SNe.

  While He is present in all our models, it can be difficult to identify in spectra. Due to blending, the 1.1\,$\mu$m region
  must be used with caution when trying to determine whether the ejecta contains He. In our type Ic model
  lines due to C\one\ and Mg\two\ are very important contributors to the P~Cygni profile seen near 1.09\,$\mu$m,
  while in the type IIb model Mg\two\ is an important contributor.
  Other lines, such as He\one\ 2.0581\,$\mu$m, can be used to confirm the presence of He\one\ in spectra, 
although  the absence of this line does not preclude the presence of $\lambda 10830$.\footnote{
 The lower levels of the $\lambda$10830 transition (1s 2s $^3S$) and of the $\lambda$20581 
transition (1s 2s $^1S$) are metastable but the former has a longer lifetime (the 1s 2s $^1S$ level can decay 
by two-photon emission). 
Hence, the associated line optical depth can be significantly higher
  for the $\lambda$10830 transition than for the $\lambda$20581 transition.}
  Paradoxically, in both our Ib and Ic models, He\one\,10830\,\AA\ is strongest at late times ($\approx$\,70\,d).
  Another method of distinguishing the type of
  progenitor is to use C\one\ lines. We obtain weak or no C\one\ lines in the  9000--11000\,\AA\ region prior to
  maximum in our SNe IIb/Ib models, which have a sizeable He-rich shell, while the SN Ic model,
  with its outer layers of He, C, and O, shows strong C\one\ lines in this region.

  Because a mass buffer rich in carbon and oxygen separates the helium-rich shell from the
  original site of \iso{56}Ni, the production of He\one\ lines in our models requires mixing. Without
  mixing, none of our models are of type Ib (or IIb), but would instead be of type Ic (or IIc) --- see \citet{dessart_etal_12}
  for discussion. Furthermore, varying the mixing strength from moderate to strong (models x1 and x2)
  did not alter the SN type of these models. As done for observations, we discuss the SN type of each of our models
  based on its spectroscopic properties. However, it is clear from the characteristic width of their light curves
  that models 5p11A and 6p5A
  have a too low $E_{\rm kin}/M_{\rm e}$ to match SN Ibc observations (see, for example, \citealt{drout_11_ibc}).

  As for SN Ia,  a true continuum is only seen at very early times. After the spectrum formation zone recedes
  into the metal rich layers of the ejecta,
  metal line opacity and emissivity completely dwarf continuum processes. Line overlap then conspires to produce a quasi continuum.
  Consequently, the mass absorption coefficient is greater than that due to electron scattering, and varies with ionisation --- it
  is not a constant in time or in depth.
  Furthermore, the ubiquity of lines in the spectrum formation region makes the notion of a well-defined photosphere obsolete.

  Despite the complexities of spectrum formation, useful inferences about the ejecta can still be made. For example, the location of
  maximum absorption in various lines (e.g., H$\alpha$ or He\one\ lines) can serve as a diagnostic of chemical stratification.
  We also discussed how the mean ejecta expansion rate, defined by $v_m = \sqrt{2E_{\rm kin}/M_{\rm e}}$ correlates
  with the P~Cygni profile absorption velocity for various lines around the time of maximum.
  With a measure of both the mean expansion rate, and the ejecta mass,  the energy of the SN explosion may be inferred.

  Finally, we computed the correction factors to invoke when approximating the synthetic spectra with a blackbody.
  Because SN colours are affected by line opacity and emissivity, the correction depends on the selected band and
  on the adequacy of the model for a particular observation. We find that our models compare well to observations
  around the time of maximum (see Paper\three) so the use of these correction factors should be limited to such epochs.

\begin{table*}
\begin{center}
    \caption{Coefficients of the second-order polynomial used to fit the distribution
of correction factors for different optical-band combinations and for models
3p65Ax1, 5p11Ax1, and 6p5Ax1 over the time span [$t_{\rm max}$/5, 1.1$\times t_{\rm max}$ ].
The polynomial is of the form
$\xi(T) = \sum_{i=0}^2 a_i (1/T_4)^i$, where $T_4$ is the temperature in
units of 10$^4$\,K.
\label{tab_corfac_fit}}
\begin{tabular}{
l|
r@{\hspace{3mm}}r@{\hspace{3mm}}r@{\hspace{3mm}}|
r@{\hspace{3mm}}r@{\hspace{3mm}}r@{\hspace{3mm}}|
r@{\hspace{3mm}}r@{\hspace{3mm}}r@{\hspace{3mm}}|
}
\hline
Coef. & \multicolumn{3}{c}{3p65Ax1} & \multicolumn{3}{c}{5p11Ax1} &  \multicolumn{3}{c}{6p5Ax1} \\
\hline
          &      $(B,V)$   &   $(B,V,I)$       &              $(V,I)$    &    $(B,V)$ &  $(B,V,I)$ & $(V,I)$ &  $(B,V)$  & $(B,V,I)$  &  $(V,I)$   \\
\hline
$a_0$  & 1.5399   & 0.1629   &  0.2439 &  0.8836    &       $-$0.0089   &  $-$0.2657 &   $-$7.2109   &  1.3178          &     $-$0.9906   \\
$a_1$  &  $-$2.2443 & $-$0.1098 & $-$0.1454   & $-$0.9726   & 	0.3429      & 0.9011  &  5.0819     &   $-$1.2462 	&	 2.1061    \\
$a_2$  & 1.2820   & 0.4347   &  0.4417   &  0.6640   & 	0.0924  &     $-$0.2389  & $-$0.3316  &    0.6604 	&	 $-$0.6237 \\
\hline
\end{tabular}
\end{center}
\end{table*}

\section*{acknowledgments}

LD acknowledges financial support from the European Community through an
International Re-integration Grant, under grant number PIRG04-GA-2008-239184,
and from ``Agence Nationale de la Recherche" grant ANR-2011-Blanc-SIMI-5-6-007-01.
DJH acknowledges support from STScI theory grant HST-AR-12640.01, and NASA theory grant NNX14AB41G.
SW was supported by NASA (NNX14AH34G) and the DOE High Energy Physics
Program (DE-SC-00010676).
S.-C. Y was supported by the Basic Science Research (2013R1A1A2061842) program through
the National Research Foundation of Korea (NRF).
This work was granted access to the HPC resources of CINES under the
allocation c2013046608 made by GENCI (Grand Equipement
National de Calcul Intensif).

\appendix

\section{Correction factors for the three reference models}

  In this section, we tabulate the correction factors for each of the three models during
the photospheric phase. These SNe IIb/Ib/Ic models correspond to explosions
of stars from relatively modest progenitor radii, and thus with intrinsic colours that are
relatively red prior to peak.

  In the future, we will study explosions of WR stars with a more extended radius
at death. For these, we expect bluer colours prior to maximum, and consequently
different correction factors.

\begin{table*}
\begin{center}
    \caption{Corrections factors for the 3p65Ax1 model sequence covering the photospheric phase.
See Section~\ref{sect_corfac} for discussion, as well as \citet{DH05b} for a
description of the tabulated quantities and a presentation of the method used to calculate them.
\label{tab_corfac_365Ax1}}
\begin{tabular}{
l@{\hspace{3mm}}
c@{\hspace{3mm}}c@{\hspace{3mm}}c@{\hspace{3mm}}
c@{\hspace{3mm}}c@{\hspace{3mm}}c@{\hspace{3mm}}
c@{\hspace{3mm}}c@{\hspace{3mm}}c@{\hspace{3mm}}
}
\hline
Age & $T_{BV}$ &  $\xi_{BV}$   &   $T_{BVI}$     &   $\xi_{BVI}$   &   $T_{VI}$     & $\xi_{VI}$
& \tphot\     &     \rphot\     &     \vphot\    \\
\hline
[day] &    [K]      &  &    [K]      &  &    [K]      &   & [K] & [cm]  & [\kms]   \\
\hline
%      AGE  &     T_BV        xi_BV        T_BVI       xi_BVI         T_VI        xi_VI            TPHOT        RPHOT      VPHOT
     2.360  &   7760.0   &     0.961 &   11170.0   &     0.505 &   18040.0  &      0.326  &      7426.022 &   3.863(14)  & 18943.610   \\
     2.600  &   6460.0   &     1.258 &    9590.0   &     0.561 &   16330.0  &      0.326  &       6990.782 &   4.064(14)  & 18093.070   \\
     2.860  &   5600.0   &     1.637 &    8270.0   &     0.664 &   13940.0  &      0.361  &       6638.097 &   4.216(14)  & 17060.560   \\
     3.150  &   5040.0   &     2.072 &    7400.0   &     0.781 &   12310.0  &      0.408  &       6493.325 &   4.328(14)  & 15901.340   \\
     3.470  &   4680.0   &     2.528 &    6870.0   &     0.899 &   11360.0  &      0.454  &       6570.802 &   4.373(14)  & 14586.0   \\
     3.820  &   4420.0   &     3.015 &    6500.0   &     1.017 &   10730.0  &      0.500  &       6711.424 &   4.388(14)  & 13295.980   \\
     4.200  &   4250.0   &     3.424 &    6280.0   &     1.104 &   10400.0  &      0.530  &       6837.968 &   4.441(14)  & 12236.960   \\
     4.620  &   4130.0   &     3.783 &    6100.0   &     1.191 &   10120.0  &      0.561  &       6948.506 &   4.496(14)  & 11263.120   \\
     5.080  &   4060.0   &     4.049 &    5960.0   &     1.273 &    9760.0  &      0.602  &       7072.313 &   4.535(14)  & 10331.840   \\
     5.590  &   3960.0   &     4.264 &    5770.0   &     1.340 &    9310.0  &      0.633  &       7014.261 &   4.625(14)  & 9575.254   \\
     6.150  &   3860.0   &     4.474 &    5590.0   &     1.396 &    8910.0  &      0.659  &       6944.160 &   4.765(14)  & 8967.916   \\
     6.760  &   3810.0   &     4.500 &    5480.0   &     1.417 &    8620.0  &      0.674  &       6863.244 &   4.981(14)  & 8527.549   \\
     7.440  &   3790.0   &     4.485 &    5440.0   &     1.411 &    8510.0  &      0.674  &       6801.329 &   5.268(14)  & 8194.578   \\
     8.180  &   3820.0   &     4.351 &    5480.0   &     1.381 &    8550.0  &      0.664  &       6780.774 &   5.615(14)  & 7944.819   \\
     9.0    &   3920.0   &     4.044 &    5610.0   &     1.319 &    8700.0  &      0.648  &       6804.780 &   6.029(14)  & 7753.137   \\
     9.900  &   4050.0   &     3.727 &    5810.0   &     1.242 &    9060.0  &      0.618  &       6865.623 &   6.505(14)  & 7604.410   \\
    10.890  &   4230.0   &     3.358 &    6080.0   &     1.155 &    9550.0  &      0.582  &       6956.454 &   7.035(14)  & 7477.017   \\
    11.980  &   4450.0   &     2.974 &    6430.0   &     1.053 &   10170.0  &      0.541  &       7055.816 &   7.629(14)  & 7370.751   \\
    13.180  &   4720.0   &     2.574 &    6820.0   &     0.956 &   10970.0  &      0.495  &       7154.051 &   8.282(14)  & 7273.257   \\
    14.500  &   5030.0   &     2.200 &    7290.0   &     0.853 &   11710.0  &      0.459  &       7241.006 &   8.974(14)  & 7163.545   \\
    15.950  &   5340.0   &     1.867 &    7750.0   &     0.756 &   12610.0  &      0.413  &       7128.274 &   9.891(15)  & 7177.419   \\
    17.550  &   5640.0   &     1.662 &    8130.0   &     0.710 &   13200.0  &      0.397  &       7332.842 &   1.044(15)  & 6882.652   \\
    19.310  &   5870.0   &     1.499 &    8420.0   &     0.664 &   13520.0  &      0.382  &       7314.003 &   1.118(15)  & 6699.158   \\
    21.240  &   6010.0   &     1.391 &    8510.0   &     0.643 &   13270.0  &      0.382  &       7248.240 &   1.189(15)  & 6478.925   \\
    23.360  &   6010.0   &     1.355 &    8390.0   &     0.643 &   12680.0  &      0.392  &       7136.445 &   1.256(15)  & 6222.907   \\
    25.700  &   5750.0   &     1.432 &    7890.0   &     0.684 &   11490.0  &      0.423  &       6985.835 &   1.323(15)  & 5959.559   \\
    28.270  &   5450.0   &     1.545 &    7350.0   &     0.740 &   10350.0  &      0.464  &       6795.695 &   1.376(15)  & 5631.556   \\
    31.100  &   5120.0   &     1.693 &    6730.0   &     0.828 &    9070.0  &      0.530  &       6550.404 &   1.400(15)  & 5208.424   \\
\hline
\end{tabular}
\end{center}
\end{table*}

\begin{table*}
\begin{center}
    \caption{Corrections factors for the 5p11Ax1 model sequence covering the photospheric phase.
See Section~\ref{sect_corfac} for discussion, as well as \citet{DH05b} for a
description of the tabulated quantities and a presentation of the method used to calculate them.
\label{tab_corfac_5p11Ax1}}
\begin{tabular}{
l@{\hspace{3mm}}
c@{\hspace{3mm}}c@{\hspace{3mm}}c@{\hspace{3mm}}
c@{\hspace{3mm}}c@{\hspace{3mm}}c@{\hspace{3mm}}
c@{\hspace{3mm}}c@{\hspace{3mm}}c@{\hspace{3mm}}
}
\hline
Age & $T_{BV}$ &  $\xi_{BV}$   &   $T_{BVI}$     &   $\xi_{BVI}$   &   $T_{VI}$     & $\xi_{VI}$
& \tphot\     &     \rphot\     &     \vphot\    \\
\hline
[day] &    [K]      &  &    [K]      &  &    [K]      &   & [K] & [cm]  & [\kms]   \\
\hline
%      AGE  &     T_BV        xi_BV        T_BVI       xi_BVI         T_VI        xi_VI            TPHOT        RPHOT      VPHOT
 3.27       &      3820.0     &        2.660      &      5590.0      &       0.799      &      9070.0    &         0.369      &         4895.1     &    3.593(14)    &       12717.7              \\
 3.60       &      3710.0     &        3.040      &      5340.0      &       0.931      &      8340.0    &         0.442      &         5322.9     &    3.462(14)    &       11129.6              \\
 3.96       &      3600.0     &        3.227      &      5150.0      &       0.982      &      7900.0    &         0.470      &         5245.4     &    3.628(14)    &       10603.8              \\
 4.36       &      3520.0     &        3.383      &      4990.0      &       1.037      &      7560.0    &         0.497      &         5267.4     &    3.786(14)    &       10050.8              \\
 4.80       &      3460.0     &        3.428      &      4900.0      &       1.037      &      7350.0    &         0.502      &         5085.7     &    4.072(14)    &        9817.7              \\
 5.28       &      3430.0     &        3.543      &      4810.0      &       1.096      &      7070.0    &         0.543      &         5226.7     &    4.177(14)    &        9155.7              \\
 5.81       &      3390.0     &        3.680      &      4760.0      &       1.123      &      7040.0    &         0.547      &         5218.4     &    4.411(14)    &        8786.7              \\
 6.39       &      3380.0     &        3.758      &      4780.0      &       1.119      &      7110.0    &         0.543      &         5221.0     &    4.703(14)    &        8517.7              \\
 7.03        &     3420.0      &       3.625       &     4840.0       &      1.087       &     7290.0     &        0.520       &        5122.0      &   5.135(14)     &       8453.9               \\
 7.73        &     3480.0      &       3.584       &     4960.0       &      1.069       &     7530.0     &        0.511       &        5262.9      &   5.453(14)     &       8164.8               \\
 8.50        &     3570.0      &       3.314       &     5100.0       &      1.005       &     7810.0     &        0.479       &        5111.8      &   6.092(14)     &       8295.6               \\
 9.35        &     3700.0      &       3.140       &     5290.0       &      0.977       &     8170.0     &        0.470       &        5340.3      &   6.448(14)     &       7981.9               \\
10.29        &     3830.0      &       2.912       &     5500.0       &      0.922       &     8510.0     &        0.451       &        5390.7      &   7.034(14)     &       7911.9               \\
11.32        &     3980.0      &       2.591       &     5720.0       &      0.849       &     8910.0     &        0.419       &        5281.0      &   7.866(14)     &       8042.9               \\
12.45       &      4130.0     &        2.441      &      5910.0      &       0.831      &      9240.0    &         0.415      &         5503.4     &    8.357(14)    &        7769.5              \\
13.70       &      4280.0     &        2.244      &      6120.0      &       0.790      &      9570.0    &         0.401      &         5565.2     &    9.084(14)    &        7674.5              \\
15.07       &      4430.0     &        2.079      &      6350.0      &       0.748      &      9930.0    &         0.387      &         5622.4     &    9.826(14)    &        7546.9              \\
16.58       &      4590.0     &        1.915      &      6550.0      &       0.716      &     10200.0    &         0.378      &         5685.1     &    1.062(15)    &        7414.2              \\
18.24       &      4730.0     &        1.796      &      6740.0      &       0.689      &     10500.0    &         0.369      &         5756.9     &    1.142(15)    &        7247.7              \\
20.06       &      4860.0     &        1.700      &      6930.0      &       0.666      &     10710.0    &         0.364      &         5827.7     &    1.224(15)    &        7062.5              \\
22.07       &      5000.0     &        1.599      &      7090.0      &       0.648      &     10810.0    &         0.364      &         5893.5     &    1.308(15)    &        6859.3              \\
24.28       &      5110.0     &        1.535      &      7190.0      &       0.643      &     10930.0    &         0.364      &         5971.6     &    1.388(15)    &        6614.2              \\
26.71       &      5200.0     &        1.480      &      7230.0      &       0.643      &     10790.0    &         0.373      &         6018.1     &    1.467(15)    &        6358.7              \\
29.38       &      5120.0     &        1.521      &      7000.0      &       0.680      &     10100.0    &         0.405      &         6010.5     &    1.541(15)    &        6072.5              \\
32.32       &      5040.0     &        1.567      &      6790.0      &       0.716      &      9520.0    &         0.438      &         5996.6     &    1.600(15)    &        5729.8              \\
35.55       &      4900.0     &        1.654      &      6530.0      &       0.762      &      8940.0    &         0.474      &         5962.6     &    1.637(15)    &        5330.4              \\
39.10       &      4720.0     &        1.750      &      6170.0      &       0.826      &      8250.0    &         0.520      &         5853.2     &    1.662(15)    &        4918.9              \\
43.01       &      4520.0     &        1.841      &      5800.0      &       0.890      &      7520.0    &         0.575      &         5695.2     &    1.676(15)    &        4510.5              \\
47.31        &     4350.0      &       1.855       &     5440.0       &      0.941       &     6820.0     &        0.630       &        5539.1      &   1.696(15)     &       4148.1               \\
52.04       &      4210.0     &        1.796      &      5120.0      &       0.973      &      6170.0    &         0.684      &         5386.5     &    1.720(15)    &        3826.0              \\
57.24       &      4210.0     &        1.590      &      4950.0      &       0.950      &      5730.0    &         0.716      &         5282.2     &    1.759(15)    &        3556.5              \\
62.96       &      4300.0     &        1.343      &      4890.0      &       0.890      &      5450.0    &         0.721      &         5233.8     &    1.819(15)    &        3344.1              \\
69.26        &     4480.0      &       1.091       &     4910.0       &      0.822       &     5290.0     &        0.707       &        5212.4      &   1.887(15)     &       3153.6               \\
\hline
\end{tabular}
\end{center}
\end{table*}

\begin{table*}
\begin{center}
    \caption{Corrections factors for the 6p5Ax1 model sequence covering the photospheric phase.
See Section~\ref{sect_corfac} for discussion, as well as \citet{DH05b} for a
description of the tabulated quantities and a presentation of the method used to calculate them.
\label{tab_corfac_6p5Ax1}}
\begin{tabular}{
l@{\hspace{3mm}}
c@{\hspace{3mm}}c@{\hspace{3mm}}c@{\hspace{3mm}}
c@{\hspace{3mm}}c@{\hspace{3mm}}c@{\hspace{3mm}}
c@{\hspace{3mm}}c@{\hspace{3mm}}c@{\hspace{3mm}}
}
\hline
Age & $T_{BV}$ &  $\xi_{BV}$   &   $T_{BVI}$     &   $\xi_{BVI}$   &   $T_{VI}$     & $\xi_{VI}$
& \tphot\     &     \rphot\     &     \vphot\    \\
\hline
[day] &    [K]      &  &    [K]      &  &    [K]      &   & [K] & [cm]  & [\kms]   \\
\hline
%      AGE  &     T_BV        xi_BV        T_BVI       xi_BVI         T_VI        xi_VI            TPHOT        RPHOT      VPHOT
  3.89      &      4140.0     &       3.865     &      5980.0     &       1.288     &      9510.0     &       0.634      &        7456.4   &     2.991(14)    &       8899.9       \\
  4.28      &      4030.0     &       3.982     &      5740.0     &       1.347     &      8900.0     &       0.670      &        7297.6   &     3.126(14)    &       8452.5            \\
  4.71      &      3880.0     &       3.995     &      5330.0     &       1.442     &      8170.0     &       0.702      &        6731.3   &     3.317(14)    &       8152.2            \\
  5.18      &      3790.0     &       4.000     &      5110.0     &       1.496     &      7790.0     &       0.720      &        6531.2   &     3.518(14)    &       7860.7            \\
  5.70      &      3740.0     &       4.000     &      5000.0     &       1.514     &      7600.0     &       0.724      &        6403.5   &     3.762(14)    &       7639.7            \\
  6.27      &      3710.0     &       3.995     &      4960.0     &       1.505     &      7490.0     &       0.724      &        6318.4   &     4.036(14)    &       7450.4            \\
  6.90      &      3690.0     &       4.000     &      4950.0     &       1.491     &      7420.0     &       0.724      &        6265.5   &     4.337(14)    &       7274.3            \\
  7.59      &      3690.0     &       4.000     &      4980.0     &       1.464     &      7450.0     &       0.715      &        6242.4   &     4.660(14)    &       7105.5        \\
  8.35       &     3700.0      &      4.000      &     5030.0      &      1.437      &     7490.0      &      0.711       &       6244.4    &    5.009(14)     &      6943.2             \\
  9.19       &     3720.0      &      3.995      &     5100.0      &      1.405      &     7580.0      &      0.702       &       6268.5    &    5.390(14)     &      6788.7             \\
 10.11       &     3740.0      &      4.000      &     5200.0      &      1.365      &     7690.0      &      0.693       &       6306.2    &    5.808(14)     &      6649.1             \\
 11.12       &     3800.0      &      3.896      &     5310.0      &      1.324      &     7850.0      &      0.679       &       6359.4    &    6.245(14)     &      6500.4             \\
 12.23       &     3890.0      &      3.689      &     5420.0      &      1.288      &     8010.0      &      0.665       &       6410.8    &    6.726(14)     &      6365.4             \\
 13.45       &     4000.0      &      3.450      &     5560.0      &      1.239      &     8150.0      &      0.656       &       6471.9    &    7.213(14)     &      6207.1             \\
 14.80      &      4100.0     &       3.251     &      5690.0     &       1.193     &      8320.0     &       0.643      &        6524.5   &     7.733(14)    &       6047.1       \\
 16.28      &      4200.0     &       3.057     &      5800.0     &       1.157     &      8420.0     &       0.634      &        6550.7   &     8.278(14)    &       5885.1            \\
 17.91      &      4300.0     &       2.867     &      5910.0     &       1.121     &      8510.0     &       0.625      &        6576.9   &     8.860(14)    &       5725.8            \\
 19.70      &      4390.0     &       2.700     &      5980.0     &       1.094     &      8540.0     &       0.620      &        6585.9   &     9.458(14)    &       5556.5            \\
 21.67      &      4460.0     &       2.574     &      6040.0     &       1.072     &      8520.0     &       0.620      &        6585.2   &     1.008(15)    &       5381.5            \\
 23.84      &      4520.0     &       2.470     &      6090.0     &       1.054     &      8470.0     &       0.625      &        6574.4   &     1.069(15)    &       5192.2            \\
 26.22      &      4580.0     &       2.371     &      6100.0     &       1.049     &      8370.0     &       0.634      &        6555.7   &     1.130(15)    &       4989.7            \\
 28.84      &      4610.0     &       2.312     &      6090.0     &       1.049     &      8230.0     &       0.647      &        6522.3   &     1.193(15)    &       4789.5       \\
 31.72      &      4620.0     &       2.276     &      6070.0     &       1.049     &      8120.0     &       0.656      &        6474.0   &     1.256(15)    &       4583.8            \\
 34.89      &      4630.0     &       2.213     &      6010.0     &       1.054     &      7900.0     &       0.675      &        6394.5   &     1.324(15)    &       4393.7            \\
 38.38      &      4490.0     &       2.290     &      5750.0     &       1.108     &      7420.0     &       0.720      &        6240.1   &     1.397(15)    &       4214.1            \\
 42.22      &      4400.0     &       2.299     &      5590.0     &       1.126     &      7150.0     &       0.733      &        6090.4   &     1.472(15)    &       4034.7            \\
 46.44      &      4240.0     &       2.358     &      5330.0     &       1.162     &      6660.0     &       0.778      &        5863.9   &     1.550(15)    &       3862.5            \\
 51.08      &      4100.0     &       2.367     &      5110.0     &       1.175     &      6320.0     &       0.792      &        5665.2   &     1.622(15)    &       3676.1            \\
 56.19       &     4020.0      &      2.222      &     4920.0      &      1.153      &     5950.0      &      0.801       &       5453.5    &    1.699(15)     &      3499.6       \\
 61.81      &      4020.0     &       1.924     &      4780.0     &       1.090     &      5580.0     &       0.805      &        5268.7   &     1.778(15)    &       3330.0            \\
 67.99      &      4090.0     &       1.586     &      4690.0     &       1.013     &      5260.0     &       0.805      &        5140.7   &     1.855(15)    &       3158.2            \\
 74.79      &      4230.0     &       1.270     &      4660.0     &       0.927     &      5040.0     &       0.792      &        5063.5   &     1.933(15)    &       2992.2            \\
\hline
\end{tabular}
\end{center}
\end{table*}

\label{lastpage}

\end{document}